\documentclass[longauth]{aa} % for the long lists of affiliations 
\usepackage{graphicx}
\usepackage[varg]{txfonts}
\usepackage{natbib}
\usepackage[colorlinks=true,linkcolor=blue,citecolor=blue]{hyperref}

\begin{document} 

\title{A planetary system with a sub-Neptune planet \\ in the habitable zone of TOI-2093}

   \author{J.~Sanz-Forcada\inst{\ref{i:CAB}}
   \thanks{\email{jsanz@cab.inta-csic.es}}
     \and E.~Gonz\'alez-\'Alvarez\inst{\ref{i:CAB},\ref{i:UCM}}
     \and M.\,R.~Zapatero~Osorio \inst{\ref{i:CAB}}
     \and J.\,A.~Caballero\inst{\ref{i:CAB}}
     \and V.\,J.\,S.~B\'ejar  \inst{\ref{i:IAC}, \ref{i:ULL}}
     \and E.~Herrero \inst{\ref{i:ICE2}}
     \and C.~Rodr\'{\i}guez-L\'opez\inst{\ref{i:IAA}}
     \and K.\,R.~Sreenivas  \inst{\ref{i:ArielU}, \ref{i:USyd}}
     \and L.\,Tal-Or \inst{\ref{i:ArielU}, \ref{i:ArielU2}}
     \and S.~Vanaverbeke\inst{\ref{i:VVS}, \ref{i:IRIS}}
     \and A.\,P.~Hatzes \inst{\ref{i:TLS}}
     \and R.~Luque\inst{\ref{i:UCI}, \ref{i:NHFP}}
     \and E.~Nagel \inst{\ref{i:IAG}}
     \and F.\,J.~Pozuelos \inst{\ref{i:IAA}}
     \and D.~Rapetti \inst{\ref{i:Ames}, \ref{i:USRA}}
     \and A.~Quirrenbach \inst{\ref{i:LSW}}
     \and P.\,J.~Amado \inst{\ref{i:IAA}}
     \and M.~Blazek \inst{\ref{i:CAHA}}
     \and I.~Carleo \inst{\ref{i:OAT}}
     \and D.~Ciardi \inst{\ref{i:NASA}}
     \and C.~Cifuentes  \inst{\ref{i:CAB}}
     \and K.~Collins \inst{\ref{i:CfA}}
     \and Th.~Henning \inst{\ref{i:MPIA}}
     \and D.\,W.~Latham\inst{\ref{i:CfA}}
     \and J.~Lillo-Box \inst{\ref{i:CAB}}
     \and E.~Marfil \inst{\ref{i:UPM}}
     \and D.~Montes \inst{\ref{i:UCM}}
     \and J.\,C.~Morales \inst{\ref{i:ICE1},\ref{i:ICE2}}
     \and F.~Murgas \inst{\ref{i:IAC}}
     \and G.~Nowak \inst{\ref{i:NCU}}
     \and E.~Pall\'e\inst{\ref{i:IAC}, \ref{i:ULL}}
     \and S.~Reffert \inst{\ref{i:LSW}}
     \and A.~Reiners \inst{\ref{i:IAG}}
     \and I.~Ribas \inst{\ref{i:ICE1},\ref{i:ICE2}}
     \and R.\,P.~Schwarz \inst{\ref{i:CfA}}
     \and A.~Schweitzer \inst{\ref{i:HS}}
   }

   \institute{
     Centro de Astrobiolog\'{i}a, CSIC-INTA, Camino bajo
     del Castillo s/n, 
     28692 Villanueva de la Ca\~nada, Madrid, Spain \label{i:CAB}
     \and
     {Departamento de F\'isica de la Tierra y Astrof\'isica \& IPARCOS 
     Instituto de F\'isica de Part\'iculas y del Cosmos, Facultad de Ciencias
	F\'isicas, Universidad Complutense de Madrid, Plaza de Ciencias 1, 
	28040 Madrid, Spain} \label{i:UCM}
     \and
     {Instituto de Astrof\'isica de Canarias, 
	38205 La Laguna, Tenerife, Spain} \label{i:IAC}
     \and
	{Departamento de Astrof\'isica, Universidad de La Laguna, 
	38206 La Laguna, Tenerife, Spain} \label{i:ULL}
     \and
     {Institut d'Estudis Espacials de Catalunya, 
	08860 Castelldefels, Barcelona, Spain} \label{i:ICE2}
     \and
     {Instituto de Astrof\'isica de Andaluc\'ia, CSIC, 
       Glorieta de la Astronom\'ia s/n, 18008 Granada, Spain} \label{i:IAA}
     \and Department of Physics, Ariel University, Ariel 40700, Israel \label{i:ArielU}
     \and Sydney Institute for Astronomy, School of Physics, University of Sydney, Sydney, NSW 2006, Australia \label{i:USyd}
     \and
     Astrophysics, Geophysics, And Space Science Research Center, Ariel University, Ariel 40700, Israel\label{i:ArielU2}
     \and  Vereniging Voor Sterrenkunde, Oostmeers 122 C, 8000 Brugge, Belgium \label{i:VVS}
     \and Public observatory ASTROLAB IRIS, Provinciaal Domein “De Palingbeek”, Verbrandemolenstraat 5, 8902 Zillebeke, Ieper, Belgium \label{i:IRIS}
     \and {Th\"uringer Landessternwarte Tautenburg, Sternwarte 5, 
       07778 Tautenburg, Germany} \label{i:TLS}
     \and
     Department of Astronomy \& Astrophysics, University of Chicago, Chicago, IL 60637, USA \label{i:UCI}
     \and
     NHFP Sagan Fellow \label{i:NHFP}  
     \and {Institut f\"ur Astrophysik und Geophysik, Georg-August-Universit\"at,
       Friedrich-Hund-Platz 1, 37077 G\"ottingen, Germany} \label{i:IAG}
     \and NASA Ames Research Center, Moffett Field, CA 94035, USA \label{i:Ames}
     \and Research Institute for Advanced Computer Science, Universities Space Research Association, Washington, DC 20024, USA \label{i:USRA}
     \and
     {Landessternwarte, Zentrum f\"ur Astronomie der Universit\"at Heidelberg,
	   K\"onigstuhl 12, 69117 Heidelberg, Germany} \label{i:LSW}
     \and {Centro Astron\'omico Hispano en Andaluc\'ia, 
	Observatorio Astron\'omico de Calar Alto, 
	Sierra de los Filabres, 04550 G\'ergal, Almer\'ia, Spain} \label{i:CAHA}
     \and INAF -- Osservatorio Astrofisico di Torino, Via Osservatorio 20, 10025, Pino Torinese, Italy \label{i:OAT}
     \and    
        NASA Exoplanet Science Institute, Caltech/IPAC, Mail Code 100-22, 1200 E. California Blvd., Pasadena, CA 91125, USA \label{i:NASA}
     \and
        Center for Astrophysics \textbar \ Harvard \& Smithsonian, 60 Garden Street, Cambridge, MA 02138, USA \label{i:CfA}
     \and Max-Planck-Institut f\"ur Astronomie, 
     K\"onigstuhl 17, 69117 Heidelberg, Germany \label{i:MPIA}
        \and
     Departamento de Ingeniería Topográfica y Cartografía, E.T.S.I. en Topografía, Geodesia y Cartografía, Universidad Politécnica de Madrid, 28031 Madrid, Spain \label{i:UPM}
     \and
     {Institut de Ci\`encies de l'Espai, CSIC, 
	c/ de Can Magrans s/n, Campus UAB, 
	08193 Bellaterra, Barcelona, Spain} \label{i:ICE1}
     \and
     Institute of Astronomy, Faculty of Physics, Astronomy and Informatics, Nicolaus Copernicus University, Grudzi\c{a}dzka 5, 87-100 Toru\'n, Poland \label{i:NCU}
     \and {Hamburger Sternwarte, Gojenbergsweg 112, 21029 Hamburg, Germany} \label{i:HS}
     }
    
   \date{Received 11 July 2025 / Accepted 12 September 2025}

  \abstract
  % context heading (optional)
  % {} leave it empty if necessary  
   {}
  % aims heading (mandatory)
 {We aim to confirm and measure the mass of the transiting planet candidate around the K5\,V star TOI-2093, previously announced by the Transiting Exoplanet Survey Satellite (TESS) project.}
   {We combined photometric data from 32 sectors between 2019 and 2024 with 86 radial velocity measurements obtained with the CARMENES spectrograph over a period of 2.4 years, along with a series of ground-based, broadband photometric monitoring campaigns to characterize the host star and the transiting planet candidate, as well as to search for additional planets in the system. Our data indicate that TOI-2093 is a main-sequence star located at a distance of 83\,pc, with solar metallicity, and a rotation period of 43.8$\pm$1.8\,d.}
  % results heading (mandatory)
{We have confirmed the planetary nature of the TESS transiting planet candidate, named TOI-2093\,c, through the detection of its Keplerian signal in the spectroscopic data. We measured a planetary radius of 2.30$\pm$0.12\,R$_\oplus$, a Neptune-like mass of $15.8^{+3.6}_{-3.8}$\,M$_\oplus$, and an orbital period of 53.81149$\pm$0.00017\,d. This makes TOI-2093\,c the smallest exoplanet known in the habitable zone of a main-sequence FGK star. Given its size and relatively high density, TOI-2093\,c belongs to a class of planets with no analog in the Solar System. In addition, the CARMENES data revealed the presence of a second planet candidate with a minimum mass of 10.6$\pm$2.5\,M$_\oplus$ and an orbital period of 12.836$\pm$0.021\,d. This inner planet, which we designated TOI-2093\,b, shows no detectable photometric transit in the TESS light curves. The orbital planes of the two planets are misaligned by more than 1.6$\degr$ despite the near 4:1 mean-motion resonance of their orbital periods.}
   {}

   \maketitle
%
%-------------------------------------------------------------------
\nolinenumbers

%vvvvvvvvvvvvvvvvvvvvvvvvvvvvvvvvvvvvvvvvvvvvvvvvvvvvv
\section{Introduction}
The search for exoplanets, particularly those with temperate conditions, is crucial for understanding planetary formation and the potential for habitability beyond our Solar System. Identifying such planets requires precise detection methods, especially for low-mass planets, where high-precision radial-velocity (RV) measurements are essential. The combination of the NASA Transiting Exoplanet Survey Satellite (TESS), which detects transits, and the fiber-fed Calar Alto high-Resolution search for M dwarfs with Exoearths with Near-infrared and optical \'Echelle Spectrographs (CARMENES), which provides precise RV measurements, greatly improves our ability to characterize these planets. Previous studies combining transit and RV methods have successfully identified and characterized temperate exoplanets, demonstrating the effectiveness of this approach in the refinement of planetary parameters and the assessment of their atmospheric properties \citep[e.g.,][]{luq19,tri21,kuz24}.

The habitability of planets can potentially be tested in planets located in the so-called habitable zone (HZ) of their host stars. 
Twenty-three planets in the HZ of main-sequence FGK stars have been reported to date, representing a very small fraction ($\sim$3.1\%) of all planets with a known mass and radius (see further discussion in Sect.~\ref{sec:discussion}).
Among them only two have a radius of $R_{\rm p}<5$\,R$_\oplus$, and only six have $M_{\rm p}<24$\,M$_\oplus$ (Neptune has 3.9\,R$_\oplus$ and 17.1\,M$_\oplus$).
The discovery of new planets orbiting the HZ of FGK stars is thus of great
importance.

A planet candidate orbiting \object{TOI-2093} (K5\,V, Table~\ref{tab:starpar}) was reported by the TESS project on 15 July, 2020. The transits indicated a size of $\sim$2.2\,R$_\oplus$ and its orbital period implied an equilibrium temperature of $\sim$329\,K (for a Bond albedo of 0.3). With this temperature we considered the candidate planet to be located within the HZ of the star, increasing the interest in determining its mass and improving its orbital parameters.
We started an RV campaign to calculate its mass, and we complemented the TESS observations with a number of photometric observations from different ground-based telescopes to better characterize the stellar photometric variability.

Details of the photometric and RV observations used in this research are described in Sect.~\ref{sec:Observations}. The host star and its photometric and spectroscopic analyses are introduced in Sect.~\ref{sec:The star}. RV and TESS photometric analyses and results on the TOI-2093 planetary system are detailed in Sect.~\ref{sec:TOI-2093 planetary system}. The results are discussed in the framework of current knowledge in Sect.~\ref{sec:discussion}. In Sect.~\ref{sec:conclusions} we wrap up the conclusions.

%
%----------------------------------  Table 1
\begin{table}
\caption[]{Stellar parameters of TOI-2093.}\label{tab:starpar}
\begin{tabular}{lcr}
  \hline \hline
  \noalign{\smallskip}
%--------------------------------------------------------------
Parameter & Value & Reference \\
\hline
%--------------------------------------------------------------
\multicolumn{3}{c}{\em Basic identifiers and data}\\
TYC & 4450-1440-1 & Tycho-2 \\
2MASS & J20105711+7052098 & 2MASS \\
TIC & 402898317 & Stas18 \\
Karmn & J20109+708 & Cab16 \\
Sp. type & K5\,V & This work \\
$G$ (mag) & $11.3640 \pm 0.0028 $ &  \textit{Gaia} DR3\\
$J$ (mag) & $9.715 \pm 0.021$ & 2MASS\\
\noalign{\smallskip}
\multicolumn{3}{c}{\em Astrometry and kinematics}\\
\noalign{\smallskip}
$\rm \alpha$ (J2000) & 20:10:57.12 &  \textit{Gaia} DR3\\
$\rm \delta$ (J2000) & +70:52:09.9 &  \textit{Gaia} DR3\\    
$\mu _{\alpha} \cos \delta$ ($\rm mas\,yr^{-1}$) & $-36.834 \pm 0.017$
& \textit{Gaia} DR3\\
$\mu _{\delta}$ ($\rm mas\,yr^{-1}$) & $ 55.115 \pm 0.018$ & \textit{Gaia} DR3\\
$\varpi$ (mas) & $12.104 \pm 0.012 $ &  \textit{Gaia} DR3 \\
$d$ (pc) & $\rm 82.615 \pm 0.084$  &  \\
$\gamma$ (km s$^{-1}$)   & $\rm 14.67 \pm 0.33$ & \textit{Gaia} DR3\\
$U$ (km s$^{-1}$)  & $-13.637 \pm 0.077$ & Cor24 \\
$V$ (km s$^{-1}$)  & $  3.12 \pm 0.30$ & Cor24 \\
$W$ (km s$^{-1}$)  & $ 26.34 \pm 0.11$ & Cor24 \\
Galactic population  & Thin disk & Cor24 \\
\noalign{\smallskip}
\multicolumn{3}{c}{\em Fundamental parameters and abundances}\\
\noalign{\smallskip}
$T_{\rm eff}$ (K)  & $\rm 4426  \pm 85 $ & This work \\
$\log g$ (cgs) & $\rm 4.87 \pm 0.06$ & This work \\
$\rm [Fe/H]$ (dex) & $\rm -0.08\pm 0.02$ & This work\\
$L$ $(\rm L_{\odot})$ & $\rm 0.1838 \pm 0.0029$ & This work\\
$R$ ($\rm R_{\odot}$) & $\rm 0.729 \pm 0.029$ & This work\\
$M$ ($\rm M_{\odot}$) & $\rm 0.745 \pm 0.034$ & This work\\  
\noalign{\smallskip}
\multicolumn{3}{c}{\em Activity and age}\\
\noalign{\smallskip}
$v \sin i_\star$ ($\mathrm{km\,s^{-1}}$) & $<2.0$  & This work\\
$P_{\rm rot}$ (d) & $43.8 \pm 1.8$ &  This work \\
$\log{L_{\rm X}}$ (erg\,s$^{-1}$)  & $\sim$27.3 & This work \\
Age (Gyr) & $\sim$6.6 & This work \\
%--------------------------------------------------------------
\hline
\end{tabular}
\tablebib{
  2MASS: \cite{skr06};
  Cab16: \cite{cab16}; 
  \textit{Gaia} DR3: \cite{gaia};
  Stas18: \cite{sta18};  
  Tycho-2: \cite{tyc00};
  Cor24: \cite{cor24}.
}
\end{table}
%--------------------------------------------------------------

%vvvvvvvvvvvvvvvvvvvvvvvvvvvvvvvvvvvvvvvvvvvvvvvvvvvvv
\section{Observations}
\label{sec:Observations}

\subsection{TESS photometry}
TESS \citep{ric15} is an all-sky
transit survey whose principal goal is the detection of planets
smaller than Neptune orbiting bright stars that can be followed up
with ground-based observations. These can then lead to the determination of
planetary masses and atmospheric compositions.
All TESS observations are made available to the community
including calibrated light curves from the simple aperture photometry \citep[SAP][]{twi10,mor20} fluxes, and presearch data conditioned simple aperture photometry (PDCSAP) fluxes
\citep{smi12,stu12,stu14}. The SAP flux is the flux after summing the calibrated pixels within the TESS optimal photometric aperture, and the PDCSAP flux corresponds to the SAP flux values corrected for instrumental and environmental variations. 
The optimal photometric aperture is defined such that the stellar signal has a high signal-to-noise ratio ($S/N$), with minimum contamination from the background. The TESS detector bandpass spans from 600 to 1000\,nm and is centered on the Cousins $I$ band (786.5\,nm), approximately.

%
%----------------------------------  Table 2
\begin{table}
\caption[]{{\em TESS} sectors containing TOI-2093 observations and their elapsed time\tablefootmark{a}.}\label{tab:tesslog}
  \tabcolsep 1.5 pt
  \begin{small}
    \begin{tabular}{cccc|cccc}
\hline \hline
\noalign{\smallskip}
%--------------------------------------------------------------
Sector & Starting date & $\Delta t$ (d) & Transit &
Sector & Starting date & $\Delta t$ (d) & Transit \\
\hline
\noalign{\smallskip}
%--------------------------------------------------------------
15 & 2019-08-15 & 26.1 & \dots & 55 & 2022-08-05 & 27.2 & \dots \\
16 & 2019-09-12 & 24.7 & (\checkmark) & 56 & 2022-09-02 & 27.9 & \checkmark  \\
17 & 2019-10-08 & 25.0 & \dots & 57 & 2022-09-30 & 28.8 & \checkmark  \\
18 & 2019-11-03 & 24.4 & \checkmark  & 58 & 2022-10-29 & 27.7 & \dots \\
19 & 2019-11-28 & 25.1 & \dots & 59 & 2022-11-26 & 26.4 & \checkmark  \\
20 & 2019-12-24 & 26.3 & \checkmark  & 60 & 2022-12-23 & 25.7 & \dots \\
22 & 2020-02-19 & 27.2 & (\checkmark) & 73 & 2023-12-07 & 26.9 & \checkmark  \\
23 & 2020-03-19 & 26.8 & \dots &  75 & 2024-01-30 & 27.7 & \checkmark \\
24 & 2020-04-16 & 26.5 & \checkmark  &  76 & 2024-02-27 & 27.1 & \dots \\
25 & 2020-05-14 & 25.7 & \dots & 77 & 2024-03-25 & 28.1 & \checkmark \\
40 & 2021-06-25 & 28.2 & \checkmark  & 78 & 2024-05-03 & 18.1 & \dots \\ 
48 & 2022-01-28 & 28.1 & \checkmark  & 79 & 2024-05-22 & 27.1 & \checkmark \\
49 & 2022-02-26 & 26.8 & \dots & 83 & 2024-09-05 & 25.0 & \checkmark \\
50 & 2022-03-26 & 26.2 & (\checkmark)  & 84 & 2024-10-01 & 25.8 & \dots \\
52 & 2022-05-19 & 24.4 & (\checkmark)  & 85 & 2024-10-27 & 25.5 & \checkmark  \\
53 & 2022-06-13 & 25.0 & \dots & 86 & 2024-11-21 & 26.6 & \dots \\
%--------------------------------------------------------------
\hline
\end{tabular}
\end{small}
 \tablefoot{
   \tablefoottext{a}{Sectors including a TOI 2093 b transit are
     marked with a \checkmark. The parentheses indicate the transits affected by instrumental problems or observing gaps, excluded from the analysis.}}
\end{table}
%--------------------------------------------------------------

TESS observed TOI-2093 in 2 min short-cadence integrations in 32 different sectors between 2019 and 2024 (Table~\ref{tab:tesslog}). The image data were reduced and analyzed by the Science Processing Operations Center \citep[SPOC,][]{jen16} at NASA Ames Research Center.
In a first analysis, TOI-2093 was announced as a TESS object of interest (TOI) by the TESS Science Office (TSO) via the dedicated Massachusetts Institute of Technology TESS data alerts public website\footnote{\url{https://tess.mit.edu/toi- releases/}}, where a planet candidate was identified at an orbital period of 53.8108$\pm$0.0002\,d.
The SPOC conducted a multi-sector transit search of Sectors 15-23 on 6 May, 2020 with an adaptive, noise-compensating matched filter \citep{jen02,jen10,jen20}, producing a threshold crossing event (TCE) for which an initial limb-darkened transit model was fit \citep{li19} and a suite of diagnostic tests were conducted to help make or break the planetary nature of the signal \citep{twi18}. The TSO reviewed the vetting information and issued an alert on 15 July, 2020 \citep{gue21}. The signal was repeatedly recovered as additional observations were made in following searches of sectors 15-25, 15-40, 15-50, 15-55, 15-60, 15-78, and 15-86, and the transit signature passed all the diagnostic tests presented in the data validation reports. Based on the latter search, the host star is located within $1.67\pm3.15\arcsec$ of the source of the transit signal, the estimated radius of the planet is 2.45$\pm$0.40\,R$_{\oplus}$, and the transit duration is 4.61$\pm$0.30\,h with a depth of 980.5$\pm$46.6 ppm.
Due to the long orbital period of the transiting planet, the 27\,d observational baseline of each TESS sector, and the presence of observational gaps, transits are not present in all TESS sectors.

The light curve files for the different sectors were downloaded from the Mikulski Archive for Space Telescopes. 
As a preliminary validation step, we examined the SAP and PDCSAP fluxes generated by the TESS pipeline to verify their reliability for subsequent analysis. This was accomplished by inspecting potential contamination from nearby stellar sources within the photometric aperture. Fig.~\ref{fig:tess_tpf_plot} displays the target pixel files (TPFs) of one sector of TOI-2093, overplotted with Gaia Data Release 3 (DR3) sources \citep{Gaia18} using the {\tt tpfplotter}\ tool \citep{Aller20}. The analysis confirmed the absence of additional {\em Gaia} sources within the pipeline-defined aperture, supporting the conclusion that the extracted TESS light curves are free from contamination from nearby stars and are thus suitable for accurate photometric modeling. The PDCSAP light curves with detected transits are displayed in Fig.~\ref{fig:tesslc_sectors}.

%
%----------------------------------  Table 3
\begin{table}
\caption[]{RV and ground-based photometric observations used in this work.}\label{tab:obslog}
    \tabcolsep 2.8 pt
\begin{tabular}{lccccc}
  \hline \hline
  \noindent{\smallskip}
%--------------------------------------------------------------
Instrument & Starting date & $\Delta t$ (d) & $N_{\rm obs}$ & Filter & {\it rms}\\
\hline
%--------------------------------------------------------------
CARMENES & 2021-04-03 & 904 & 86 & VIS & 6.8\,m\,s$^{-1}$ \\
CARMENES & 2021-04-03 & 904 & 86 & NIR & 27.8\,m\,s$^{-1}$ \\
OSN & 2022-04-30 & 237 & 1740 & $R$, $V$ & 6.6\,mmag \\
TJO & 2021-02-25 & 687 & 782 & $R$ & 7.3\,mmag \\
LCOGT-1m & 2022-05-05 & 542 & 417 & $B$ & 6.7\,mmag \\
LCOGT-0.4m & 2022-05-07 & 227 & 1422 & $V$ & 21.5\,mmag \\
%--------------------------------------------------------------
\hline
\end{tabular}
\end{table}
%--------------------------------------------------------------

\subsection{Keck adaptive optics imaging}
To further assess the contamination by nearby companions, we used
the KeckII~10\,m telescope to obtain a Near Infrared Camera (NIRC2)
adaptive optics image in the $K$-band of TOI-2093.
Our target was observed on 23 June, 2022. 
No companions were detected with a contrast of 7.2~mag at $0.5\arcsec$ (Fig.~\ref{fig:nirspec}).
TOI-2093 is located at a distance of 82.615\,pc (Table~\ref{tab:starpar}), and it has absolute magnitudes $M_G$=6.7787$\pm$0.0036, $M_J$=5.130$\pm$0.021, and $M_{K_s}$=4.390$\pm$0.018. 
Assuming that the interstellar extinction towards TOI-2093 is negligible since its optical and near-infrarred colors are consistent with those of mid-K dwarfs, and applying the mass--magnitude relations from \citet{pec13}, we found no evidence for a companion warmer than spectral type L0--L2 in the field between $0.5\arcsec$ and $3.0\arcsec$ (41--250\,au) from TOI-2093.
That is, there is no stellar companion with mass larger than 0.075\,M$_\odot$ at these separations.
Brown dwarf companions cannot be excluded based on these high-spatial resolution data.

\subsection{Other photometric monitoring}
To characterize the stellar variability and determine the stellar rotation period, we monitored TOI-2093 using broadband optical filters and ground-based telescopes, and also explored various photometric catalogs (Table~\ref{tab:obslog}).
We discarded photometric $V$-band data acquired within the All-Sky Automated Survey
for Supernovae (ASAS-SN) project \citep{asassn}, covering 1700\,d starting on 18 March, 2014, because the data have too large error bars.
We also discarded 85 e-EYE (shorthand for Entre Encinas y Estrellas)\footnote{\url{https://www.e-eye.es/}} $V$-filter data covering 122\,d starting on 17 May, 2022 because of their large dispersion as compared with data from other telescopes. e-EYE is a telescope-hosting facility located in Fregenal de la Sierra, Badajoz, Spain.

\subsubsection{OSN}
The T90 telescope at OSN (Observatorio de Sierra Nevada, Spain) was employed to collect photometric data of TOI-2093. T90 is a 90 cm Ritchey-Chrétien telescope equipped with a CCD camera Andor Ikon-L DZ936N-BEX2-DD 2k$\times$2k with a resulting field of view (FOV) of $13.2\arcmin \times 13.2\arcmin$.  The camera is based on a back-illuminated CCD chip, with high quantum efficiency from ultraviolet to near-infrared. This camera also includes thermo-electrical cooling down to $-100\,\degr$C for negligible dark current.

Our set of observations, collected in Johnson $V$ and $R$ filters,
typically consisted of 20 exposures in each filter per night, of 60\,s
and 50\,s, respectively. All CCD measurements were obtained by the method of aperture photometry and no detector binning was applied during the observations.
Each CCD frame
was corrected in a standard way for bias and flat-fielding. Different
aperture sizes were also tested to choose the best ones for
our observations. A number of nearby and relatively bright stars
within the frames were selected as check stars to choose the
best ones to be used as reference stars. A $3\sigma$ clipping to remove
outliers was applied to the light curves.

\subsubsection{TJO}
We observed TOI-2093 with the 0.8\,m Telescopi Joan Or\'o
\citep[TJO,][]{col10} at the Observatori del Montsec in Lleida,
Spain. Images with exposure times of 60\,s were obtained using the
Johnson $R$ filter of the LAIA imager, a 4k$\times$4k CCD with a FOV
of $30\arcmin$, and a scale of $0.4\arcsec$ $\rm pixel^{-1}$.
Raw frames were corrected for dark current and bias and were
flat-fielded using the {\tt ICAT} pipeline \citep{col06} of the TJO.
The differential photometry was extracted with the {\tt AstroImageJ}
software \citep{col17} by using an optimal aperture size that
minimized the root mean square ({\it rms}) of the resulting relative fluxes. To derive the
differential photometry of TOI-2093, we selected the 10 brightest
comparison stars in the field that did not show any variability. Then,
we employed our own pipelines to remove outliers and measurements
affected by poor observing conditions or with a low $S/N$.

\subsubsection{LCOGT}
Las Cumbres Observatory Global Telescope \citep[LCOGT,][]{bro13} 1-m telescopes 
were employed to observe
TOI-2093 in the $B$ photometric filter. An exposure time of 12\,s was
used for each image. The aperture photometry was also performed using
{\tt AstroImageJ}. Using trial and error, we adopted an aperture size of 10
pixels, selected four reference stars of similar counts, and performed
photometry. 
The 0.4\,m telescopes of the network were equipped with the 
$V$ photometric filter. The data from nights with bad photometric conditions in $B$ or $V$ were excluded.

%-------------------------------------
\subsection{CARMENES spectroscopy}
\label{subsec:CARMENES spectroscopy}
We used CARMENES \citep{spie14} 
to spectroscopically observe TOI-2093, to measure the mass
of its planetary companions. We obtained a total of 86  high-resolution spectra from 3 April, 2021, through 24 September, 2023.
CARMENES is installed at the
3.5\,m telescope of the Calar Alto Observatory in Almer\'{i}a, 
Spain. It was specifically designed to deliver high-resolution
spectra at optical (resolving power $\mathcal{R} \approx$ 94\,600) and
near-infrared ($\mathcal{R} \approx$ 80\,400) wavelengths from 520 to
1710\,nm. CARMENES has two different channels, one for the
optical (the VIS channel) and the other for the near-infrared (the NIR
channel), with a break at 960\,nm \citep{spie14}. All
data were acquired with integration times of 1800\,s (which is the
maximum exposure employed for precise RV measurements) and
followed the data flow of the CARMENES GTO program
\citep{rib23}. CARMENES raw data are automatically
reduced with the {\tt caracal} pipeline
\citep{caracal}. A correction from the presence of telluric lines was then applied following \citet{nag23}. Relative RVs were extracted separately for
each \'echelle order using the {\tt serval} software
\citep{serval}. The final VIS and NIR RVs for each epoch were
computed as the weighted RV mean over all \'echelle orders of the
respective spectrograph, with an {\it rms} as described in Table~\ref{tab:obslog}. 
The mean $S/N$ of the TOI-2093 spectra in the order 75 ($\lambda\lambda$8090-8271\,\AA) was 54.

In this work, we used only the CARMENES VIS-channel RVs (Table~\ref{tab:toi2093_rv_act_data}, internal precision of 1.2\,m\,s$^{-1}$, mean error bar of 3.5\,m\,s$^{-1}$) to search for planetary candidates. For stars as blue as TOI-2093 the NIR channel delivers relative RVs with significantly lower precision than the VIS channel \citep{rein18, Bau20}. The expected RV semi-amplitudes of the transiting planets ($\le \,\rm 5\,m\,s^{-1}$) are below the typical NIR precision of our data (mean error bar of 14\,$\rm m\,s^{-1}$). Nevertheless, the CARMENES NIR RVs for TOI-2093 are briefly analyzed to investigate stellar rotation rather than to search for additional planets.

At a high spectral resolution, the profile of the stellar lines may
change due to photospheric and chromospheric activity, which has an
impact on accurate RV measurements. It is therefore crucial to disentangle the
effects of stellar activity from the Keplerian signals. The CARMENES
{\tt serval} pipeline provides measurements for a number of spectral
features that are considered indicators of stellar activity (Table~\ref{tab:toi2093_rv_act_data}), such as
the differential line width (dLW), H$\alpha$,  the Ca\,{\sc ii}
infrared triplet $\lambda\lambda$8498, 8542, 8662\,\AA~(IRT), and the
chromatic index (CRX). The CRX quantifies the slope of RV measurements as a function of wavelength, and it is used as an indicator of the presence of stellar
active regions \citep{serval}. All these indices may have a chromospheric component
in active K dwarfs.

%vvvvvvvvvvvvvvvvvvvvvvvvvvvvvvvvvvvvvvvvvvvvvvvvvvvvvvvvvvvvvvvvvvvvvvvv
\section{The star TOI-2093}\label{sec:The star}

\subsection{Stellar parameters}\label{sect:starproperties}
All relevant parameters collected from the literature, and derived in this work, are listed in Table~\ref{tab:starpar}. 
The stellar atmospheric parameters ($T_{\rm eff}$, $\log{g}$, and [Fe/H]) of TOI-2093 were analyzed using CARMENES VIS and NIR template spectra by \cite{mar21} through the {\tt SteParSyn}\footnote{\url{https://github.com/hmtabernero/SteParSyn/}} code
\citep{tab22}, using the line list and model grid described by \citet{mar21}. 
The derived values, presented in Table~\ref{tab:starpar}, are $T_{\rm eff}$ = 4482\,K, $\log{g}$ = 4.87~dex, and [Fe/H] = $-0.08$~dex. To determine the stellar radius, we employed the spectroscopic $T_{\rm eff}$, the bolometric luminosity calculated following \cite{cifu20}, and the Stefan-Boltzmann law, obtaining $R = 0.7291$\,R$_\sun$. The mass of TOI-2093 was then estimated using the mass-radius relationship established by \cite{sch19} for eclipsing binary stars, resulting in $M = 0.745$\,M$_\sun$. These parameters indicate that TOI-2093 is a K5\,V star, consistent with spectral features in the CARMENES spectra.

We compared the mass and radius adopted for the star (Table~\ref{tab:starpar}) with values obtained using various mass–luminosity relations available in the literature for low-mass stars \citep[e.g.,][]{mann15,mann19,ben16}. Discrepancies of up to 15\% were observed. Our mass estimate is consistent within 1$\sigma$ of the value inferred from PARSEC evolutionary models for solar metallicity \citep{chen14}. According to these models, based on the bolometric luminosity and effective temperature of TOI-2093, the star is predicted to have a mass of 0.72\,$\pm$\,0.04 M$_\odot$ and an age between 40 and 60 Myr. However, this age appears too young given the observed stellar activity level and the long rotation period (Sect~\ref{subsec:Stellar variability}). To investigate further, we examined the CARMENES VIS template spectrum around 6708\,\AA\ to search for lithium absorption, which would be expected if TOI-2093 were as young as predicted by the evolutionary models \citep{ran21}. No lithium was detected in the spectrum, with a 3$\sigma$ upper limit on the equivalent width of the resonance doublet of 10 m\AA. We thus ruled out a young age for TOI-2093. This lithium upper limit implies an age greater than 300\,Myr based on the relation between lithium equivalent width and age for stars of similar temperature established by \citet{jef23}.

The CARMENES template spectrum of the star shows a small amount of
emission near the core of the \ion{Ca}{ii}~IRT lines, and some
filling-in of H$\alpha$. This indicates the presence of a rather low,
but not negligible, chromospheric activity. No X-rays or UV
observations are available to date, except for a marginal GALEX NUV detection.
The low level of activity is consistent
with the rotational $v \sin i$ upper limit of 2\,km\,s$^{-1}$ measured from
the spectrum (Table~\ref{tab:starpar}). Using the stellar radius and assuming a rotation axis inclination of $\sim$90$\degr$, we derive a rotation period $P_{\rm rot}\ga$18.4\,d.

\subsection{Stellar rotation period}\label{subsec:Stellar variability}
Before examining the CARMENES RVs to search for or confirm planets, we first analyzed all available photometric and spectroscopic (activity index) time series. This allowed us to identify the characteristic frequencies of the star's variability and to determine the rotation period of TOI-2093.

The analysis of TESS light curves (with an observational baseline of $\sim$27\,d per sector) does not allow for a proper 
determination of such long rotation periods. The ground-based photometric
data (Table~\ref{tab:obslog}) span larger time windows, allowing
for a better search of the period. Despite that, the rotation signal
is rather small, and we needed to combine photometry with the 
activity indicators from the CARMENES spectra to better determine the
rotation period.

\subsubsection{LCOGT 1-m, LCOGT 0.4-m, OSN, and TJO light curve analysis}
\begin{figure}
  \centering
  \includegraphics[width=0.48\textwidth]{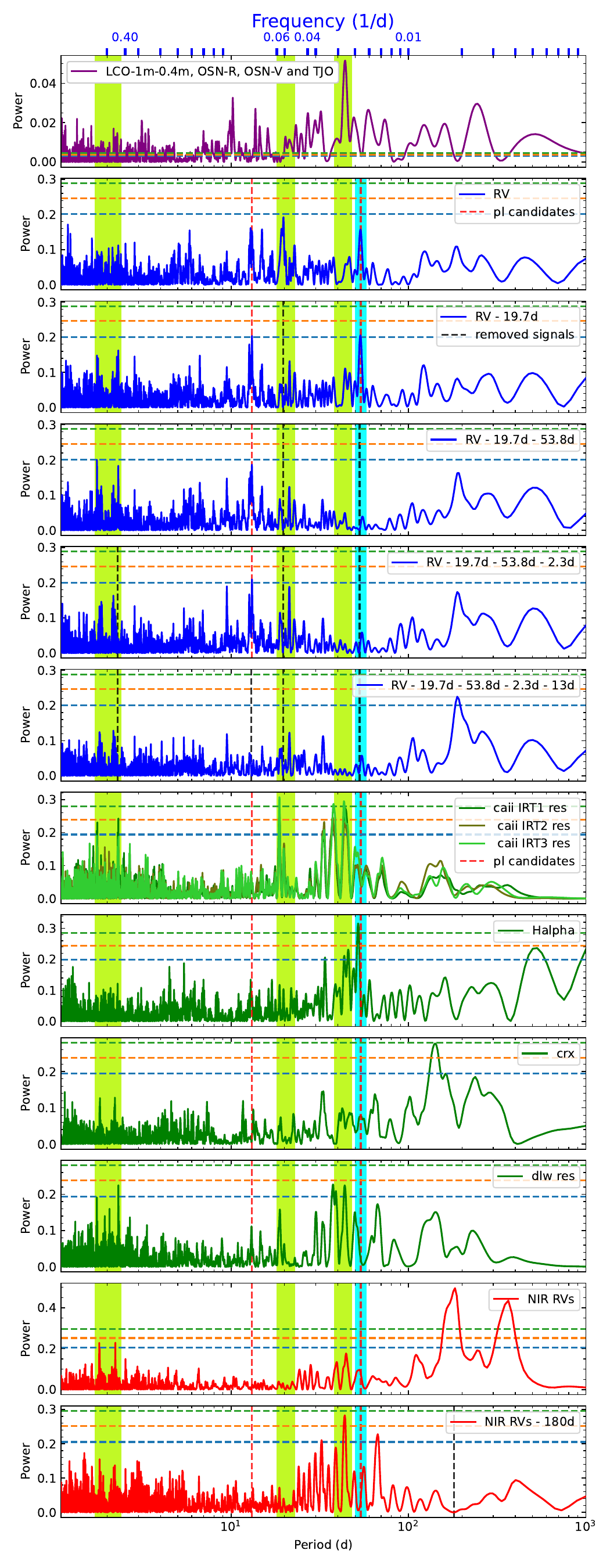}
  \caption{GLS periodograms of all photometric and spectroscopic data. Vertical dashed lines mark planetary orbital periods (transiting planet on a cyan band), and vertical green bands indicate stellar activity timescales (see text). Horizontal dashed lines show FAP levels (0.1\,\% in green, 1\,\% in orange, and 10\,\% in blue) across all panels.}\label{fig:glsgeneral}  
\end{figure}

To determine the photometric rotation period of TOI-2093, we analyzed all available photometric data and applied Gaussian process (GP) modeling. As a preliminary step, we first combined the OSN, TJO, and LCOGT light curves, accounting for the offsets between instruments, and generated the GLS periodogram. Our photometric follow-up spans approximately 1000\,d, and the GLS periodogram reveals a long-term trend. We then removed this trend and generated the GLS periodogram of the residuals, which is shown in Fig.~\ref{fig:glsgeneral}. The next peak corresponding to a signal of 43.8$\pm$1.8\,d (Table~\ref{tab:starpar}) is identified in the GLS periodogram, where the error corresponds to the half width of the peak. This rotation period is consistent with the
$v \sin i$ upper limit, and the observed low level of activity (see below).

Unlike GLS periodograms, which are based on static sinusoidal models, GPs are capable of capturing the quasi-periodic nature of stellar activity. We used a kernel composed of two simple harmonic oscillators, commonly referred to as the double simple harmonic oscillator (dSHO) kernel. All light curves shared the GP hyperparameters with the only exception of the amplitude of the variability, which was allowed to vary from filter to filter. The physical theory underlying the dSHO kernel, along with the empirical knowledge accumulated in the literature \citep[e.g.,][]{for17,ang18,nic22,sto23} and practical applications \citep[e.g.,][]{dav19,gil20,gon23}, supports the interpretation of its hyperparameters.

The inferred photometric rotation period using the dSHO kernel is $P\rm_{rot} =42.8 \pm$0.2\,d, in good agreement with the period detected
from the photometric GLS periodogram. The fit quality factor ($Q_0$ = 2.22$\pm$0.08) indicates a low coherent signal, consistent with the quasi-periodic nature of stellar rotation modulated by evolving active regions. The relative contribution of the secondary oscillation ($f$ = 0.79$\pm$0.03) suggests that the second mode contributes considerably to the profile, probably due to asymmetries and evolving spots. The difference in quality factors ($dQ$ = 0.95$\pm$0.41) points to both modes having a smooth damping without overfitting. The GP amplitudes vary across instruments (Fig.~\ref{fig:photometrylc}), with stronger signals observed in the OSN-$V$ and OSN-$R$ filters, suggesting wavelength-dependent variability likely driven by stellar activity.

%----------------------------------  Fig. 1
%----------------------------------------------
%

%
%----------------------------------  Fig. 2
\begin{figure*}
  \centering
  \includegraphics[width=0.45\textwidth]{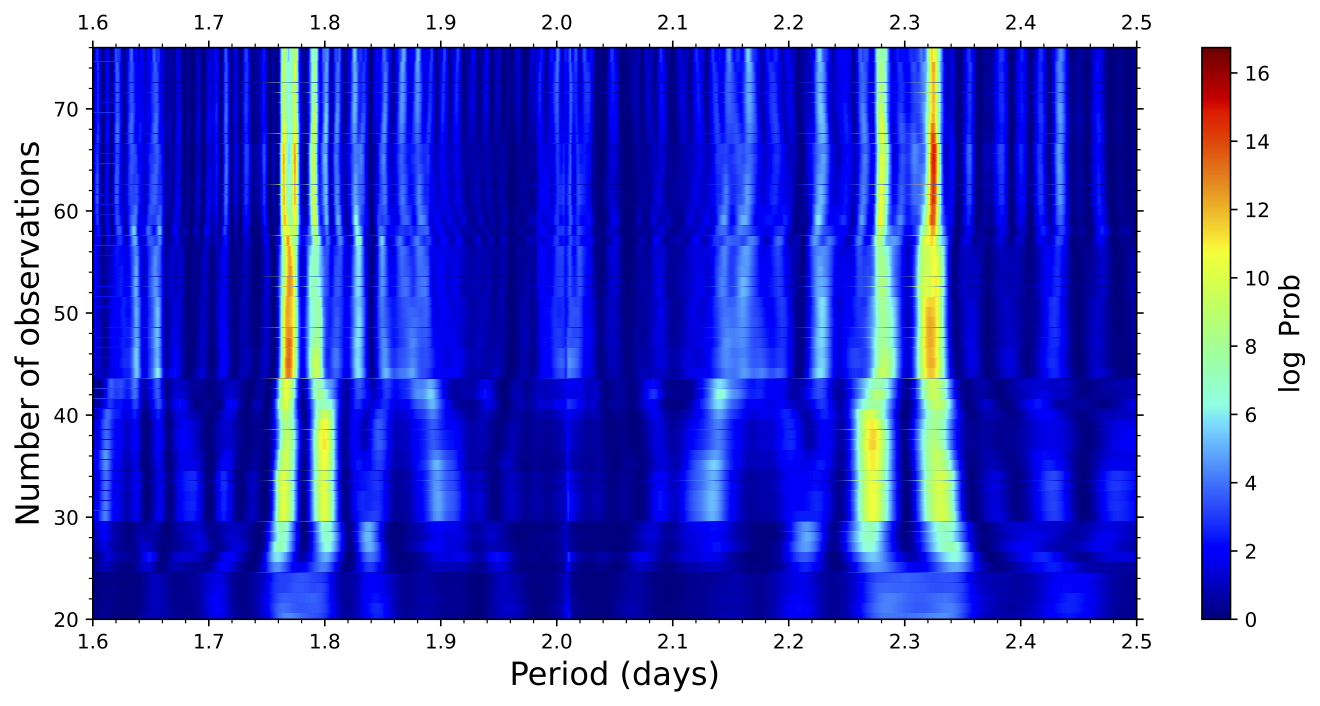}
  \includegraphics[width=0.45\textwidth]{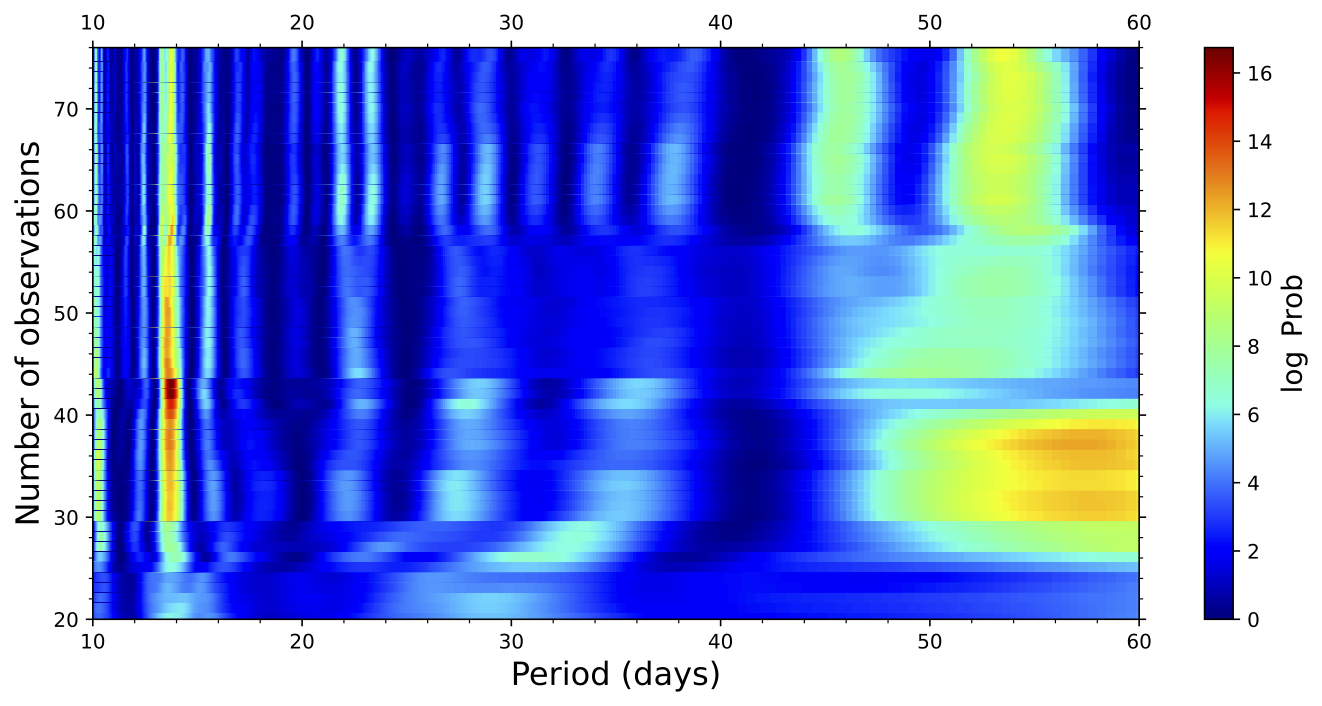}
  \caption{Evolution of the s-BGLS periodogram of the CARMENES RV data
    around 2\,d ({\em left}) and in the region between 10 and 60\,d ({\em right}). Both s-BGLS periodograms have the 1-year alias of the first harmonic of the stellar rotation period (19.7\,d) removed.}\label{fig:toi2093_color_s-BGLS}
\end{figure*}
%----------------------------------------------
%

\subsubsection{CARMENES activity indicators and NIR RV}
\label{subsubsec:CARMENES activity indicators}

We calculated the GLS periodograms for several stellar activity indicators provided by the CARMENES {\tt serval} pipeline. The spectroscopic GLS periodograms are displayed in Fig. \ref{fig:glsgeneral}: they cover the interval 1.1--1000\,d, limits related to the Nyquist frequency (0.5\,d$^{-1}$) and the duration of the observations.
The highest peaks of the CARMENES observations window function are seen at $\sim$1\,d and approximately one year (Fig. \ref{fig:window_carmenes_vis}).
There are three \ion{Ca}{ii}~IRT indices, one for each atomic line, whose corresponding GLS periodograms are presented together in Fig.~\ref{fig:glsgeneral}. This periodogram exhibits peaks around $\sim$2, $\sim$19, and $\sim$40\,d. The $\sim$19, and $\sim$40\,d signals reach false alarm probability (FAP) levels above 0.1\%. These two signals are related to the stellar $P_{\rm rot}$ and $P_{\rm rot}$/2. A similar situation is seen in the dLW GLS periodogram with a lower significance. 
The GLS of the H$\alpha$ index is not so conclusive, showing some peaks at the 1\% FAP level around $\sim$40\,d and the highest one close to the orbital period of the transiting planet at $\sim$54\,d. Meanwhile, the GLS periodogram of the CRX index do not display any significant peaks exceeding the defined FAP thresholds at $\sim$19 and $\sim$40\,d signals, but shows a broad peak greater than 150\,d. The two shorter period signals ($\sim$2\,d) observed in the \ion{Ca}{ii}~IRT, CRX, and dLW residual GLS periodograms are also detected in the RV data, indicating a stellar origin.

For a given rotation period, we can expect to find signals at the value of $P_{\rm rot}$ and its harmonics, such as $P_{\rm rot}/2$, as well as for some aliases (Sect.~\ref{subsec:GLS periodograms}). 
In this case a rotation period of $P_{\rm rot}=43.8$\,d is significant, with
a 1-year alias at $\sim$39.1\,d and its half period at $\sim$19.6\,d (this can also be interpreted as the 1-year alias of $P_{\rm rot}/2$): the three are identifed in the \ion{Ca}{ii}~IRT lines, and the
latter also in the VIS RV domain.

We fit a dSHO kernel to the different activity indicators showing stellar activity signals (\ion{Ca}{ii}~IRT, dLW, and H$\alpha$), using the same procedure as for photometric data. The amplitude of the kernel was fit independently for each dataset, with the rest of hyperparameters shared across all indicators (Fig.~\ref{fig:activityindicators}). The inferred spectroscopic rotation period using the dSHO kernel is $P_{\rm rot} = 42.8^{+1.4}_{-1.3}$\,d.

As is explained in Sect. \ref{subsec:CARMENES spectroscopy}, 
the CARMENES NIR channel data were used to investigate stellar rotation, 
as stellar activity may also impact the NIR wavelengths.
The two bottom panels of Fig.~\ref{fig:glsgeneral} show the GLS periodogram of the original CARMENES NIR RVs and the residuals after subtracting the first harmonic of one year at 180\,d. The highest peak in the residuals corresponds to a 43.7$\pm$1.2\,d signal,
consistent with the photometric and stellar activity indicators analysis, as is listed in Table~\ref{tab:rotationperiods}.

We roughly estimated the stellar age by means of the relation between rotation period and stellar age provided by \citet{mam08}. We derived a value of $6.63^{+0.89}_{-0.77}$\,Gyr for TOI-2093, but since this relation was calibrated with faster rotators (thus younger stars) than TOI-2093, its age determination is quite uncertain.
We estimated the expected X-ray luminosity using the rotation
period and the relations of \citet{wri11}\footnote{The alternative
  relations of \citet{rei14} led to $\log L_{\rm X}$(erg\,s$^{-1}$) =27.1.},
resulting in $\log L_{\rm X}$(erg\,s$^{-1}$)=27.3. We also
estimated the extreme ultraviolet (EUV) stellar luminosity using the
\citet{san25} relations. This led to
$\log L_{\rm  EUV}$(erg\,s$^{-1}$)=28.10 in the 100--920~\AA\ range,
and 27.63 in the 100--504~\AA\ range.

%vvvvvvvvvvvvvvvvvvvvvvvvvvvvvvvvvvvvvvvvvvvvvvvvvvvvv
\section{The planetary system TOI-2093}\label{sec:TOI-2093 planetary system}

\subsection{TESS light curve analysis}\label{sec:tesslc}
The transit least squares (TLS) algorithm \citep{hip19}, implemented via the {\tt astropy.timeseries} Python package, was used iteratively to detect transiting exoplanets in TESS PDCSAP time-series data with possible orbital periods between 0.5 and 60\,d. The strongest signal identified in the first TLS iteration correspond to the 53.8\,d planet candidate previously reported by TESS. After removing this planet candidate, no additional transit signals were detected in the residuals. A total of 32 TESS sectors were analyzed, with useful planetary transits detected in 14 of them, as summarized in Table~\ref{tab:tesslog}.
The TLS algorithm yielded preliminary estimates of the central transit time ($T_0$), transit duration ($t_{\rm 14}$), and transit depth ($\delta$), all of which are in agreement with the parameters reported in the TESS alerts for TOI-2093.

We searched for possible transit timing variations (TTVs) in the TESS data by modeling the individual transits of the 53.8\,d planet with {\tt juliet} \citep{juliet}. We did not find any periodic variability or linear trend with time in the TTVs (Fig.~\ref{fig:ttvs}). We set a 3$\sigma$ upper limit of 33.6\,min on the TTVs spanning $\sim$1840\,d of observation.

\subsection{CARMENES VIS RV analysis}
\subsubsection{GLS periodograms}\label{subsec:GLS periodograms}

We searched for the Keplerian signal of the 53.8\,d planetary candidate in our CARMENES RV time series by fitting and subtracting sinusoidal components corresponding to activity-related signals or other planetary candidates. First, we analyzed the spectral window function of the CARMENES VIS RVs (Fig. \ref{fig:window_carmenes_vis}) to identify potential aliasing and artifacts in the RV periodograms \citep[e.g.,][]{daw10,sto20}. The strong peaks of the window function may introduce alias 
peaks, sometimes stronger than the true signals, in the RV periodogram at frequencies according 
to the expression $f_{\rm alias}$ = $f_{\rm true} \pm m f_{\rm
  window}$, where $m$ is an integer, $f_{\rm true}$ is the frequency
identified in the RV periodogram, and $f_{\rm window}$ is the
frequency from the window function \citep{dee75}. 
Typical aliases 
affecting ground-based observations are associated with the year,
synodic month, sidereal day, and solar day. In our spectroscopic
window function, the highest peaks occur at $\sim$1\,d and $\sim$2\,d,
close to half a year and close to a year.

The first iteration of the GLS periodogram of the original CARMENES VIS RVs, shown in the top panel of Fig.~\ref{fig:glsgeneral}, reveals the strongest periodic signal at 19.6\,d, which we attribute to half the stellar rotation period (Sect.~\ref{subsec:Stellar variability}). After modeling and subtracting this signal using a sinusoidal function, the second iteration of the GLS periodogram applied to the RV residuals reveals two peaks with a FAP lower than 10\% (second panel of Fig.~\ref{fig:glsgeneral}), at approximately 13\,d and 53.8\,d. The latter matches the orbital period of the transiting planet detected by TESS. The CARMENES data thus provide spectroscopic confirmation of a new planet orbiting TOI-2093 with an orbital period of 53.8\,d.

After removing the 53.8\,d signal, the third iteration of the GLS periodogram of the RV residuals (the third panel of Fig.~\ref{fig:glsgeneral}) shows that the 13\,d  peak persists, while two new prominent peaks emerge at 1.75\,d and 2.31\,d. The 13\,d peak is actually a doublet, consisting of signals at 12.8 and 13.2\,d, which are 1\,yr aliases of each other. None appears to be associated with either known stellar activity signals or aliasing effects, suggesting that the system may be in a 4:1 mean-motion resonance (or commensurability).
We named the two planets \object{TOI-2093\,b} (the 13\,d RV planet candidate) and \object{TOI-2093\,c} (the 53.8\,d transiting confirmed planet).
In the final iteration (fifth panel of Fig.~\ref{fig:glsgeneral}), after removing the 13\,d RV signal associated with TOI-2093\,b, no additional signals exceed the 10\% FAP threshold. 

We verified the stability of the RV signals at 1.75, 2.31\,d, and
13\,d over the entire observational time baseline by producing the
stacked Bayesian generalized Lomb-Scargle periodogram
\citep[s-BGLS,][]{mor15} shown in
Fig.~\ref{fig:toi2093_color_s-BGLS}. The significance or probability
of these signals increases with time until a stable level is reached
at a certain number of observations. For this, we employed the
CARMENES RVs free of the signal at 19.6\,d related to stellar rotation.
The short-period signals at 1.75 and 2.31\,d are related by the 1\,d alias and are also present in the \ion{Ca}{ii}~IRT periodograms, indicating that they are likely associated with stellar activity. Therefore, they are not considered further in this study until Sect.~\ref{sect:constraints}. 

%vvvvvvvvvvvvvvvvvvvvvvvvvvvvvvvvvvvvv
\subsubsection{RV models}\label{sec:Radial velocity models}
To address the reality of a multiplanetary system around TOI-2093 using the CARMENES VIS RV data, we computed several models using the Python-based {\tt juliet} code and compared them using log-evidence statistics. The {\tt juliet} code internally employs the {\tt radvel} package \citep{ful18} to model Keplerian RV signals. Stellar activity-induced variability was accounted for using GPs, specifically with a dSHO kernel.

%
%----------------------------------  Table 4
\begin{table}[]
\centering
\tabcolsep 0.5pt
\renewcommand{\arraystretch}{1.2}
\begin{small}
\caption{Comparison of different {\tt juliet} RV models for TOI-2093, only using the CARMENES RV data.}
\label{tab:toi2093_rv_model_comparison}
\begin{tabular}{lccc}
\hline \hline
%--------------------------------------------------------------
Model  & $P_{\rm rot,GP}$ & $ \Delta \ln \mathcal{Z}$ & $P_{\rm orb}$ \\ 
       &  (d)  & & (d) \\
\hline
%--------------------------------------------------------------
GP  ($\ln \mathcal{Z} = -281.4$)  & 19.8$\pm$0.6 & 0 &  \dots\\
1pl$_{\rm(13\,d)}+$GP               & 19.8$\pm$0.6 & 1.6 & 12.84$^{+0.05}_{-0.12}$ \\
1pl$_{\rm(53\,d)}+$GP                & 19.7$\pm$0.4 & 3.9 & 53.810$\pm$0.009 \\
2pl$_{\rm(13\,d,\,53\,d)}+$GP          & 19.5$\pm$0.4 & 5.6 & 12.84$\pm$0.04, 53.809$\pm$0.009 \\
3pl$_{\rm(1-300\,d,\,13\,d,\,53\,d)}+$GP & 19.6$\pm$0.4 & 5.6 & 12.84$\pm$0.03, 53.809$\pm$0.09, 190$^{+50}_{-5}$ \\
\noalign{\smallskip}    
%--------------------------------------------------------------
\hline
\end{tabular}
\end{small}
\renewcommand{\arraystretch}{1.}
\tablefoot{All models were run with $P_{\rm rot,GP}$ prior $\mathcal{U}(15, 100)$.}
\end{table}
%--------------------------------------------------------------

We based the selection of the best model on the rules defined by
\cite{tro08} for the Bayesian model log-evidence, $\ln{\mathcal{Z}}$:
if $\Delta \ln{\mathcal{Z}}<3,$ the two models are
indistinguishable and none is preferred, while if
$\rm \Delta \ln{\mathcal{Z}}>3,$ the model with the largest Bayesian log-evidence
is favored. The different approaches that we performed are summarized
in Table~\ref{tab:toi2093_rv_model_comparison}. All included a base model (BM) consisting of RV offset and jitter plus a GP with $P_{\rm rot}$ in the interval 15--100\,d to simulate the stellar activity due to stellar rotation. Since stellar activity in the CARMENES VIS RVs appears at $P_{\rm rot}/2$ rather than at $P_{\rm rot}$ (see Sect. \ref{subsec:GLS periodograms}), we adopted a wide uniform prior on the GP period that included both values, allowing the model to identify the most likely stellar modulation in the VIS RVs.
The other ingredients for the different RV models are the following:
        ({\em i}) one Keplerian signal (1pl$+$GP);
        ({\em ii}) two Keplerian signals at $\sim$13\,d, and 53.8\,d (2pl$+$GP); and 
        ({\em iii}) three Keplerian signals at $\sim$13\,d, 53.8\,d, and an uninformed search of another Keplerian signal (3pl$+$GP). 

The dSHO kernel consistently converged to a mean $P_{\rm rot}/2$ value of 19.7$\pm$0.5\,d in all models (Table~\ref{tab:toi2093_rv_model_comparison}). 
We tested multiple kernels (exponential-sine-squared, quasi-periodic, and dSHO), all of which produced consistent results. We adopted the dSHO kernel as it offers a more physically motivated and accurate representation of stellar activity through its ability to model a damped, periodic variability.
We also tested two different GP models assuming a normal prior on $P_{\rm rot}$ centered at $\sim$20\,d and $\sim$44\,d. We found that the dSHO kernel with a GP period of 19.7\,d is clearly favored by the VIS RV data instead of the $\sim$44\,d period.
The GP modeling quality factor, $Q_0$, was moderately high ($Q_0 \sim$5.3), indicating a coherent and recurrent signal. Additionally, the shape parameters of the kernel ($f \sim$0.70 and $dQ \sim$ 1.08) suggest a relatively broad secondary mode and low damping difference, which is consistent with active regions evolving over time and producing quasi-periodic variability. The high amplitude (GP$_{\sigma}\sim$5.87\,m\,s$^{-1}$) indicates that the stellar activity signal in RVs is higher than the Keplerian signature of the transiting planet.
The stellar variability fundamental period was independently inferred to be $\sim$44\,d from photometric and spectroscopic activity indicators, and also corroborated from the CARMENES NIR RV data.

The resulting log-evidence for each model is provided in Table~\ref{tab:toi2093_rv_model_comparison}. The third column shows the difference in Bayesian log-evidence relative to the stellar activity–only model. Including one or two planetary signals improves the statistical evidence, with the two-planet model (orbital periods of 12.8 and 53.8 d) being the most favored. However, the improvement over the one-planet model is modest. Adding a third Keplerian signal with a free orbital period between 1 and 300 d does not make the fit to the RV data better. This analysis yields planetary parameters for TOI-2093\,b and c that are fully consistent with those reported for the joint study in the next section. We also remark that introducing a second planet does not affect the parameters of the first.

Our analysis of the CARMENES VIS RV data provides that there is
statistical evidence supporting the presence of at least one planet around TOI-2093. The confirmed outer planet is transiting, while the inner planet candidate does not show any transits in the TESS light curve. TESS data folded in phase with the orbital period of TOI-2093\,b, after subtracting the 53.8\,d signal previously identified in the light curve, is displayed in Fig.~\ref{fig:lc_vs_phase_12.8}. No further transiting planet candidates are present in the TESS light curves with a radius larger than 0.87\,R$_\oplus$ (assuming a 10$\sigma$ detection threshold).

%vvvvvvvvvvvvvvvvvvvvvvvvvvvvvvvvvvvvv
\subsection{Joint photometric and spectroscopic analysis}
\label{sec:jointphotometryradvel}
To determine the planetary parameters of TOI-2093\,b and c, we carried out a combined photometric and spectroscopic analysis using TESS and CARMENES VIS RV data.
The fit was performed using the {\tt juliet} code to model Keplerian RV signals, and  the {\tt batman} package \citep{Krei15} to model the transit light curves, and a total of 5000 live points.

To reduce computational time, given the large number of TESS sectors and the absence of detectable transits for TOI-2093\,b, we limited the photometric dataset by selecting time intervals of 26\,h centered around the expected mid-transit times of TOI-2093\,c. We then removed photometric outliers using a sigma-clipping algorithm, and flattened the resulting light curve segments by applying a Savitzky-Golay filter with a third-order polynomial. The planetary transits were excluded from this flattening process.

%
%----------------------------------  Fig. 3
\begin{figure}
  \centering
  \includegraphics[width=0.45\textwidth]{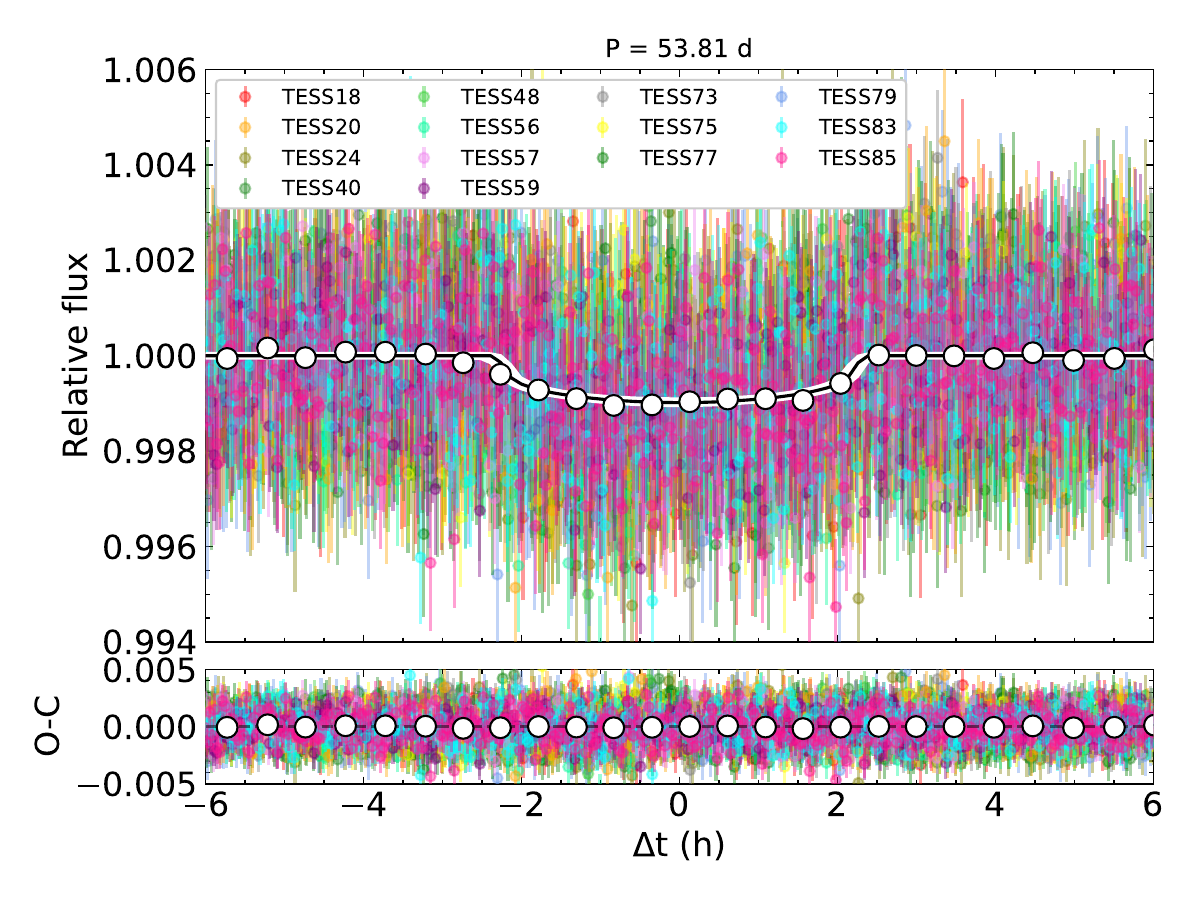}
  \caption{TESS light curves folded in phase with the orbital period of TOI-2093\,c assuming $e=0$. Binned data are plotted as white circles. The best transit model is shown with a black line (uncertainties in white).}\label{fig:tesslc} 
\end{figure}
%----------------------------------------------
%

We used the $r_{1}$ and $r_{2}$ parameterization \citep{espin18} instead of determining the relative radii and the impact parameters ($b$) of the planet directly: $r_{1}$ and $r_{2}$ can vary between 0 and 1 and are defined to explore all the physically meaningful ranges for $R_{p}/R_{\star}$ and $b$. We defined a prior on the stellar density, $\rho_{\star}$ instead of the scaled semimajor axis of the planets, $a$. In this way, a single value of $\rho_{\star}$ is defined for the system. 
Based on the results obtained so far, we modeled the dataset with two planetary Keplerian signals, the 53\,d transiting planet and the 13\,d RV-only planet signal.
We also employed a dSHO GP kernel to account for stellar activity in the RV data centered at 19.7\,d. We assumed that both planets are in circular orbits. 
The model was called 2pl+GP. The posteriors from the 2pl+GP fit that we adopted as planetary parameters for the TOI-2093 system are presented in Table \ref{tab:planets_params}. For clarity, the posteriors of the remaining fit parameters can be found in Table \ref{tab:priors+posteriors}. 

\begin{figure*}[]
	\centering
	\includegraphics[width=0.9\textwidth]{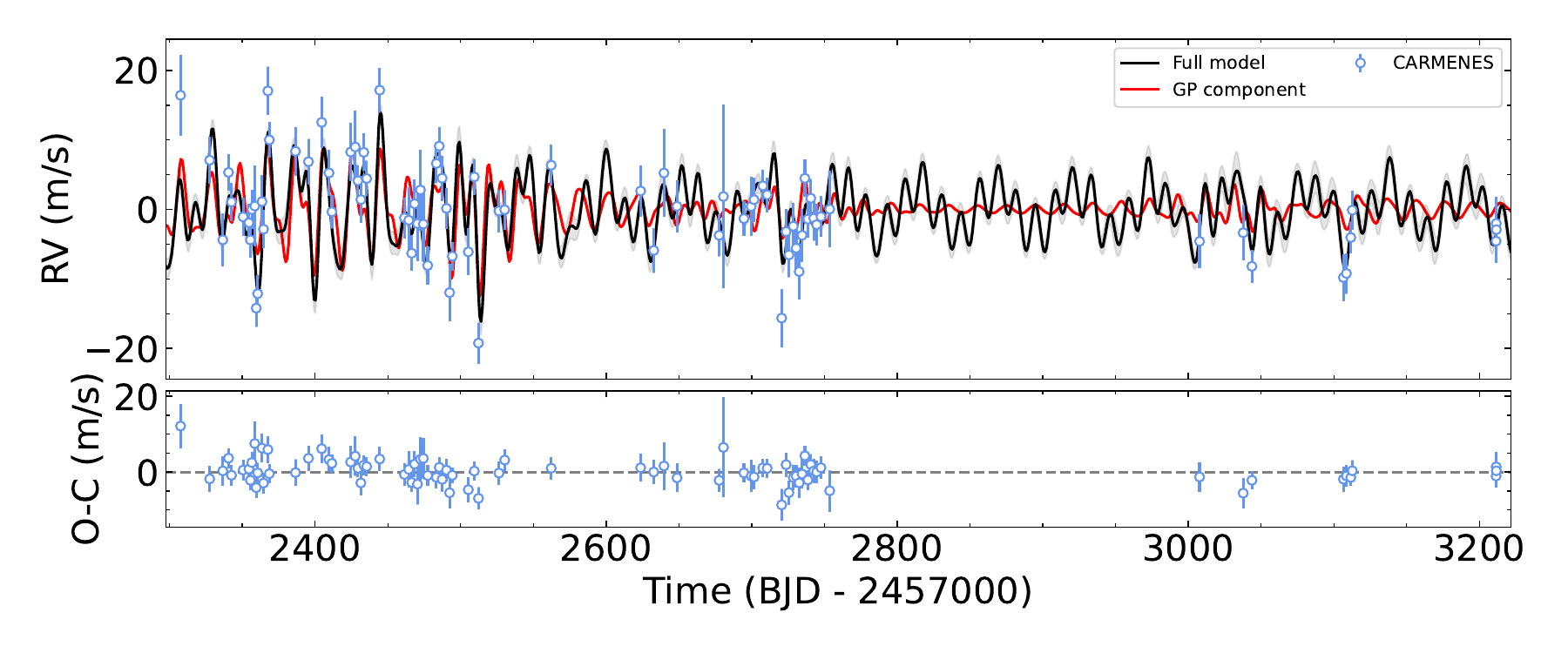}
	\caption{TOI-2093 CARMENES RVs (blue dots) and the best model assuming two planets in the system (black line) obtained from the combined photometric and spectroscopic fit. The top panel shows the entire RV time series as a function of the time. The red line shows the GP component that models the stellar activity. The bottom panel shows the RV residuals after subtracting the full model.}
    \label{fig:toi2093_RVmodel_vs_time}
\end{figure*}

\begin{figure}[]
  \centering
  \includegraphics[width=\columnwidth]{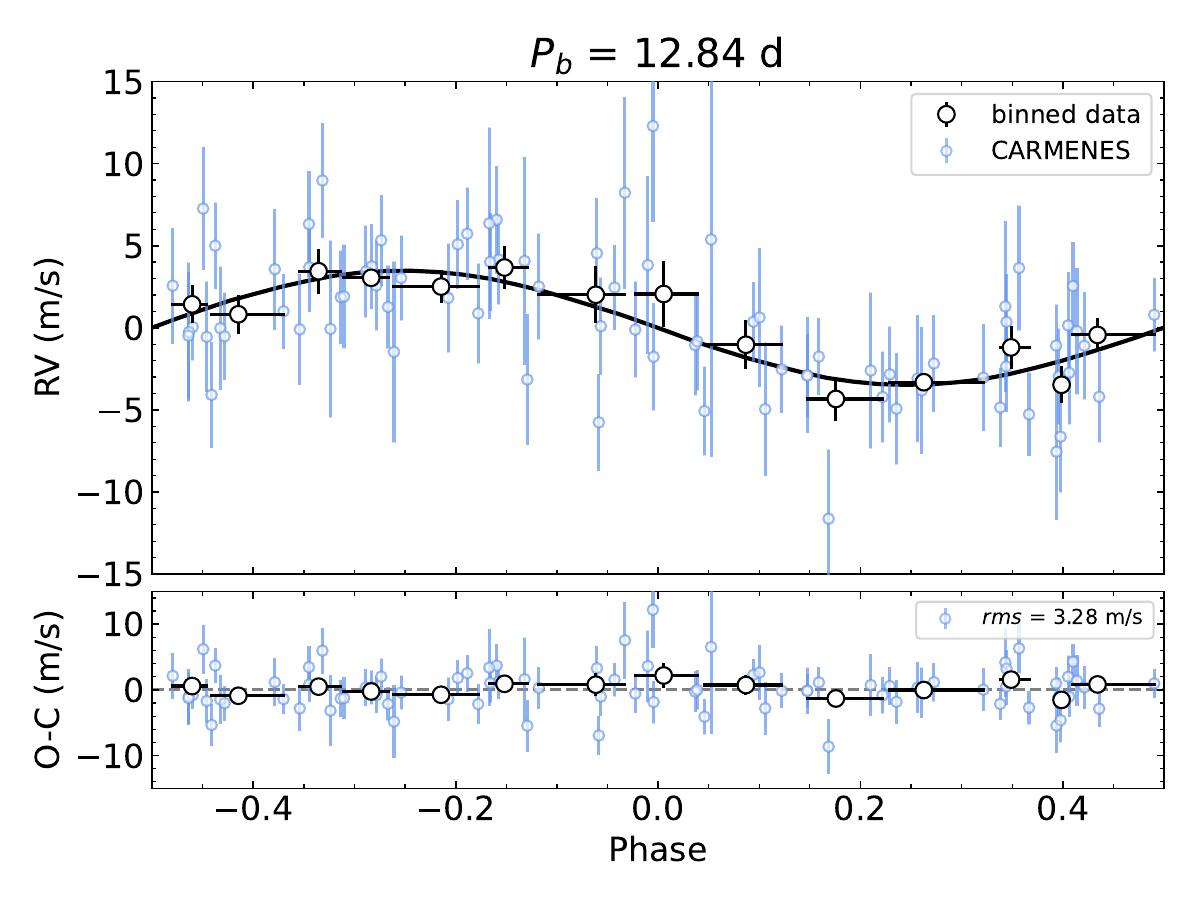}
  \includegraphics[width=\columnwidth]{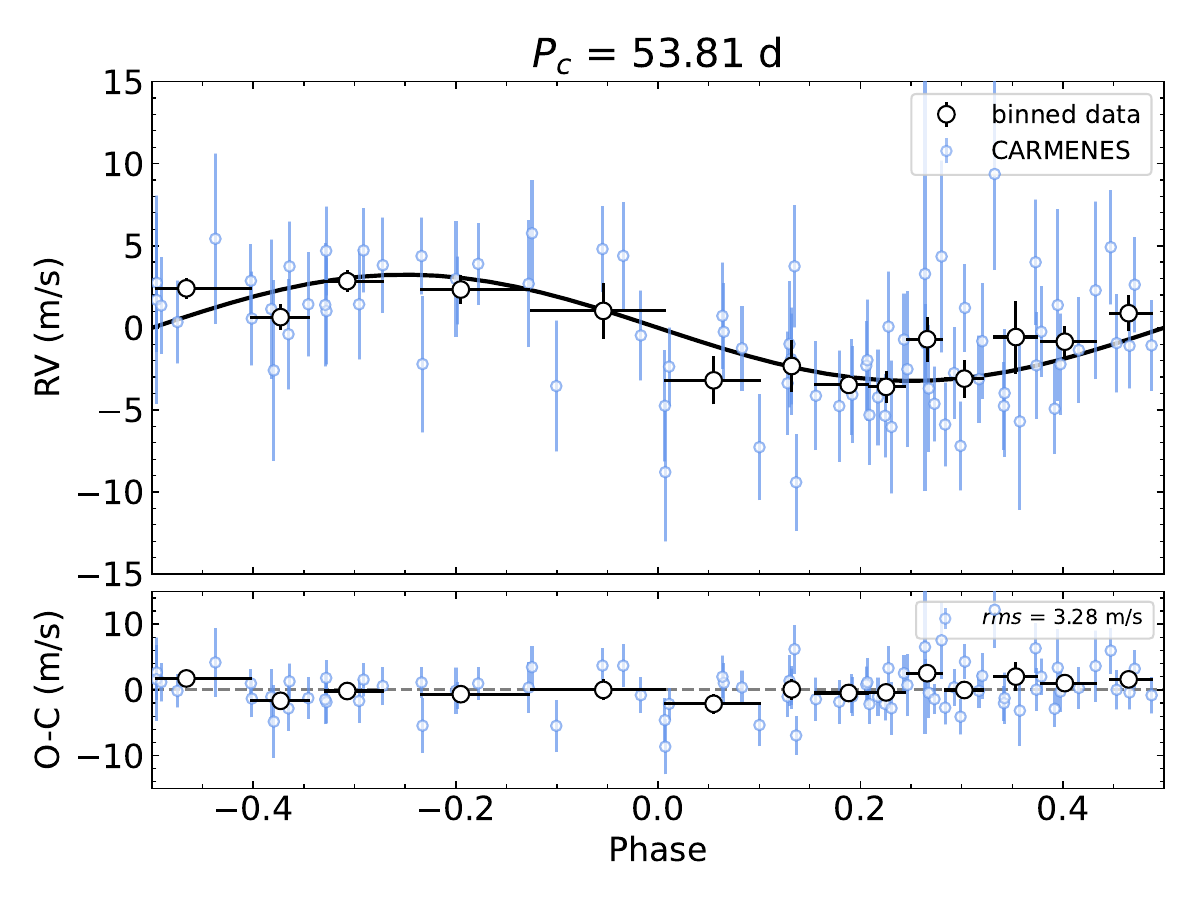}
  \caption{TOI-2093 CARMENES RVs (blue dots) and the best model (black
    line) from the joint photometric and spectroscopic fit folded in phase with the orbital
    period of TOI-2093\,b, and c, assuming $e=0$. The GP component has been removed from the data. The binned data are plotted as open black symbols. }
  \label{fig:toi2093_RVmodel_vs_phase}
\end{figure}

The folded light curves, with all sectors combined and phased with the orbital period of the transiting planet, are shown in Fig.~\ref{fig:tesslc}. For completeness, the corner plot displaying the posterior distributions of some of the planetary parameters obtained from the joint fit is presented in Fig.~\ref{fig:toi2093_cornerplot}. The resulting RV model is shown in Fig.~\ref{fig:toi2093_RVmodel_vs_time}, while the RV curves folded in phase are displayed in Fig.~\ref{fig:toi2093_RVmodel_vs_phase}. With an RV amplitude of 3.26$^{+0.74}_{-0.80}$\,$\rm m\,s^{-1}$, the transiting planet has a mass of 15.8$^{+3.6}_{-3.8}$\,M$_\oplus$ with a significance of 4.3\,$\rm \sigma$. The {\it rms} of the RV residuals (i.e., observed RVs minus the best fit) is 3.3\,$\rm m\,s^{-1}$, which is very similar to, but slightly lower than, the mean value of the CARMENES VIS RV errors (3.5\,$\rm m\,s^{-1}$). This suggests that detecting additional components in the system could be challenging, as we are already close to the noise level of our data.

We attribute the signal found at 12.836$\pm$0.021\,d to a planet, TOI-2093\,b, with
$M_{\rm p} \sin i=10.6 \pm 2.5$\,M$_\oplus$. This period is close to a resonance 4:1 with TOI-2093\,c (which would fall at 13.45\,d). Current data favor the solution of TOI-2093\,b with 12.84\,d instead of the alternative solution at 13.2\,d, which would be closer to the 4:1 resonance with the transiting planet. 

We also investigated whether the planets have eccentric orbits by allowing this parameter to vary freely in our joint fit (Fig.~\ref{fig:toi2093_cornerplot_ecc}). The eccentricities of TOI-2093\,b and TOI-2093\,c are consistent with zero within their uncertainties ($e_{\rm b}$ = 0.17$^{+0.27}_{-0.12}$, $e_{\rm b}$ = 0.23$\pm$0.10). We adopted the results for circular orbits, while Table~\ref{tab:planets_params} includes the results of both joint analyses. 

We investigated whether the planetary system is dynamically stable using the angular momentum deficit (AMD) stability criterion \citep{las97,Lask17}.
The AMD can be interpreted as a measure of the excitation of the orbits that limits close encounters among the planets, and ensures long-term stability. The main ingredients are the semimajor axes, masses, and orbital eccentricities. The result of the analysis 
yielded stable AMD solutions only for those posterior distributions with low eccentricity values: $e_{\rm b} < 0.55$, and $e_{\rm c} < 0.40$, in agreement with our findings.

%
%----------------------------------  Fig. 4
%----------------------------------------------
%

%
%----------------------------------  Fig. 5
%----------------------------------------------
%

% Table with parameters for the two planets
%--------------------------------------------------------------
\begin{table}[]
\renewcommand{\arraystretch}{1.3}
\centering
\caption{Selected parameters for TOI-2093\,b and TOI-2093\,c.}
\label{tab:planets_params}
\tabcolsep 1.6 pt
\begin{scriptsize}
  \begin{tabular}{lcccc}
  \hline \hline
%--------------------------------------------------------------
Parameter & \multicolumn{2}{c}{TOI-2093\,b} & \multicolumn{2}{c}{ TOI-2093\,c} \\
   & $e = 0$ & free $e$ & $e = 0$ & free $e$  \\
%--------------------------------------------------------------
\hline
\noalign{\smallskip}
\multicolumn{5}{c}{\textit{Fit planet parameters}} \\
\noalign{\smallskip}
$P$ (d) & 12.836$\pm$0.021 & $12.801^{+0.036}_{-0.042}$ & 53.81149$\pm$0.00017 & $53.81149^{+0.00019}_{-0.00016}$ \\
$T_0$ (BJD$-$2450000) & $9320.53^{+0.43}_{-0.50}$ & $9320.74^{+0.57}_{-0.48}$ & $8751.6122^{+0.0040}_{-0.0042}$ & $8751.6124^{+0.0044}_{-0.0048}$\\
$K$ (m\,s$^{-1}$) & $3.52^{+0.79}_{-0.84}$ & $3.71^{+1.01}_{-0.85}$ & $3.26^{+0.74}_{-0.80}$ & $3.83^{+0.78}_{-0.81}$ \\
$e$ & 0 (fixed) & $0.17^{+0.27}_{-0.12}$ & 0 (fixed) & $0.23 \pm 0.10$ \\
$\omega$ & 90 (fixed) & $119^{+146}_{-67}$ & 0 (fixed) & $148 \pm 34$ \\

\noalign{\smallskip}
\multicolumn{5}{c}{\textit{Derived planet parameters} } \\
\noalign{\smallskip}
$R_{\rm p}/R_{\star}$ & \multicolumn{2}{c}{\dots}& $0.0290^{+0.0010}_{-0.0011}$ & $0.0282^{+0.0013}_{-0.0011}$  \\
$R_{\rm p}$ (R$_\oplus$) & \multicolumn{2}{c}{\dots}& $2.30\pm0.12$ & $2.25^{+0.14}_{-0.13}$ \\
$a$ (au) & $0.097^{+0.002}_{-0.002}$ & $0.097^{+0.002}_{-0.002}$ & $0.257^{+0.017}_{-0.018}$  & $0.252^{+0.017}_{-0.018}$ \\
b = $(a/R_{\star}) \cos i$ & \multicolumn{2}{c}{\dots}& $0.560^{+0.070}_{-0.095}$  & $0.35^{+0.22}_{-0.23}$ \\
$i$ (deg) & \multicolumn{2}{c}{\dots}& $89.577^{+0.088}_{-0.075}$  & $89.72^{+0.18}_{-0.20}$ \\
$t_{\rm 14}$ (h) & \multicolumn{2}{c}{\dots}& $4.68^{+0.60}_{-0.50}$  & $4.62^{+0.93}_{-1.02}$ \\
$t_{\rm 23}$ (h) & \multicolumn{2}{c}{\dots}& $4.30^{+0.98}_{-0.12}$  & $4.33^{+1.22}_{-0.73}$ \\
$\delta$ (ppm) & \multicolumn{2}{c}{\dots}& $841^{+60}_{-61}$ & $797^{+75}_{-59}$  \\
$M_{\rm p} \sin i$ (M$_{\oplus}$) & $10.6\pm2.5$ & $10.6^{+2.8}_{-2.6}$ & $15.8^{+3.6}_{-3.8}$ & $18.0^{+3.6}_{-3.8}$ \\
$M_{\rm p}$ (M$_{ \oplus}$) & \multicolumn{2}{c}{\dots}& $15.8^{+3.6}_{-3.8}$  & $18.0^{+3.6}_{-3.8}$ \\
$\rho_{\rm p}$ (g $\rm cm^{-3}$) & \multicolumn{2}{c}{\dots}& $7.0^{+2.1}_{-1.9}$  & $8.6^{+2.6}_{-2.4}$ \\
$T_{\rm eq}$ (K) ($A_{\rm Bond}=0.3$) & \multicolumn{2}{c}{$535\pm17$} & $329^{+13}_{-11}$ & $332^{+13}_{-11}$ \\
\noalign{\smallskip}
\hline
\end{tabular}
\end{scriptsize}
\renewcommand{\arraystretch}{1.}
\end{table}
%--------------------------------------------------------------

%vvvvvvvvvvvvvvvvvvvvvvvvvvvvvvvvvvvvvvvvvvvvvvvvvvvvv
\section{Discussion}\label{sec:discussion}

\subsection{The confirmed transiting planet}
\begin{figure*}
  \centering
  \includegraphics[width=0.49\textwidth]{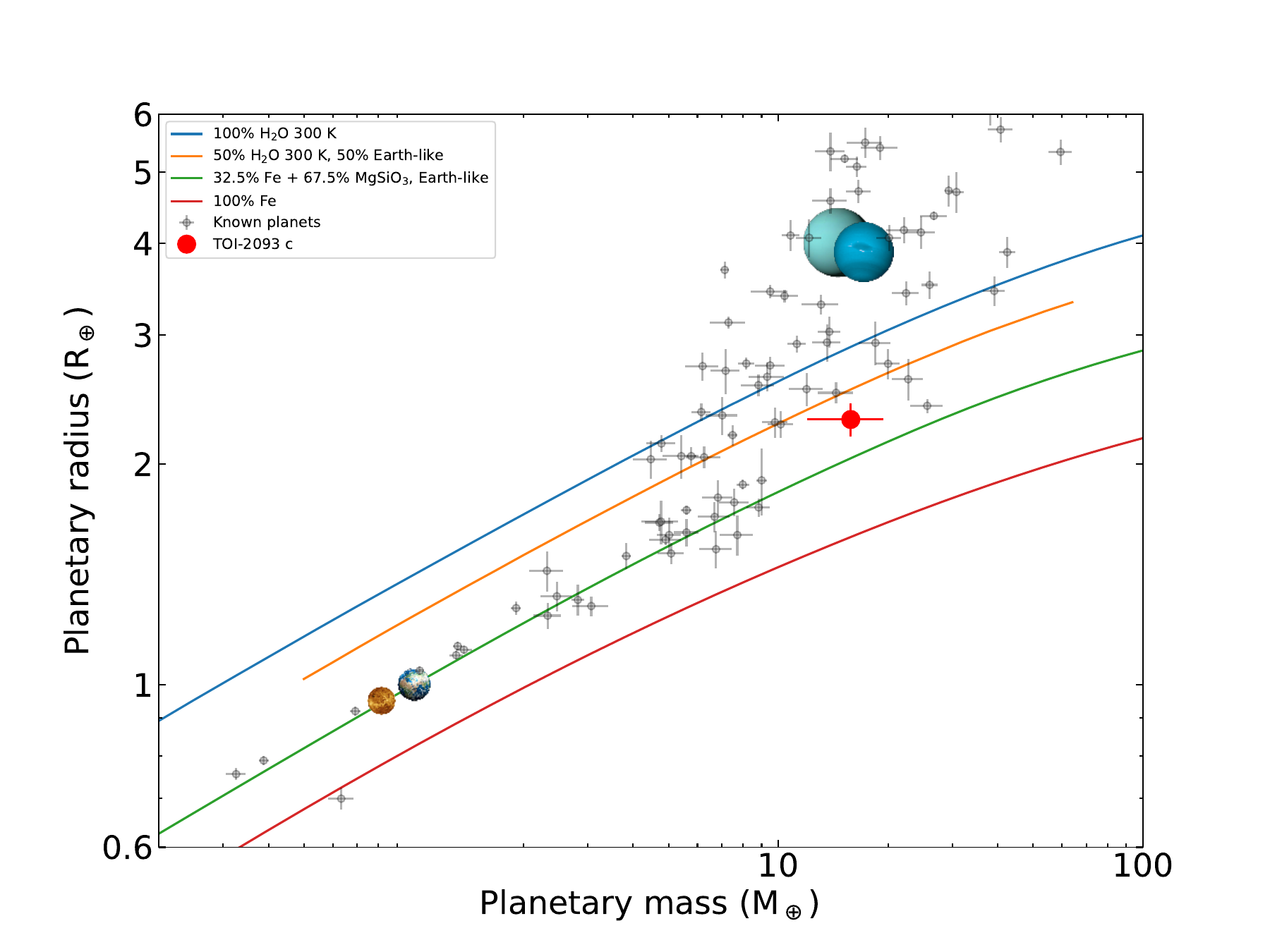}
  \includegraphics[width=0.49\textwidth]{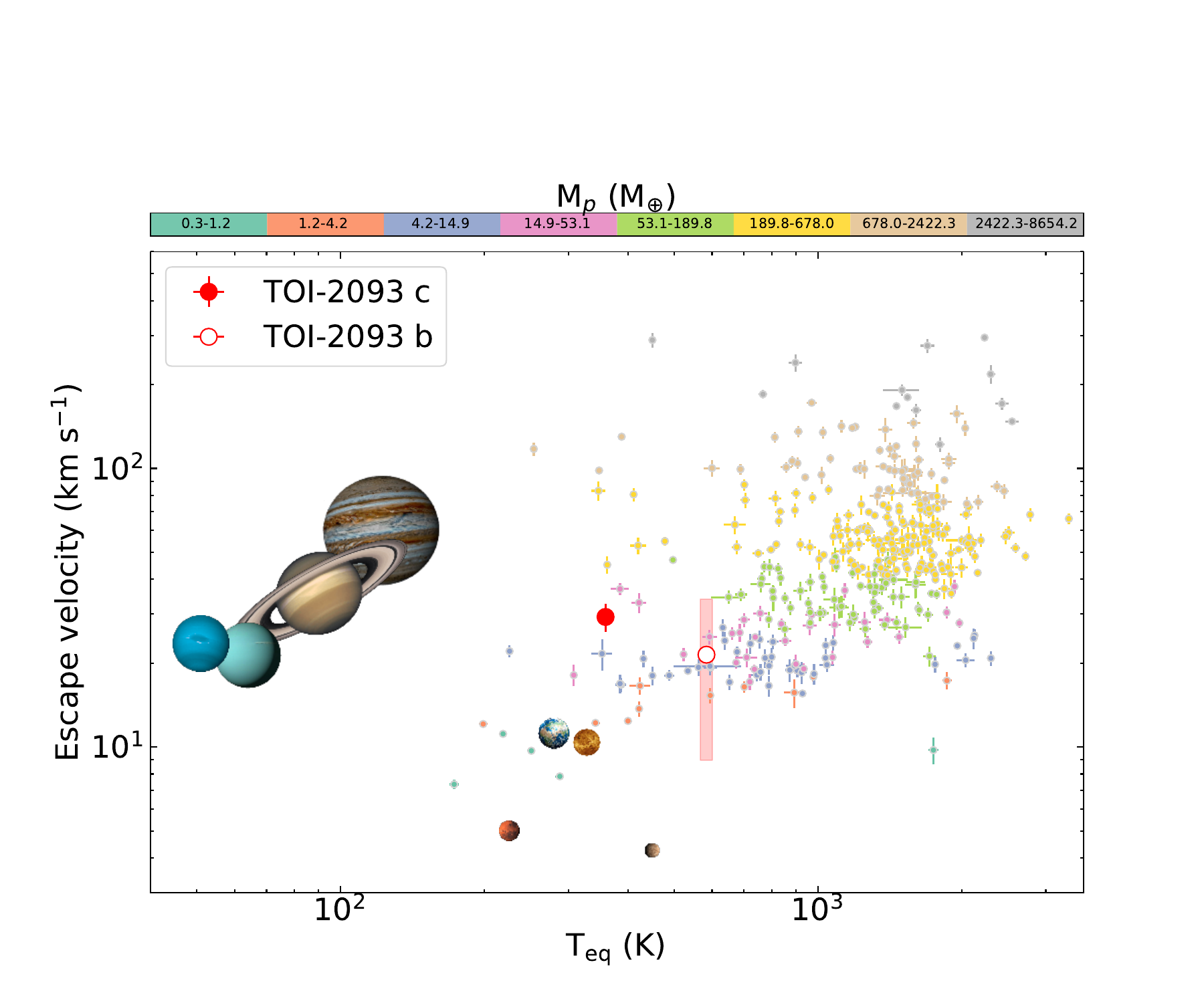}  
  \caption{TOI 2093~c in the context of exoplanets with known mass and radius with errors smaller than 12\%. {\em Left panel:} Planetary mass and radius, with theoretical models from \citet{zen16} (see Sect.~\ref{sec:TOI-2093 planetary system}). {\em Right panel:} Escape velocity against equilibrium temperature (planetary data from the NASA exoplanets database).
 TOI 2093~b is included assuming $i_{\rm b} \gtrsim 50\degr$,
      with the planet radius calculated for a bulk density between 0.5 and 10\,g\,cm$^{-3}$, with errors shown as a red band.}\label{fig:diagramaMR} 
\end{figure*}

The planet TOI-2093\,c is identified in the TESS
transits curve. The mass calculated using the RV data
confirms the planetary nature of this signal. The parameters determined
from the analysis described above are listed in
Table~\ref{tab:planets_params}. The derived density is compatible with the
Earth's bulk density (5.54\,g\,cm$^{-3}$). We compared our data
with theoretical models of the mass-radius relation which depends on the  bulk composition of the planet. We used the \citet{zen16} models as reference (Fig.~\ref{fig:diagramaMR}). The models that bracket the position of TOI-2093\,c in this diagram are those assuming a 50\% of H$_2$O at 300\,K and 50\% of Earth-like composition, and those that assume an Earth-like composition containing 32.5\% of Fe and 67.5\% of MgSiO$_3$.  
We used the X-ray and EUV (XUV) luminosity calculated in Sect.~\ref{subsubsec:CARMENES activity indicators}
to evaluate the energy-limited mass loss rate following \citet{san11}
and references therein. An expected rate of
$1.3\times 10^8$\,g\,s$^{-1}$, or $7.0\times 10^{-4}$\,M$_\oplus$\,Gyr$^{-1}$, should have a
negligible effect on the current planet's atmosphere stability. 
We used the relation between XUV stellar irradiation and the detection of the \ion{He}{i}~$\lambda$10830\,\AA\ triplet in the planet atmosphere derived by \citet{san25} to evaluate the expected emission in TOI-2093\,c. If we assume that its atmosphere has He, we expect an equivalent width of 0.5--1.5\,m\AA. Therefore it is unlikely to detect the \ion{He}{i} triplet in this star with current instrumentation.

%
%----------------------------------  Fig. 6
%----------------------------------------------
%

The equilibrium temperature of a planet is calculated by
balancing the incoming stellar irradiation with the planet
emission as a blackbody, both calculated with the
Stefan-Boltzmann law. Using the
bolometric luminosity of TOI-2093 and a Bond albedo of $A_{\rm Bond}$=0.3 (most Solar System planets have $A_{\rm Bond}\sim$0.3), TOI-2093\,c has a $T_{\rm eq}=329$\,K.
The planet should not suffer a substantial mass loss, its bulk density is compatible with the Earth's density, and its equilibrium temperature is promising. 

The definition of the HZ has been
subject of debate for a long time \citep[e.g.,][and references
  therein]{kas93,lam09,kop13,cab22}. Planet habitability is a complex
problem,
involving atmospheric composition, the existence of a planetary magnetic field,
planetary mass, tidal locking to the host star, or the presence of
large moons around the planet.
We established simple
parameters to evaluate, as a first approximation, whether a planet
falls in the HZ of its host star. 
Assuming that a planet with equilibrium temperature between 273 and
373\,K may have liquid water, TOI-2093\,c would lie in the HZ of its host star.
The NASA exoplanet database\footnote{\url{https://exoplanetarchive.ipac.caltech.edu/}}
includes the calculation of $T_{\rm eq}$ for most planets. In Fig.~\ref{fig:diagramaMR} we compare
$T_{\rm eq}$ with the escape velocity of the planets, including Solar
System planets as references, and assuming an $A_{\rm Bond}$=0 for all of them for
consistency. TOI-2093\,c shows a similar $T_{\rm eq}$ to that of Venus,
but with a larger escape velocity.

Although a more accurate calculation of the
temperature at the planetary surface would require proper knowledge of
the atmospheric composition and wind circulation, we can
use $T_{\rm eq}$ as a simple parameter to compare with other 
potentially habitable planets. With that purpose in mind,
we made our own calculations of $T_{\rm eq}$ using the values from the
Exoplanets Encyclopaedia\footnote{\url{https://exoplanet.eu}, accessed on 8 Sepember, 2025.}. We
filtered the catalog to choose only planets with known mass and
radius, at less than 1000\,pc from us,
and with $M_{\rm p} < 13$\,M$_{\rm Jup}$.
Only 47 exoplanets out of 1490 lie in their HZ.

Planets in the HZ of M stars are likely
tidally locked, purportedly making them less suitable for harboring life \citep[e.g.,][]{bar17}.
Although planets around late K stars may still be tidally locked, we
assumed that planets orbiting F, G, and K stars would have a HZ without
potential problems of orbital synchronization with their rotation. This
would reduce the sample of planets in a HZ to just 23, merely 3.2\% of
those with known mass and radius.
Among them only two have a radius $R_{\rm p}<5$\,R$_\oplus$, and only six have $M_{\rm p}<24$\,M$_\oplus$ (Neptune has 3.9\,R$_\oplus$ and 17.1\,M$_\oplus$).
None of them are smaller than TOI-2093\,c (Table~\ref{tab:planetsHZ}).
Thus TOI-2093\,c is the smallest planet in a HZ of an FGK star reported to date.
Although the mass is rather high compared to Earth, a hypothetical
moon orbiting the planet would provide favorable conditions \citep{mar19}.

\subsection{The planet candidate in the system}
We found an additional inner planet candidate, TOI-2093\,b, in the spectroscopic analysis (Table~\ref{tab:planets_params}).
The planet is also included in the right panel of Fig.~\ref{fig:diagramaMR}.
We can estimate the approximate mass loss rate expected for TOI-2093\,b: if we assume a density of 1\,g\,cm$^{-3}$, a moderate mass loss rate of
$6.9\times 10^9$\,g\,s$^{-1}$ or 0.036\,M$_\oplus$\,Gyr$^{-1}$ is calculated. Since the star was much more active in its early stages (the $L_{\rm X}$ was up to two orders of magnitude larger at an age of 100\,Myr), this planet was likely more massive than TOI-2093\,c originally, but it has suffered a larger accumulated mass loss. Current uncertainties on stellar age and XUV emission prevent one from reconstructing a more precise past evolution of both planets. 

No transit was detected for planet b. One explanation could be the misalignment between the orbits of TOI-2093\,b and c. An angle of just 1.6\degr\ between the two orbital planes can explain the lack of transits of planet b. An alternative explanation is that planet b is not large enough to be detected in the TESS light curves: this would set an upper limit of $R_{\rm b} \le 0.87$\,R$_\oplus$, assuming a 10$\sigma$ threshold detection. This size implies a bulk density of $\rho_{\rm b}$>67\,g\,cm$^{-3}$. We considered this as an implausible value, thus discarding TOI-2093\,b as a transiting planet.

Dynamical analysis of the planetary system \citep[based on the MEGNO parameter implemented in the {\tt Rebound} package,][]{cin00,reb12,meg15} shows that system stability degrades rapidly for orbital inclinations below $\approx 50\degr$ for planet b (Fig.~\ref{fig:stability}), favoring higher inclinations. This minimal inclination suggests that TOI-2093\,b's true mass is close to the $m$\,sin\,$i$ value listed in Table~\ref{tab:planets_params}. Based on these mass values, and following the method of \cite{poz23}, we predicted the TTVs expected on the outer planet due to interactions with TOI-2093\,b. They are most likely below 0.5\,min, and thus smaller than the 11.2\,min from the 1\,$\sigma$ $rms$ of the TTV analysis. This result is consistent with the non-detection of TTVs (Sect.~\ref{sec:tesslc}).

\subsection{Constraints on the presence of additional planets}\label{sect:constraints}
\begin{figure}
  \centering
  \includegraphics[width=0.49\textwidth]{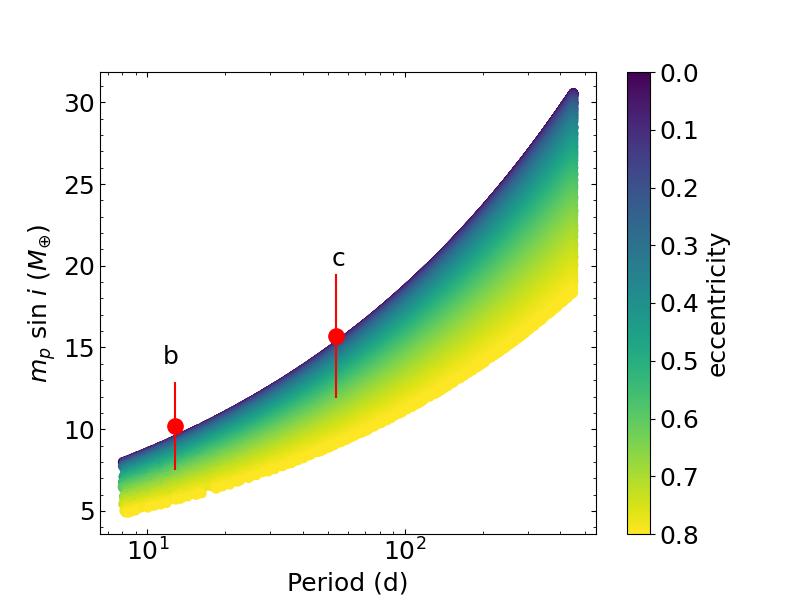}
  \caption{Detectability limits for hypothetical planetary companions to TOI-2093 as a function of orbital period. Assumed eccentricities are color-coded. The red dots correspond to the TOI-2093\,b and c planets.}\label{fig:mass_limit_detection}
\end{figure}

We estimated the minimum mass of any potential planet orbiting TOI-2093 within a period range of 8 to 450\,d, applying Kepler’s third law, the stellar mass listed in Table~\ref{tab:starpar}, and a range of eccentricity values. The lower limit of the period range corresponds to four times the Nyquist frequency of the RV dataset, while the upper limit was set to half the time span of the CARMENES observations. Fig.~\ref{fig:mass_limit_detection} shows the resulting minimum masses, assuming that the smallest detectable RV amplitude corresponds to the 1$\sigma$ level {\it rms} of the CARMENES VIS residuals after subtracting the stellar activity and the two planets. The values obtained for the GP parameters, and discussed in Sect.~\ref{sec:Radial velocity models}, indicate that the GP models the stellar activity with an appropriate balance of coherence and flexibility, suggesting minimal overfitting and limited absorption of planetary signals. With these assumptions, we ruled out planets in circular orbits with masses above $\sim$9, 18\,M$_\oplus$ and orbital periods of about 10, 100\,d, respectively. 

The signals at 1.75\,d and 2.31\,d, observed in both the stellar RVs and activity indices, do not appear to be related to the measured rotation period of the K dwarf, as a K-type star with a $\sim$2\,d rotation period would typically show a much higher level of stellar activity. An alternative, although rather speculative scenario, would be the presence of a planet detected with low confidence that induces activity in the stellar hemisphere facing the
planet. Then we can expect that the activity signal has the synodic period between those of the stellar rotation and the planet orbit. 
Taking a closer look at Fig.~\ref{fig:glsgeneral}, and the zoomed view in Fig.~\ref{fig:glszoom}, we can identify these two stellar activity signals near 2\,d. Their removal results in two peaks in the RV periodogram with low significance at 1.85\,d and 2.15\,d, with one being the 1\,d alias of the other. Although they could be related to the observational window (Fig.~\ref{fig:window_carmenes_vis}), one of them might actually correspond to a third planet.
In a prograde orbit, the synodic period must be larger than the orbital period. Thus a planet orbiting TOI-2093 with a 2.15\,d periodicity might induce an activity signal with a synodic period of $\sim$2.26\,d. In the case of a retrograde orbit, a planet with 1.85\,d would induce an activity signal at $\sim$1.78\,d instead. 
Despite the attractiveness of this scenario, there is a high probability that the signals at 1.75\,d and 2.31\,d are actually related to unknown stellar activity at timescales significantly shorter than the stellar rotation period.

%----------------------------------  Fig. 7
%----------------------------------------------

%vvvvvvvvvvvvvvvvvvvvvvvvvvvvvvvvvvvvvvvvvvvvvvvvvvvvv
\section{Conclusions}
\label{sec:conclusions}

The RV and photometric data of TOI-2093\,c allow us to confirm that this is a transiting planet, with the parameters shown in Table~\ref{tab:planets_params}. We also determined the stellar parameters, which allowed us to show that TOI-2093 is not a young star. The equilibrium temperature of the planet, 329\,K, indicates that it lies on the HZ of its host star. This is the smallest planet, with known mass and radius, orbiting in the HZ of an FGK star detected to date.
This makes this planet an interesting target for further atmospheric characterization of a potentially habitable planet.

The planetary system includes a less massive planet candidate in an inner orbit close to a 4:1 resonance with TOI-2093\,c. We demonstrated that this is not a transiting planet, which implies that the orbital plane of
  TOI-2093\,c is inclined by
  at least 1.6$\degr$ with respect to that of the inner planet.
Further observations of the system could refine better the mass and
orbital parameters of both planets b and c.

Finally, four peaks are observed in the periodogram close to $\sim$2\,d, with the two most significant being related to stellar activity, despite of being unrelated to stellar rotation. Although stellar activity is the most plausible explanation, an interesting alternative would be the presence of a third planet that is inducing some activity on the stellar hemisphere that faces the planet.

\begin{acknowledgements}
We thank the anonymous referee for the useful comments that helped improve the manuscript.
CARMENES is an instrument at the Centro Astron\'omico Hispano en Andaluc\'ia (CAHA) at Calar Alto (Almer\'{\i}a, Spain), operated jointly by the Junta de Andaluc\'ia and the Instituto de Astrof\'isica de Andaluc\'ia (CSIC).
CARMENES was funded by the Max-Planck-Gesellschaft (MPG), 
  the Consejo Superior de Investigaciones Cient\'{\i}ficas (CSIC),
  the Ministerio de Econom\'ia y Competitividad (MINECO) and the European Regional Development Fund (ERDF) through projects FICTS-2011-02, ICTS-2017-07-CAHA-4, and CAHA16-CE-3978, 
  and the members of the CARMENES Consortium (Max-Planck-Institut f\"ur Astronomie,
  Instituto de Astrof\'{\i}sica de Andaluc\'{\i}a,
  Landessternwarte K\"onigstuhl,
  Institut de Ci\`encies de l'Espai,
  Institut f\"ur Astrophysik G\"ottingen,
  Universidad Complutense de Madrid,
  Th\"uringer Landessternwarte Tautenburg,
  Instituto de Astrof\'{\i}sica de Canarias,
  Hamburger Sternwarte,
  Centro de Astrobiolog\'{\i}a and
  Centro Astron\'omico Hispano-Alem\'an), 
  with additional contributions by the MINECO, 
  the Deutsche Forschungsgemeinschaft (DFG) through the Major Research Instrumentation Programme and Research Unit FOR2544 ``Blue Planets around Red Stars'', 
  the Klaus Tschira Stiftung, 
  the states of Baden-W\"urttemberg and Niedersachsen, 
  and by the Junta de Andaluc\'{\i}a.
 The Telescopi Joan Oró (TJO) of the Montsec Observatory (OdM) is owned by the Generalitat de Catalunya and operated by the Institute for Space Studies of Catalonia (IEEC).
This paper includes data collected by the TESS mission that are publicly available from the Mikulski Archive for Space Telescopes (MAST) at the Space Telescope Science Institute (STScI). 
Funding for the TESS mission is provided by National Aeronautics and Space Administration (NASA) Science Mission Directorate.
STScI is operated by the Association of Universities for Research in Astronomy, Inc., under NASA contract NAS 5–26555.
We acknowledge the use of public TESS data from pipelines at the TESS Science Office and at the TESS Science Processing Operations Center.
This research has made use of the Exoplanet Follow-up Observation Program website, which is operated by the California Institute of Technology, under contract with the NASA under the Exoplanet Exploration Program.
Resources supporting this work were provided by the NASA High-End Computing (HEC) Program through the NASA Advanced Supercomputing (NAS) Division at Ames Research Center for the production of the SPOC data products.
This research has made use of data obtained from the portal exoplanet.eu of The Extrasolar Planets Encyclopaedia.
We acknowledge financial support from the Agencia Estatal de Investigaci\'on (AEI/10.13039/501100011033) of the Ministerio de Ciencia, Innovaci\'on y Universidades and the ERDF ``A way of making Europe'' through projects  
  PID2021-125627OB-C31,
  PID2022-137241NB-C4[1:4],
PID2023-150468NB-I00,
and the Centre of Excellence ``Severo Ochoa'' and ``Mar\'ia de Maeztu'' awards to the Instituto de Astrof\'isica de Canarias (CEX2019-000920-S), Instituto de Astrof\'isica de Andaluc\'ia (CEX2021-001131-S) and Institut de Ci\`encies de l'Espai (CEX2020-001058-M), and 
from María Zambrano contract for the attraction of international talent, funded by the Spanish Ministry of Universities and the European Union–Next Generation EU, at Universidad Complutense de Madrid.
CSIC also funded this work through the internal project 2023AT003 associated to RYC2021-031640-I.
Support for this work was also provided by the NASA Hubble Fellowship grant HST-HF2-51559.001-A awarded by the STScI, and the NASA award 80NSSC25M7110.
We received support from the Deutsches Zentrum für Luft- und Raumfahrt (DLR, project number 50OP2502).
Funding was also received from the Israel Science Foundation through grant No. 1404/22,
and the Ministry of Science and Higher Education programme the "Excellence Initiative - Research University" conducted at the Centre of Excellence in Astrophysics and Astrochemistry of the Nicolaus Copernicus University in Toru\'n, Poland. We are also grateful to the Centre of Informatics Tricity Academic Supercomputer and networK (CI TASK, Gda\'nsk, Poland) for computing resources (grant no. PT01187). 
\end{acknowledgements}

\begin{appendix} %First appendix
\onecolumn
\section{Supplementary Figures and Tables}

%--------------------------------------------------------------
%----------------------------------  Table A.1
%TOI-2093 rv and activity data
\begin{table*}[h!]
  \caption{TOI-2093 CARMENES data.}\label{tab:toi2093_rv_act_data}
\vspace{-3mm}
\begin{center}
    \setlength{\tabcolsep}{3pt}
    \begin{scriptsize}
      \begin{tabular}{lrrccccccrrrrcc}
\hline\hline
\noalign{\smallskip}
%--------------------------------------------------------------
BJD & RV & eRV & CaIRT$_1$ & eCaIRT$_1$ & CaIRT$_2$ & eCaIRT$_2$ & CaIRT$_3$ & eCaIRT$_3$ & CRX & eCRX & dLW & edLW &   H$\alpha$ & eH$\alpha$ \\
(d)  & (m\,s$^{-1}$) &  (m\,s$^{-1}$) &  &  &  &  &  &  & (m\,s$^{-1}$\,Np$^{-1}$) & (m\,s$^{-1}$\,Np$^{-1}$) & (m$^2$\,s$^{-2}$)  & (m$^2$\,s$^{-2}$)  & & \\
\noalign{\smallskip}
\hline
%--------------------------------------------------------------
\noalign{\smallskip}
2459307.6287 &  16.5 &   5.8 & 0.4901 & 0.0037 & 0.3960 & 0.0035 & 0.3674 & 0.0034 &    13 &    37 &  14.6 &   7.9 & 0.5942 & 0.0029  \\ 
2459327.6396 &   7.1 &   3.3 & 0.4894 & 0.0028 & 0.3829 & 0.0024 & 0.3632 & 0.0024 &   $-$22 &    25 &  12.3 &   5.4 & 0.5968 & 0.0020  \\ 
2459336.6571 &  $-$4.4 &   3.9 & 0.4881 & 0.0027 & 0.3769 & 0.0024 & 0.3698 & 0.0024 &    50 &    28 & $-$11.1 &   6.7 & 0.5942 & 0.0020  \\ 
2459340.5832 &   5.4 &   2.6 & 0.4883 & 0.0022 & 0.3931 & 0.0019 & 0.3701 & 0.0018 &     2 &    19 &   9.6 &   4.2 & 0.5901 & 0.0015  \\ 
2459342.6257 &   1.1 &   2.7 & 0.5038 & 0.0021 & 0.3904 & 0.0018 & 0.3735 & 0.0018 &     1 &    20 &  17.2 &   3.9 & 0.5942 & 0.0015  \\ 
2459350.6164 &  $-$1.0 &   2.8 & 0.5062 & 0.0023 & 0.4054 & 0.0020 & 0.3883 & 0.0020 &    $-$6 &    20 &  22.7 &   5.1 & 0.6009 & 0.0017  \\ 
2459354.6125 &  $-$1.9 &   2.7 & 0.5043 & 0.0018 & 0.4020 & 0.0015 & 0.3806 & 0.0015 &   $-$48 &    17 &  36.9 &   3.5 & 0.6005 & 0.0013  \\ 
2459355.6114 &  $-$4.4 &   2.5 & 0.5002 & 0.0019 & 0.4007 & 0.0016 & 0.3797 & 0.0016 &   $-$45 &    17 &  35.2 &   3.9 & 0.5999 & 0.0014  \\ 
2459356.6153 &   0.0 &   2.8 & 0.5033 & 0.0024 & 0.3931 & 0.0021 & 0.3757 & 0.0020 &   $-$36 &    20 &  19.0 &   4.4 & 0.5967 & 0.0017  \\ 
2459358.6164 &   0.5 &   5.8 & 0.4937 & 0.0043 & 0.3926 & 0.0043 & 0.3749 & 0.0041 &   $-$81 &    43 &  19.6 &   9.4 & 0.5992 & 0.0016  \\ 
2459359.6255 & $-$14.2 &   2.7 & 0.5020 & 0.0022 & 0.3902 & 0.0019 & 0.3716 & 0.0019 &   $-$45 &    19 &  20.5 &   5.9 & 0.5964 & 0.0015  \\ 
2459360.6067 & $-$12.1 &   2.6 & 0.4983 & 0.0020 & 0.3942 & 0.0018 & 0.3735 & 0.0017 &   $-$44 &    18 &  27.3 &   5.1 & 0.5889 & 0.0022  \\ 
2459363.6158 &   1.2 &   3.8 & 0.4915 & 0.0029 & 0.3895 & 0.0027 & 0.3728 & 0.0026 &   $-$50 &    28 &  21.8 &   5.0 & 0.5866 & 0.0017  \\ 
2459364.6321 &  $-$2.8 &   2.8 & 0.4943 & 0.0024 & 0.3886 & 0.0021 & 0.3711 & 0.0021 &     4 &    20 &  15.9 &   5.2 & 0.5947 & 0.0016  \\ 
2459367.6144 &  17.1 &   3.5 & 0.4932 & 0.0022 & 0.3852 & 0.0019 & 0.3747 & 0.0019 &     2 &    26 &  24.6 &   4.7 & 0.5969 & 0.0014  \\ 
2459368.6127 &  10.0 &   2.6 & 0.5002 & 0.0020 & 0.3979 & 0.0017 & 0.3750 & 0.0016 &   $-$37 &    18 &  34.2 &   4.5 & 0.5896 & 0.0020  \\ 
2459386.6017 &   8.4 &   3.5 & 0.4982 & 0.0027 & 0.3970 & 0.0024 & 0.3684 & 0.0024 &    16 &    25 &  17.0 &   6.2 & 0.5905 & 0.0018  \\ 
2459395.4971 &   6.9 &   3.3 & 0.4936 & 0.0025 & 0.3863 & 0.0022 & 0.3727 & 0.0022 &    35 &    22 &  18.8 &   5.2 & 0.5999 & 0.0017  \\ 
2459404.6088 &  12.6 &   3.7 & 0.5056 & 0.0023 & 0.3983 & 0.0020 & 0.3796 & 0.0020 &    39 &    27 &  27.0 &   4.5 & 0.6020 & 0.0018  \\ 
2459409.6041 &   5.3 &   3.3 & 0.5092 & 0.0024 & 0.4124 & 0.0022 & 0.3867 & 0.0021 &    20 &    25 &  38.1 &   5.7 & 0.5941 & 0.0014  \\ 
2459411.5887 &  $-$0.3 &   2.4 & 0.5090 & 0.0019 & 0.4096 & 0.0017 & 0.3829 & 0.0016 &    $-$2 &    17 &  37.7 &   4.1 & 0.5833 & 0.0020  \\ 
2459424.5030 &   8.3 &   4.2 & 0.4955 & 0.0027 & 0.3927 & 0.0025 & 0.3649 & 0.0024 &    44 &    32 &   8.6 &   5.8 & 0.5842 & 0.0026  \\ 
2459427.6208 &   9.0 &   5.2 & 0.4989 & 0.0035 & 0.3946 & 0.0032 & 0.3621 & 0.0031 &    $-$8 &    40 &   6.6 &   5.6 & 0.5859 & 0.0013  \\ 
2459429.5091 &   4.2 &   2.2 & 0.4984 & 0.0018 & 0.3915 & 0.0016 & 0.3703 & 0.0016 &    23 &    16 &  17.4 &   4.0 & 0.5889 & 0.0018  \\ 
2459431.5063 &   1.4 &   3.4 & 0.4935 & 0.0025 & 0.3929 & 0.0022 & 0.3702 & 0.0022 &    68 &    23 &  16.0 &   4.3 & 0.5864 & 0.0015  \\ 
2459433.5108 &   8.2 &   2.7 & 0.4906 & 0.0020 & 0.3855 & 0.0018 & 0.3678 & 0.0017 &    $-$3 &    20 &   7.2 &   4.2 & 0.5920 & 0.0015  \\ 
2459435.4996 &   4.5 &   2.6 & 0.4898 & 0.0021 & 0.3854 & 0.0018 & 0.3651 & 0.0018 &    30 &    18 &   1.6 &   4.4 & 0.5984 & 0.0018  \\ 
2459444.4585 &  17.2 &   3.2 & 0.5065 & 0.0024 & 0.4041 & 0.0021 & 0.3791 & 0.0021 &    61 &    22 &  37.3 &   4.9 & 0.5994 & 0.0014  \\ 
2459461.4330 &  $-$1.2 &   2.9 & 0.5018 & 0.0020 & 0.3958 & 0.0017 & 0.3795 & 0.0017 &    20 &    18 &  43.4 &   3.7 & 0.5980 & 0.0023  \\ 
2459464.4216 &  $-$1.5 &   4.7 & 0.4905 & 0.0031 & 0.3903 & 0.0027 & 0.3691 & 0.0027 &   $-$23 &    26 &  30.8 &   5.9 & 0.5895 & 0.0014  \\ 
2459466.4291 &  $-$6.3 &   2.5 & 0.4884 & 0.0019 & 0.3755 & 0.0017 & 0.3564 & 0.0016 &    36 &    21 & $-$10.0 &   3.9 & 0.5843 & 0.0023  \\ 
2459468.4025 &   0.8 &   3.5 & 0.4882 & 0.0032 & 0.3808 & 0.0029 & 0.3653 & 0.0029 &   $-$51 &    30 & $-$15.8 &   6.2 & 0.5916 & 0.0021  \\ 
2459470.4028 &  $-$2.0 &   5.4 & 0.4866 & 0.0027 & 0.3851 & 0.0025 & 0.3677 & 0.0024 &    $-$0 &    18 & $-$38.5 &   6.1 & 0.5961 & 0.0017  \\ 
2459472.4177 &   2.8 &   5.8 & 0.4882 & 0.0023 & 0.3837 & 0.0021 & 0.3625 & 0.0020 &    38 &    26 &  $-$7.1 &   4.8 & 0.5899 & 0.0027  \\ 
2459474.4290 &  $-$2.1 &   5.4 & 0.4847 & 0.0035 & 0.3810 & 0.0033 & 0.3668 & 0.0033 &    24 &    41 & $-$37.3 &   7.2 & 0.5860 & 0.0031  \\ 
2459477.4068 &  $-$8.0 &   2.8 & 0.4782 & 0.0040 & 0.3753 & 0.0039 & 0.3596 & 0.0038 &   $-$21 &    45 &  $-$9.4 &   8.9 & 0.5874 & 0.0029  \\ 
2459483.3695 &   6.6 &   2.8 & 0.4925 & 0.0041 & 0.3873 & 0.0039 & 0.3600 & 0.0038 &   $-$28 &    42 & $-$31.7 &   7.8 & 0.5865 & 0.0013  \\ 
2459485.3745 &   9.1 &   2.7 & 0.4808 & 0.0019 & 0.3763 & 0.0016 & 0.3622 & 0.0016 &    20 &    19 &  $-$5.7 &   3.1 & 0.5885 & 0.0016  \\ 
2459487.3379 &   4.5 &   3.3 & 0.4968 & 0.0022 & 0.3892 & 0.0019 & 0.3718 & 0.0019 &    $-$8 &    21 &  $-$0.2 &   5.4 & 0.5963 & 0.0013  \\ 
2459490.3345 &   0.2 &   2.9 & 0.5005 & 0.0019 & 0.3936 & 0.0016 & 0.3715 & 0.0016 &    $-$2 &    18 &  24.6 &   4.1 & 0.5932 & 0.0016  \\ 
2459492.4483 & $-$11.9 &   4.2 & 0.5000 & 0.0022 & 0.3983 & 0.0019 & 0.3754 & 0.0019 &   $-$50 &    23 &   2.0 &   4.5 & 0.5914 & 0.0015  \\ 
2459494.3274 &  $-$6.7 &   2.0 & 0.4956 & 0.0021 & 0.3874 & 0.0018 & 0.3706 & 0.0018 &   $-$10 &    19 &  11.2 &   4.0 & 0.5912 & 0.0022  \\ 
2459505.3370 &  $-$6.1 &   3.4 & 0.4860 & 0.0031 & 0.3845 & 0.0028 & 0.3626 & 0.0027 &   $-$27 &    30 & $-$17.6 &   6.2 & 0.5906 & 0.0014  \\ 
2459509.4343 &   4.7 &   2.6 & 0.4923 & 0.0020 & 0.3827 & 0.0018 & 0.3668 & 0.0017 &   $-$19 &    14 &  $-$6.6 &   3.8 & 0.5884 & 0.0015  \\ 
2459512.3151 & $-$19.2 &   2.9 & 0.4805 & 0.0021 & 0.3740 & 0.0018 & 0.3600 & 0.0018 &    19 &    25 &  $-$0.9 &   4.5 & 0.5948 & 0.0016  \\ 
2459526.3717 &  $-$0.2 &   3.1 & 0.4790 & 0.0024 & 0.3774 & 0.0021 & 0.3606 & 0.0020 &   $-$10 &    17 &  11.6 &   4.1 & 0.5956 & 0.0016  \\ 
2459530.3223 &  $-$0.0 &   2.9 & 0.4786 & 0.0023 & 0.3801 & 0.0020 & 0.3587 & 0.0019 &    $-$7 &    22 &  $-$2.1 &   3.6 & 0.5926 & 0.0017  \\ 
2459562.2801 &   6.4 &   3.0 & 0.4946 & 0.0025 & 0.3972 & 0.0022 & 0.3724 & 0.0021 &   $-$12 &    22 &  21.4 &   5.5 & 0.5896 & 0.0016  \\ 
2459623.7265 &   2.7 &   3.7 & 0.4889 & 0.0024 & 0.3795 & 0.0021 & 0.3613 & 0.0021 &   $-$22 &    19 &   5.5 &   4.9 & 0.5921 & 0.0017  \\ 
2459632.7141 &  $-$5.9 &   3.2 & 0.4939 & 0.0023 & 0.3906 & 0.0021 & 0.3744 & 0.0020 &    $-$6 &    23 &  34.1 &   3.6 & 0.5929 & 0.0022  \\ 
2459639.7309 &   5.3 &   6.3 & 0.4687 & 0.0031 & 0.3631 & 0.0029 & 0.3491 & 0.0029 &   $-$72 &    27 & $-$19.4 &   5.9 & 0.5905 & 0.0015  \\ 
2459648.7075 &   0.5 &   3.7 & 0.4788 & 0.0022 & 0.3728 & 0.0019 & 0.3584 & 0.0019 &    53 &    21 &   4.9 &   4.1 & 0.5900 & 0.0037  \\ 
2459677.6497 &  $-$3.7 &   3.0 & 0.4687 & 0.0046 & 0.3793 & 0.0046 & 0.3584 & 0.0045 &    11 &    50 &  $-$4.4 &  10.5 & 0.5956 & 0.0022  \\ 
2459680.6044 &   1.9 &  13.2 & 0.4817 & 0.0030 & 0.3747 & 0.0027 & 0.3624 & 0.0027 &   $-$40 &    28 &  $-$7.4 &   7.2 & 0.5918 & 0.0017  \\ 
2459694.6596 &  $-$1.3 &   2.5 & 0.4834 & 0.0024 & 0.3799 & 0.0021 & 0.3616 & 0.0020 &    $-$7 &    23 &   3.1 &   5.1 & 0.5830 & 0.0086  \\ 
2459699.6519 &   0.4 &   4.2 & 0.4893 & 0.0107 & 0.3882 & 0.0123 & 0.3594 & 0.0117 &   $-$55 &   105 & $-$57.7 &  24.5 & 0.5885 & 0.0015  \\ 
2459701.6197 &   1.4 &   3.2 & 0.4743 & 0.0021 & 0.3683 & 0.0018 & 0.3527 & 0.0017 &    $-$9 &    19 &  $-$2.4 &   3.8 & 0.5902 & 0.0024  \\ 
2459707.6424 &   3.4 &   2.3 & 0.4783 & 0.0031 & 0.3633 & 0.0028 & 0.3487 & 0.0028 &     8 &    32 & $-$36.3 &   8.4 & 0.5834 & 0.0019  \\ 
2459710.6524 &   2.1 &   2.5 & 0.4757 & 0.0025 & 0.3706 & 0.0022 & 0.3527 & 0.0022 &    57 &    22 & $-$31.0 &   4.4 & 0.5928 & 0.0013  \\ 
2459720.6022 & $-$15.6 &   4.2 & 0.4936 & 0.0017 & 0.3834 & 0.0015 & 0.3650 & 0.0014 &    22 &    16 &   7.9 &   3.6 & 0.5919 & 0.0014  \\ 
2459723.6410 &  $-$3.2 &   3.2 & 0.4944 & 0.0019 & 0.3887 & 0.0017 & 0.3681 & 0.0016 &    33 &    17 &  12.3 &   4.5 & 0.5897 & 0.0022  \\ 
2459725.6109 &  $-$6.5 &   3.2 & 0.4787 & 0.0029 & 0.3724 & 0.0026 & 0.3579 & 0.0025 &   $-$11 &    32 & $-$39.7 &   5.5 & 0.5920 & 0.0016  \\ 
2459728.6133 &  $-$2.4 &   3.3 & 0.4800 & 0.0021 & 0.3693 & 0.0018 & 0.3539 & 0.0018 &    66 &    22 & $-$22.9 &   4.2 & 0.5882 & 0.0019  \\ 
2459730.5487 &  $-$5.5 &   2.9 & 0.4777 & 0.0024 & 0.3668 & 0.0021 & 0.3460 & 0.0021 &    23 &    23 & $-$47.3 &   4.7 & 0.5873 & 0.0018  \\ 
2459732.6333 &  $-$8.9 &   4.0 & 0.4766 & 0.0024 & 0.3658 & 0.0021 & 0.3487 & 0.0020 &   $-$39 &    25 & $-$39.0 &   5.7 & 0.5884 & 0.0024  \\ 
2459734.6197 &  $-$3.7 &   3.9 & 0.4720 & 0.0021 & 0.3587 & 0.0018 & 0.3461 & 0.0018 &    42 &    20 & $-$47.2 &   5.1 & 0.5825 & 0.0021  \\ 
2459736.5379 &   4.5 &   2.7 & 0.4687 & 0.0031 & 0.3495 & 0.0028 & 0.3417 & 0.0028 &   $-$59 &    28 & $-$42.8 &   6.6 & 0.5857 & 0.0019  \\ 
2459738.6124 &  $-$1.4 &   2.7 & 0.4646 & 0.0027 & 0.3612 & 0.0024 & 0.3400 & 0.0023 &    61 &    28 & $-$60.2 &   5.3 & 0.5825 & 0.0014  \\ 
2459740.5983 &   1.6 &   2.8 & 0.4680 & 0.0024 & 0.3508 & 0.0022 & 0.3382 & 0.0021 &    17 &    20 & $-$32.5 &   4.7 & 0.5914 & 0.0015  \\ 
2459742.5986 &  $-$1.2 &   3.2 & 0.4680 & 0.0019 & 0.3555 & 0.0017 & 0.3398 & 0.0017 &   $-$14 &    19 & $-$33.0 &   4.6 & 0.5857 & 0.0020  \\ 
2459744.6076 &  $-$2.2 &   3.0 & 0.4754 & 0.0020 & 0.3620 & 0.0017 & 0.3462 & 0.0016 &   $-$11 &    21 & $-$24.5 &   3.8 & 0.5914 & 0.0018  \\ 
2459747.6096 &  $-$1.0 &   2.9 & 0.4784 & 0.0026 & 0.3664 & 0.0024 & 0.3543 & 0.0023 &    $-$5 &    24 & $-$33.0 &   6.9 & 0.5913 & 0.0017  \\ 
2459753.5963 &   0.1 &   5.5 & 0.4859 & 0.0024 & 0.3734 & 0.0021 & 0.3608 & 0.0021 &    $-$8 &    22 & $-$13.8 &   4.7 & 0.5804 & 0.0032  \\ 
2460007.7051 &  $-$4.6 &   3.9 & 0.4909 & 0.0024 & 0.3776 & 0.0021 & 0.3612 & 0.0021 &    20 &    22 &  11.3 &   4.6 & 0.5895 & 0.0022  \\ 
2460037.6750 &  $-$3.3 &   4.0 & 0.4883 & 0.0041 & 0.3772 & 0.0038 & 0.3621 & 0.0037 &    48 &    42 & $-$19.2 &   8.8 & 0.5855 & 0.0024  \\ 
2460043.6743 &  $-$8.2 &   2.4 & 0.4600 & 0.0030 & 0.3598 & 0.0028 & 0.3472 & 0.0027 &   $-$15 &    31 & $-$50.8 &   7.4 & 0.5863 & 0.0013  \\ 
2460106.5364 &  $-$9.8 &   3.4 & 0.4582 & 0.0031 & 0.3553 & 0.0028 & 0.3427 & 0.0028 &    21 &    30 & $-$61.0 &   7.7 & 0.5875 & 0.0022  \\ 
%--------------------------------------------------------------
\hline
\end{tabular}
\end{scriptsize}
\end{center}
\end{table*}
%-----------------
\setcounter{table}{0}
\begin{table*}
\caption{continued.}
\setlength{\tabcolsep}{3pt} 
\begin{center}
\begin{scriptsize}
\begin{tabular}{lrrccccccrrrrcc}
\hline \hline
%  \noalign{\smallskip}
%--------------------------------------------------------------
BJD & RV & eRV & CaIRT$_1$ & eCaIRT$_1$ & CaIRT$_2$ & eCaIRT$_2$ & CaIRT$_3$ & eCaIRT$_3$ & CRX & eCRX & dLW & edLW &   H$\alpha$ & eH$\alpha$ \\
(d)  & (m\,s$^{-1}$) &  (m\,s$^{-1}$) &  &  &  &  &  &  & (m\,s$^{-1}$\,Np$^{-1}$) & (m\,s$^{-1}$\,Np$^{-1}$) & (m$^2$\,s$^{-2}$) & (m$^2$\,s$^{-2}$) & & \\
\hline
%--------------------------------------------------------------
2460108.5952 &  $-$9.2 &   2.9 & 0.4782 & 0.0018 & 0.3619 & 0.0015 & 0.3501 & 0.0015 &    $-$7 &    18 & $-$11.8 &   4.2 & 0.5894 & 0.0019  \\ 
2460111.5971 &  $-$4.0 &   2.3 & 0.4753 & 0.0029 & 0.3655 & 0.0026 & 0.3473 & 0.0026 &     2 &    26 & $-$26.2 &   5.7 & 0.5834 & 0.0014  \\ 
2460112.6420 &  $-$0.1 &   2.8 & 0.4756 & 0.0024 & 0.3615 & 0.0021 & 0.3478 & 0.0021 &    $-$6 &    22 & $-$33.1 &   5.5 & 0.5823 & 0.0014  \\ 
2460211.4144 &  $-$4.5 &   3.1 & 0.4710 & 0.0019 & 0.3527 & 0.0016 & 0.3408 & 0.0016 &    18 &    16 & $-$23.7 &   4.4 & 0.5834 & 0.0016  \\ 
2460211.5125 &  $-$2.0 &   3.8 & 0.4704 & 0.0018 & 0.3565 & 0.0016 & 0.3374 & 0.0015 &     4 &    20 & $-$26.2 &   3.5 & 0.5907 & 0.0018  \\ 
2460211.6064 &  $-$2.9 &   3.3 & 0.4750 & 0.0022 & 0.3619 & 0.0019 & 0.3501 & 0.0019 &    $-$0 &    19 & $-$22.1 &   4.8 & 0.5893 & 0.0020  \\ 
%--------------------------------------------------------------
\hline
\end{tabular}
\end{scriptsize}
\end{center}
\end{table*}
%---------------------------------------------

%----------------------------------  Fig. A.1
\begin{figure}
  \centering
  \includegraphics[width=0.45\textwidth]{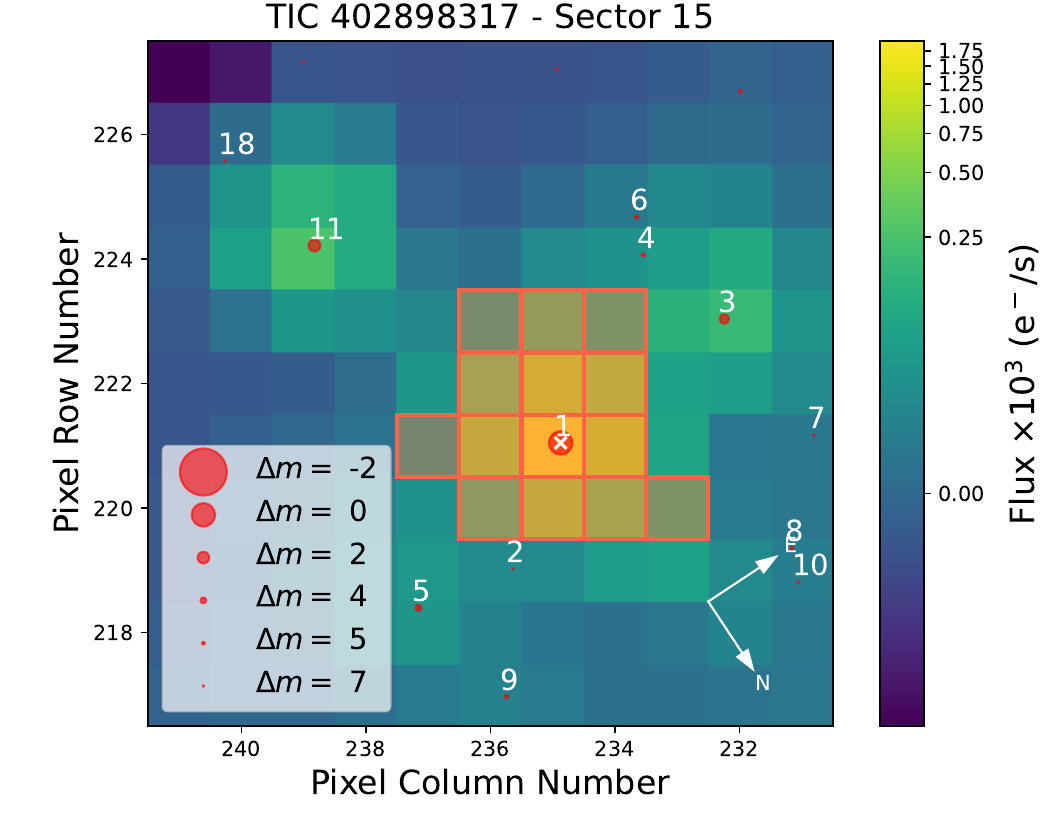}
  \caption{TPF of TOI-2093 (cross) in TESS sector 15. Electron counts are color-coded. TPFs in other sectors are similar. The TESS optimal photometric aperture for this sector used to obtain the SAP light curve is marked with red squares. The {\em Gaia} DR3 objects with $G$-band magnitudes down to 7 mag fainter than TOI-2093 are labeled with numbers (TOI-2093 corresponds to number 1), 
  and their scaled brightness based on {\it Gaia} magnitudes is shown by red circles of different sizes (inset). The pixel scale is 21 arcsec pixel$^{-1}$.}
  \label{fig:tess_tpf_plot} 
\end{figure}
%----------------------------------------------
%

%
%----------------------------------  Fig. A.2
\begin{figure*}
  \centering
  \includegraphics[width=\textwidth]{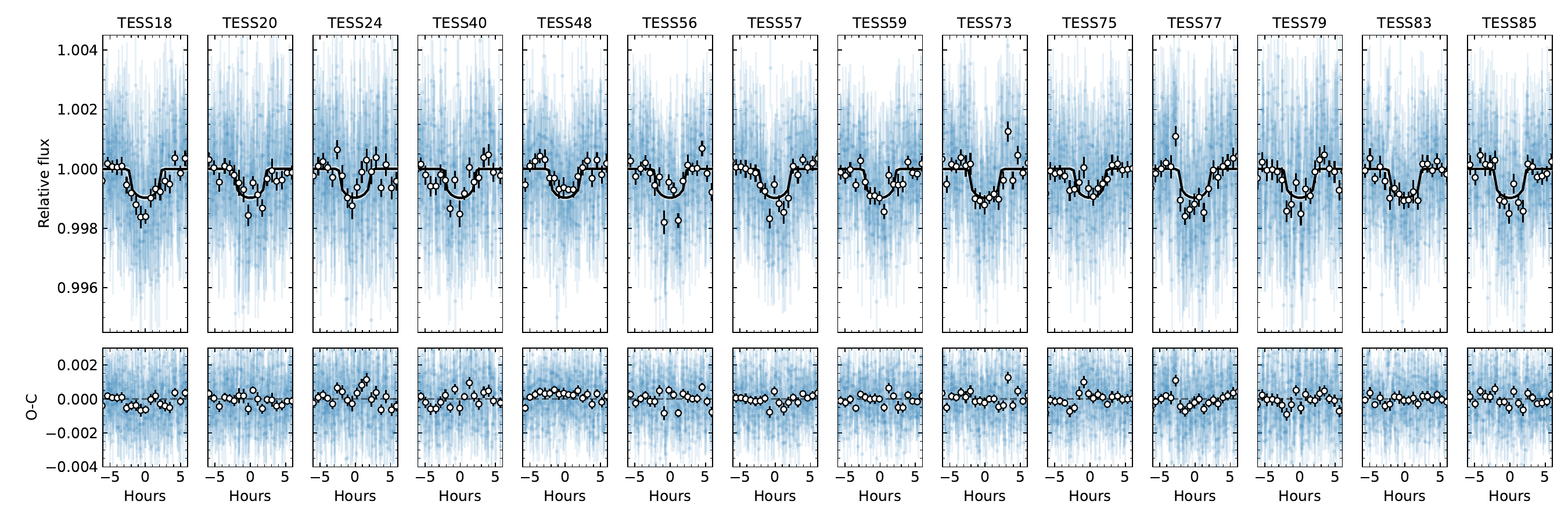}
  \caption{TESS PDCSAP light curves (blue points) folded in phase with the orbital period of the transiting planet per sector. The best joint fit solution is plotted as a black line. The white dots correspond to the binned photometric data. Time is computed from the mid-transit times as derived from the best joint photometric and spectroscopic fit. Residuals are shown at the bottom panels. The TESS sectors are indicated above each panel.}
  \label{fig:tesslc_sectors} 
\end{figure*}
%----------------------------------------------
%

\onecolumn

%
%----------------------------------  Fig. A.3
\begin{figure}
  \centering
  \includegraphics[angle=90,width=0.45\textwidth]{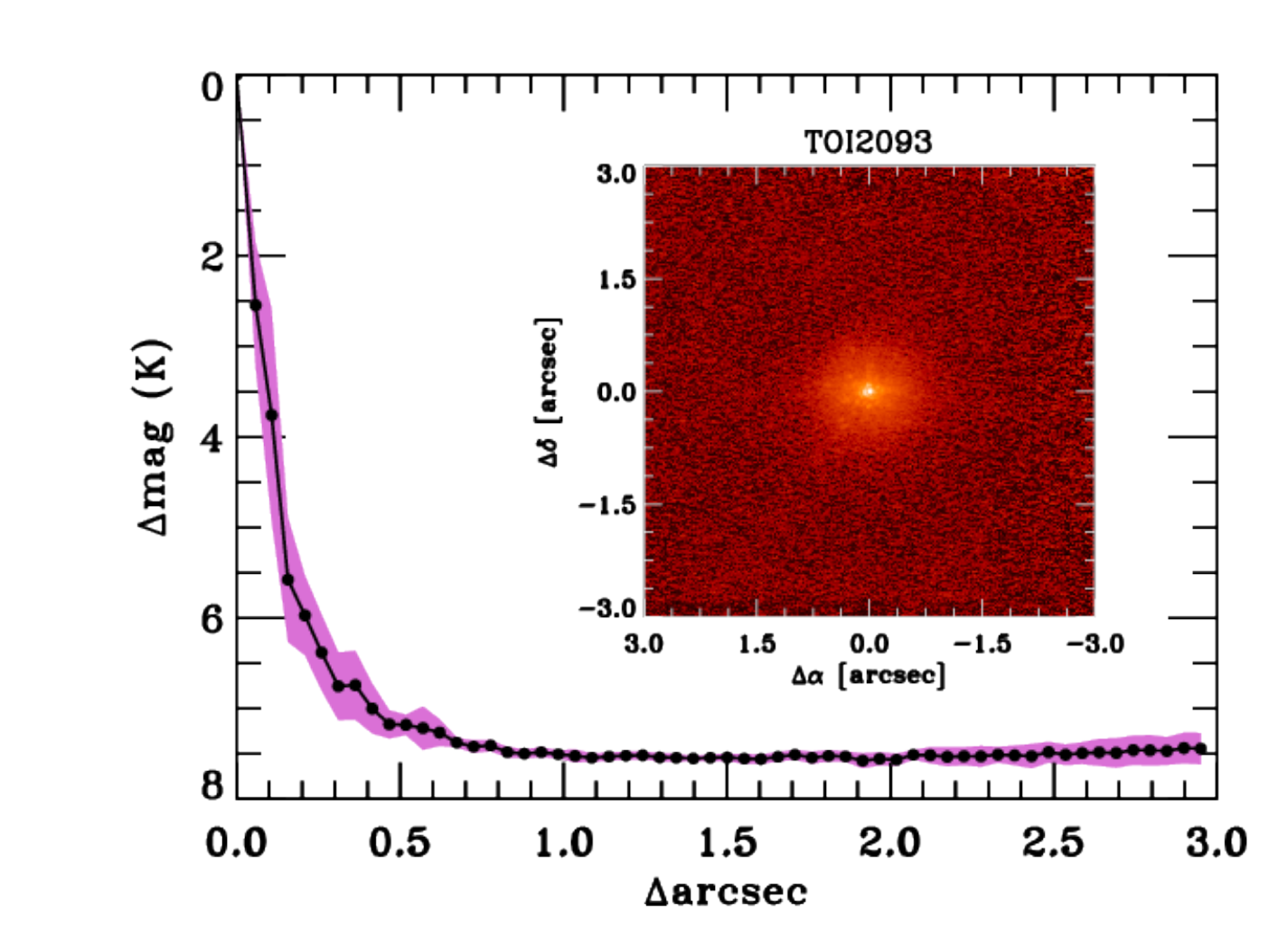}
  \caption{Keck Near Infrared Camera (NIRC2) adaptive optics image and contrast curve in
    the $K$ band of TOI-2093. The dotted curve indicates the magnitude contrast at each angular separation, with uncertainties band in magenta.}\label{fig:nirspec} 
\end{figure}
%----------------------------------------------
%

% Table with rotation periods
%----------------------------------  Table A.2
\begin{table}
\caption[]{TOI-2093 rotation period using different methods.}\label{tab:rotationperiods}
  \renewcommand{\arraystretch}{1.2}
\begin{tabular}{lr}
  \hline \hline
%--------------------------------------------------------------
Method & $P_{\rm rot}$ (d) \\
\hline
%--------------------------------------------------------------
Photometry (GLS)\tablefootmark{a}  & 43.8$\pm$1.8 \\
Photometry (dSHO)  & 42.8$\pm$0.2 \\
Spectroscopic activity indicators (dSHO) &  42.83$^{+1.4}_{-1.3}$ \\
CARMENES NIR RV (GLS) & 43.7$\pm$1.2 \\
%--------------------------------------------------------------
\hline
\end{tabular}
\renewcommand{\arraystretch}{1.}
\tablefoot{\tablefoottext{a}{Adopted value.}}
\end{table}
%--------------------------------------------------------------

\twocolumn

%
%----------------------------------  Fig. A.4
\begin{figure}
  \centering
  \includegraphics[width=0.45\textwidth]{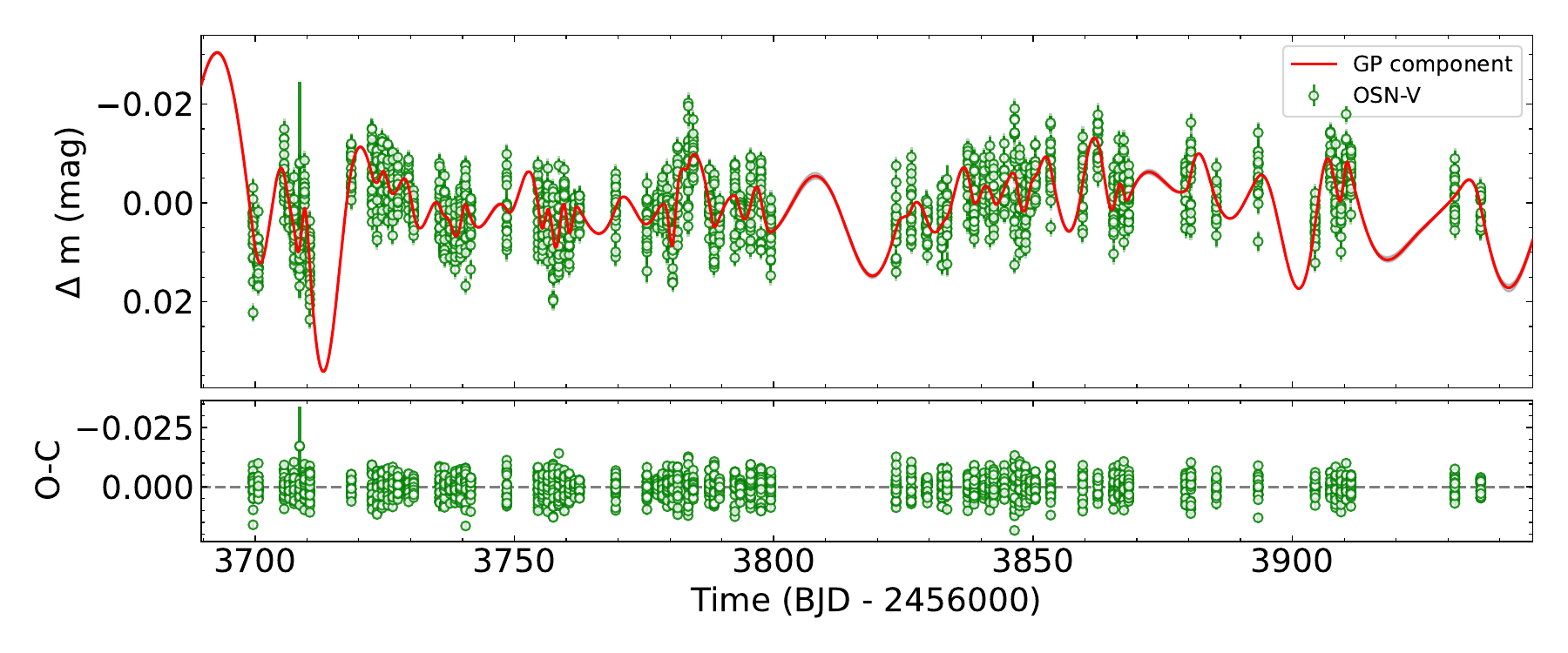}
  \hfill
  \includegraphics[width=0.45\textwidth]{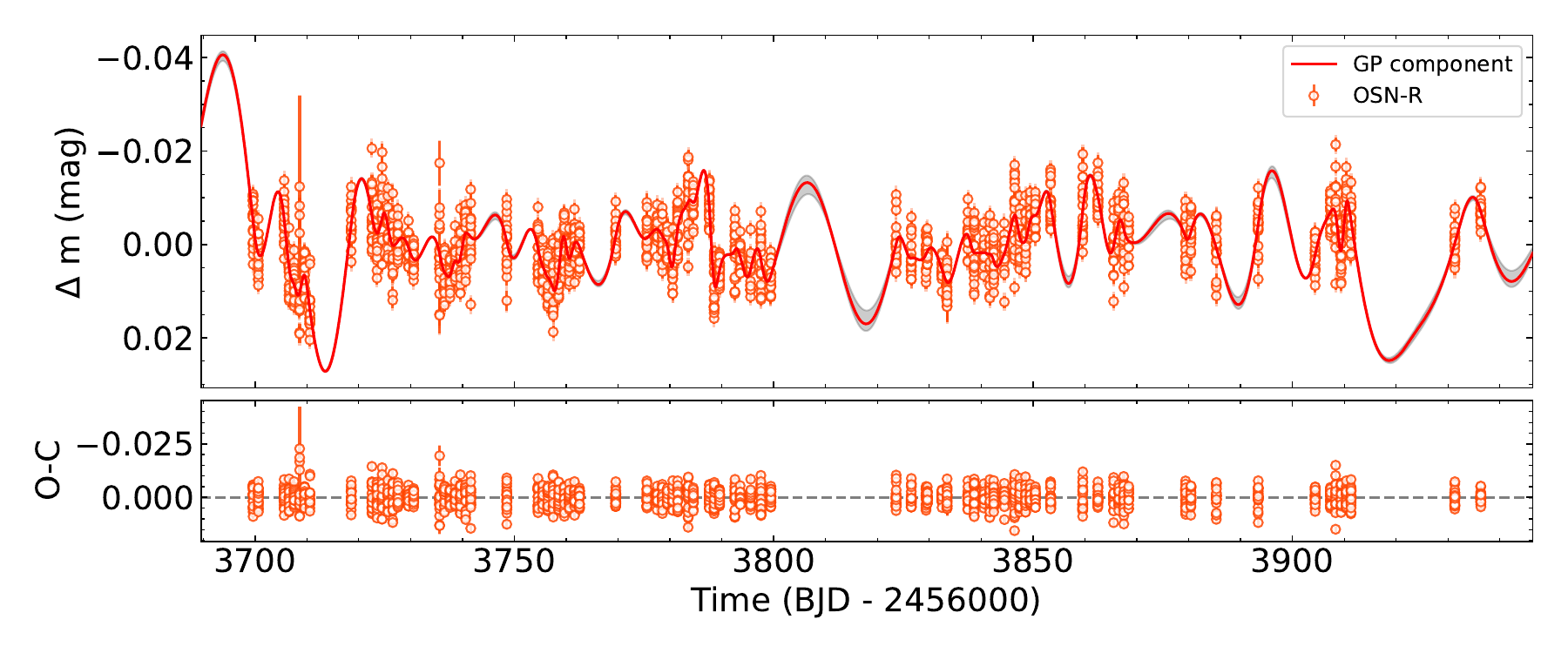}
  \vspace{0.3cm}
  \includegraphics[width=0.45\textwidth]{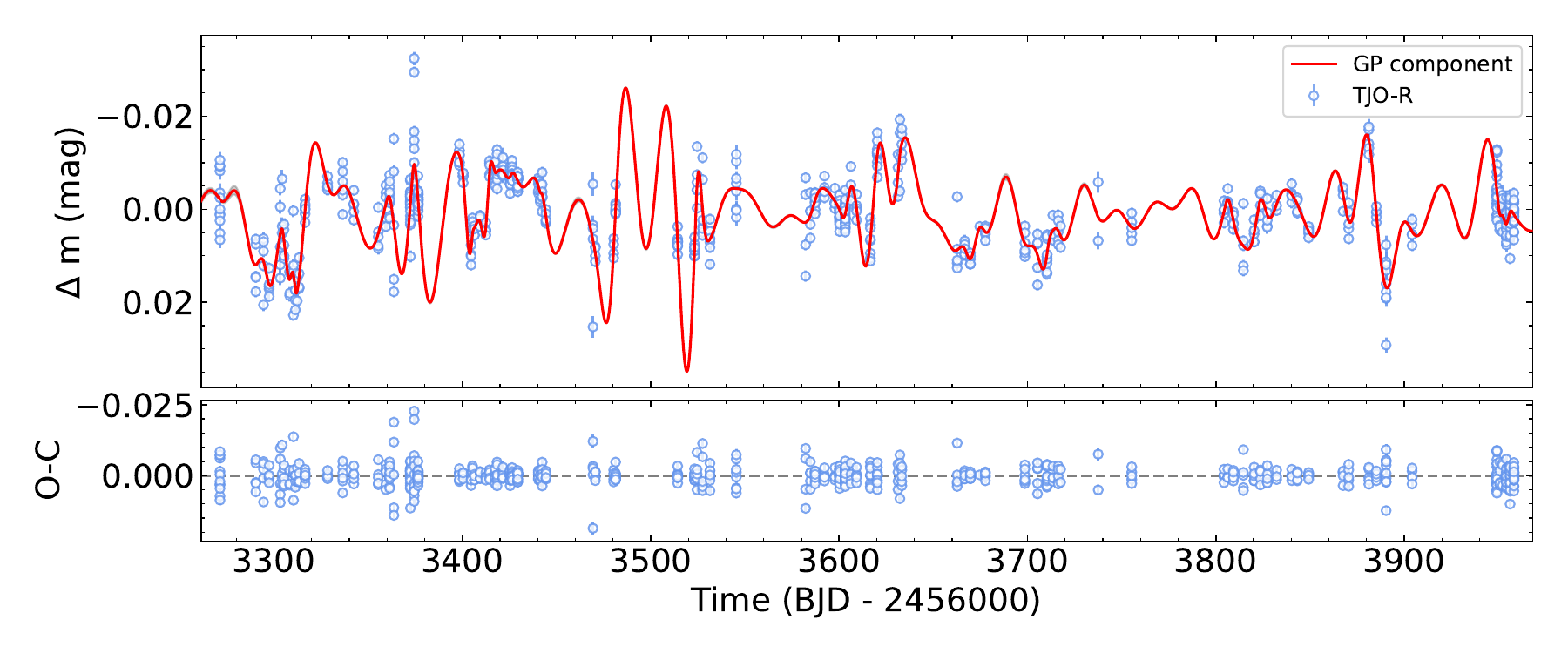}
  \hfill
  \includegraphics[width=0.45\textwidth]{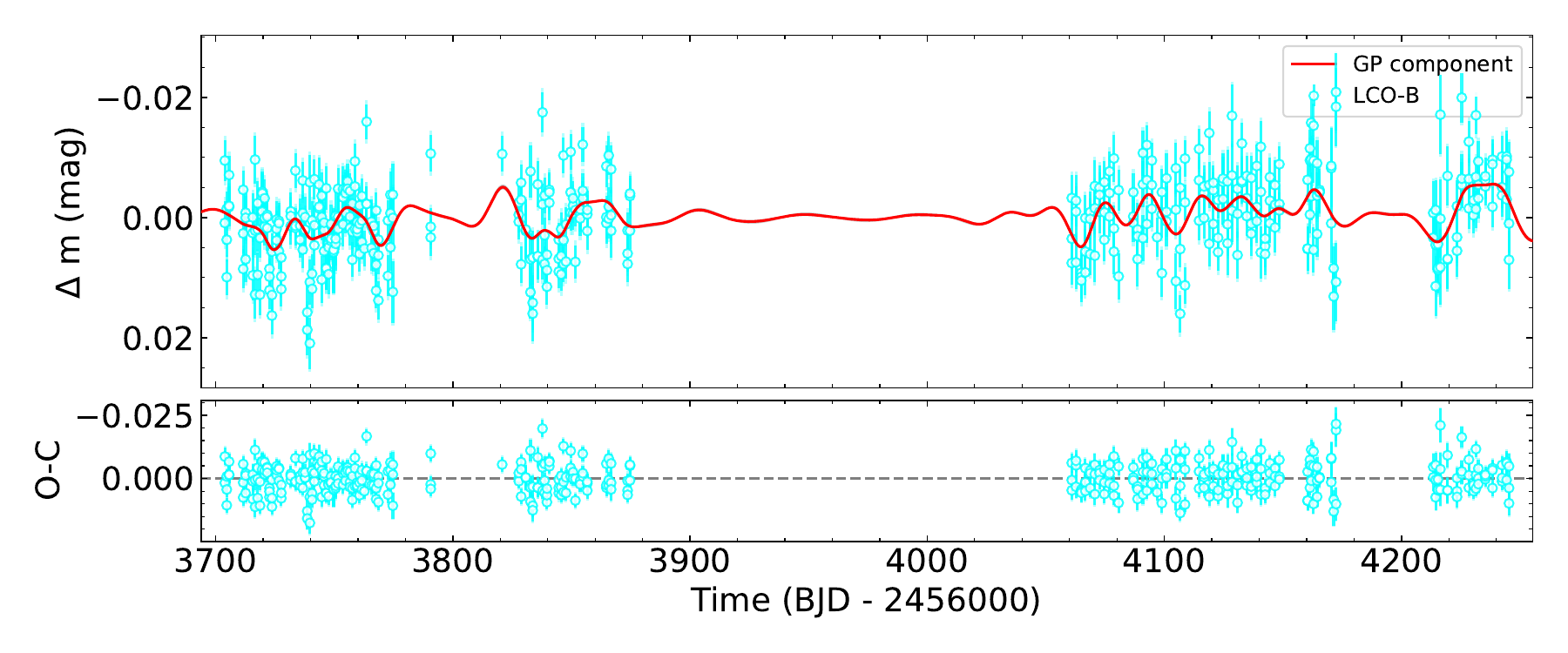}
  \includegraphics[width=0.45\textwidth]{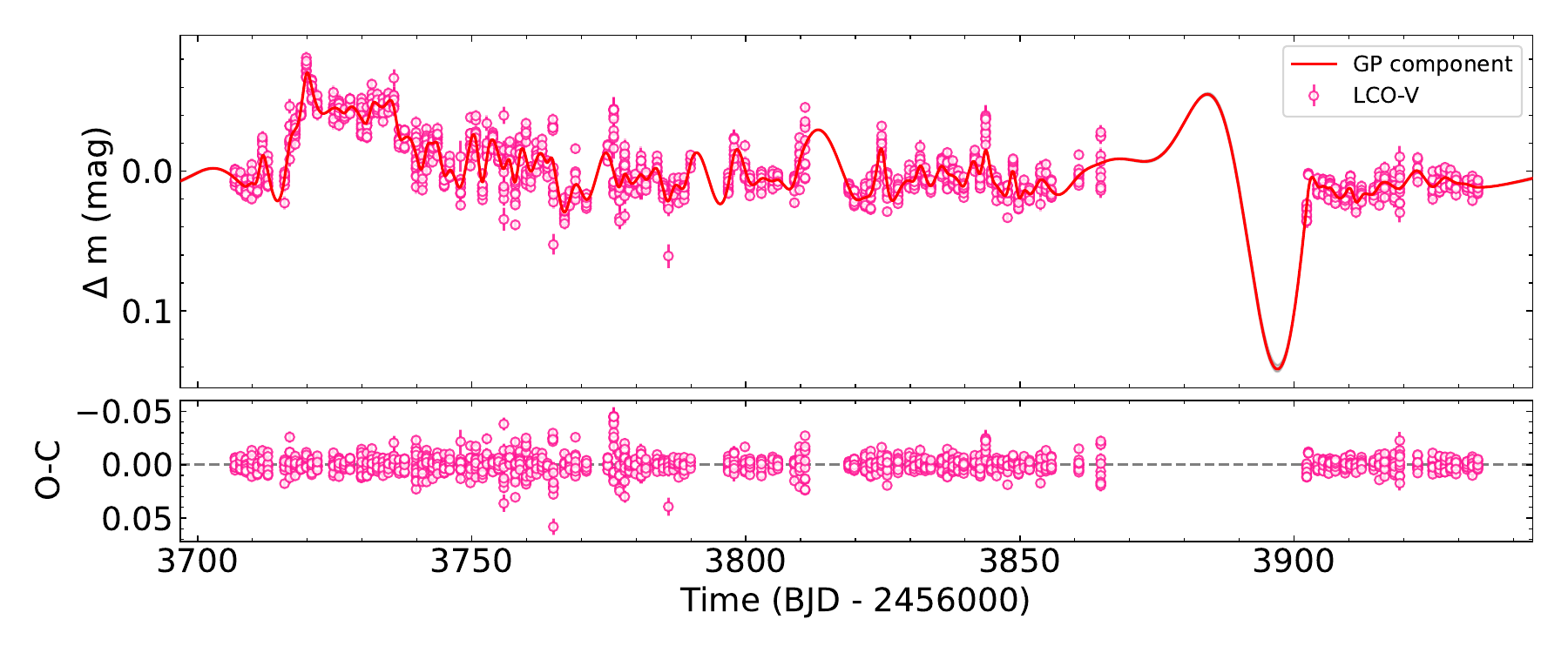}
    \caption{Photometric light curves from ground-based observatories. From top to bottom: OSN $V$, OSN $R$, TJO $R$, LCOGT $B$, and LCOGT $V$. The GP model fit (red line) places the stellar rotation period at $P_{\rm rot} =42.8 \pm$0.2\,d.}
    \label{fig:photometrylc} 
\end{figure}
%------------------------------------------------

%WINDOW RVs
%------------------------------------ Fig. A.5
\begin{figure}
  \centering
  \includegraphics[width=0.45\textwidth]{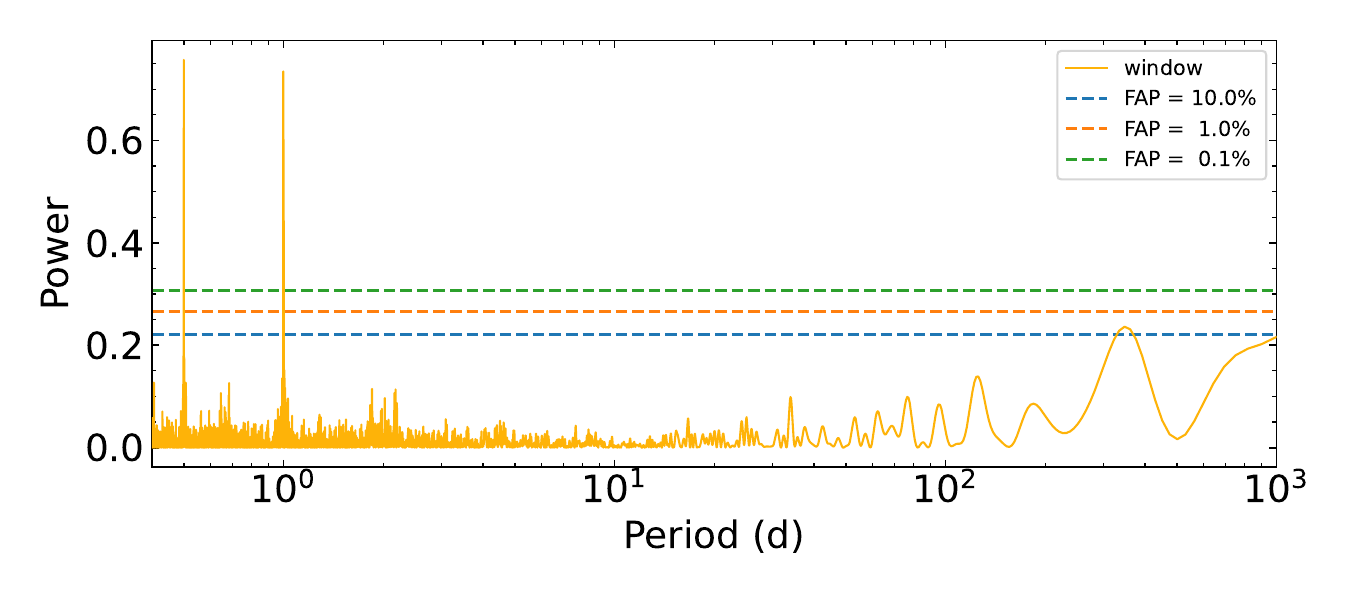}
  \caption{Window function of TOI-2093 CARMENES VIS RV data.}\label{fig:window_carmenes_vis} 
\end{figure}
%----------------------------------------------
%

%
%----------------------------------  Fig. A.6
\begin{figure}
  \centering
  \includegraphics[width=0.45\textwidth]{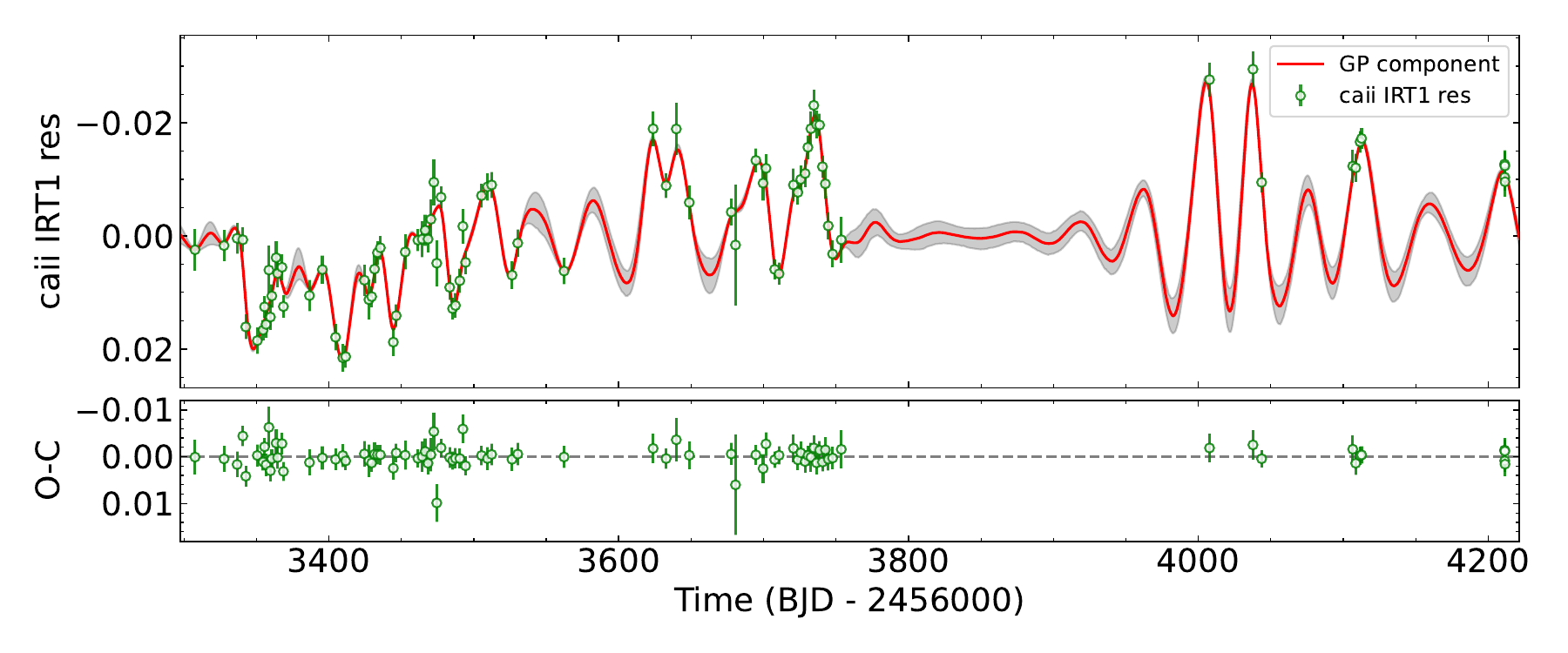}
  \hfill
  \includegraphics[width=0.45\textwidth]{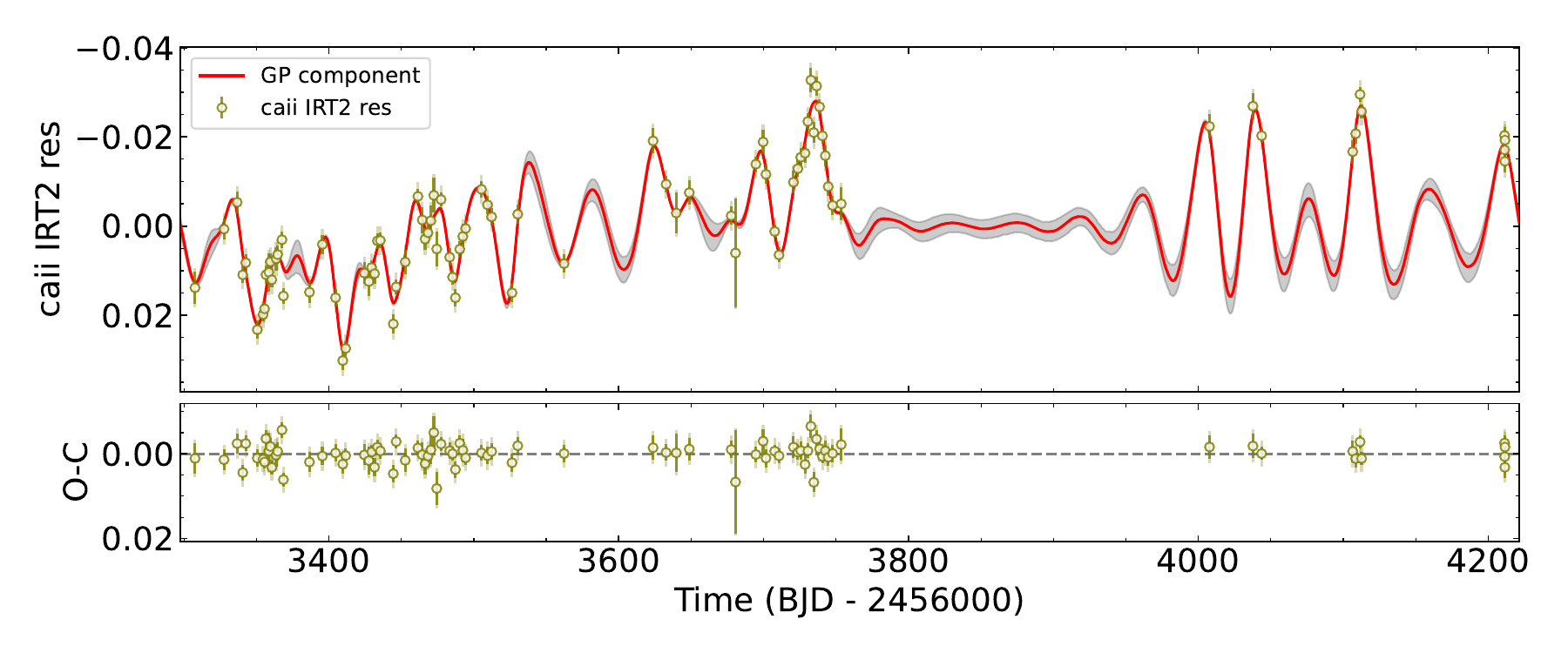}
  \vspace{0.3cm}
  \includegraphics[width=0.45\textwidth]{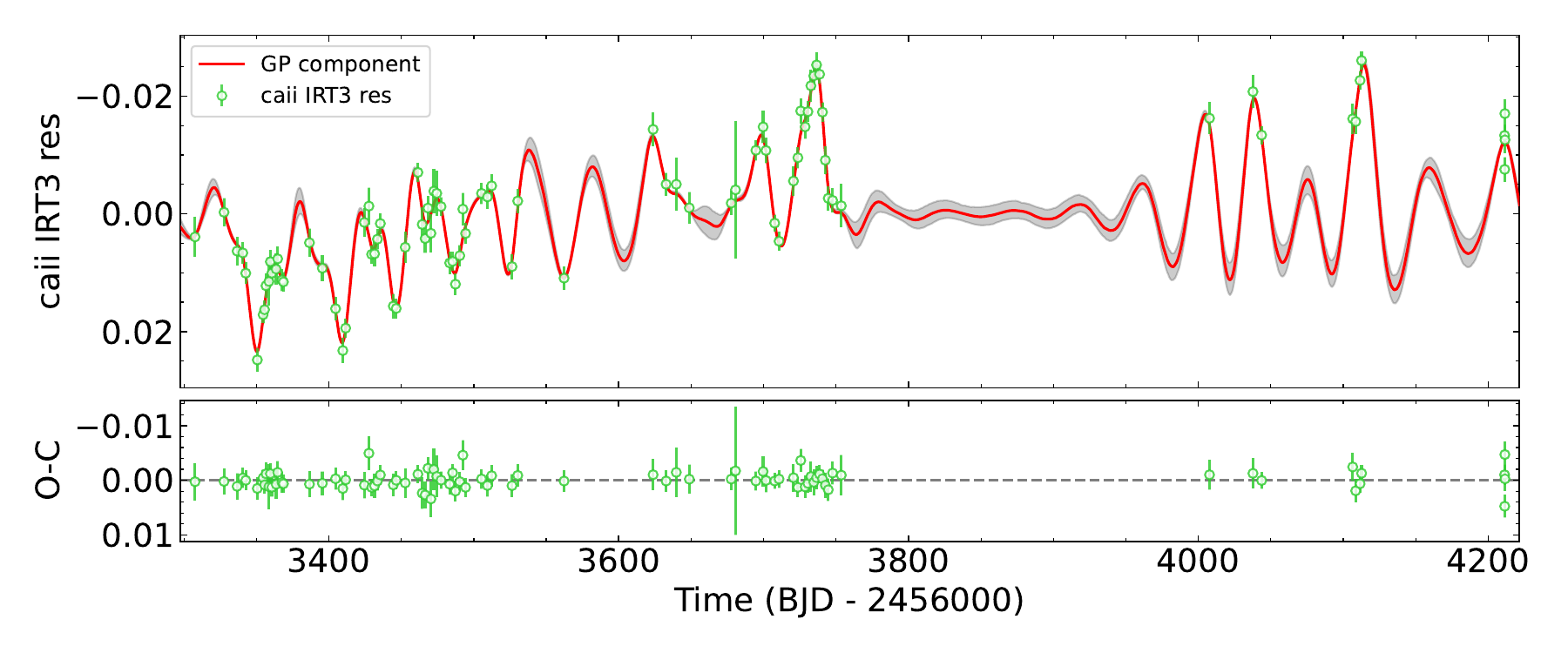}
  \hfill
  \includegraphics[width=0.45\textwidth]{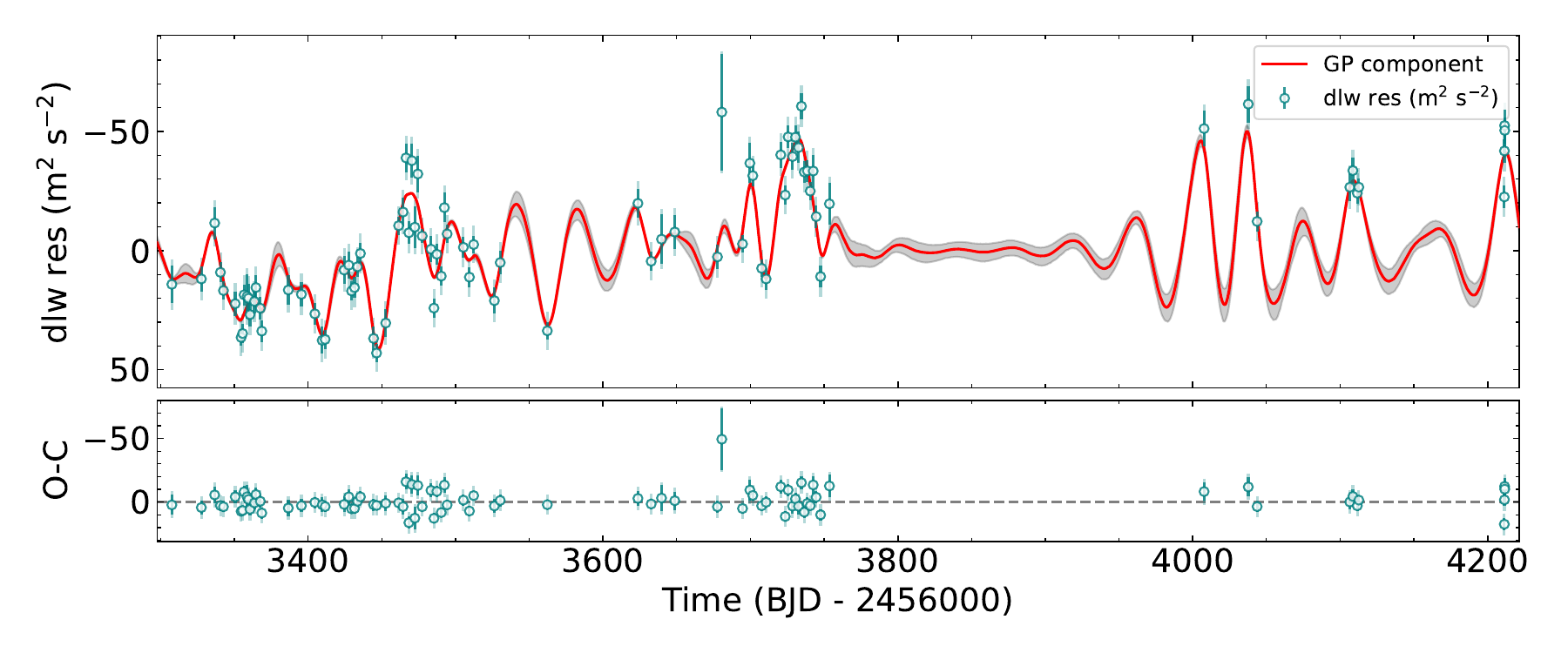}
  \includegraphics[width=0.45\textwidth]{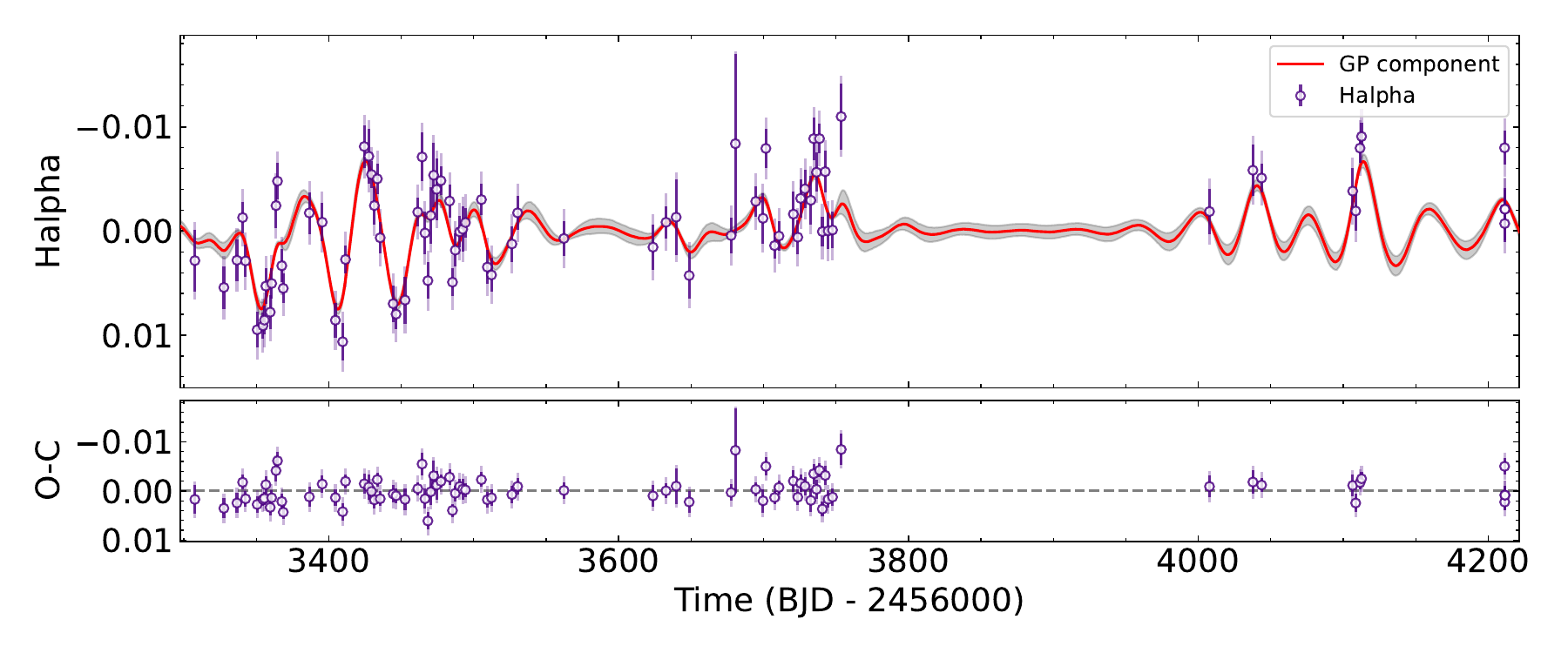}
    \caption{Photometric light curves of different CARMENES spectroscopic activity indicators. From top to bottom: \ion{Ca}{ii}~IRT, dLW, and H$\alpha$. The GP model fit (red line) places the stellar rotation period at $P_{\rm rot} =42.83^{+1.4}_{-1.3}$\,d.}
    \label{fig:activityindicators} 
\end{figure}
%------------------------------------------------

%TTVs
%------------------------------------ Fig. A.7
\begin{figure}
  \centering
  \includegraphics[width=0.45\textwidth]{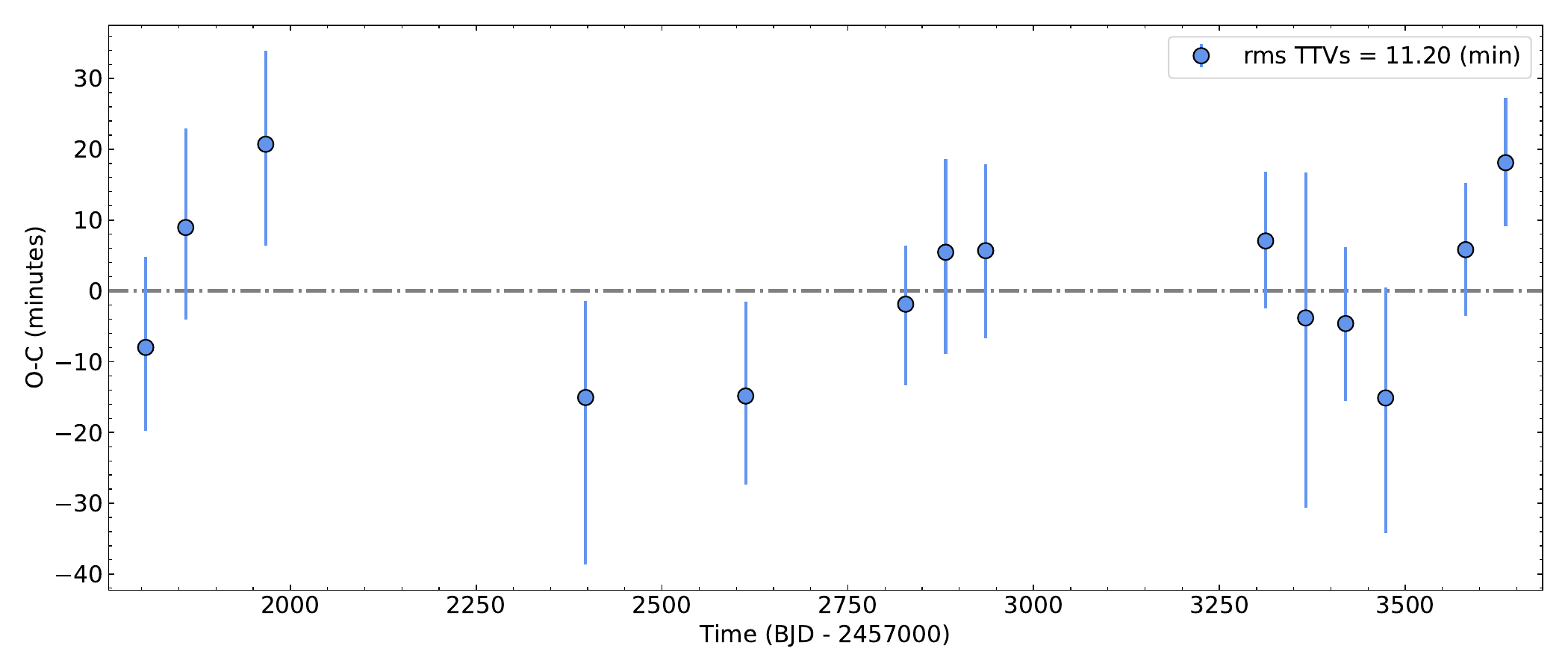}
  \caption{TTVs evolution of TOI-2093\,c over time.}\label{fig:ttvs} 
\end{figure}
%----------------------------------------------
%

% PRIORS TABLE
%------------------------------------------ Table A.3
\begin{table*}
\centering
\caption{Priors and posteriors of the joint transit and RV fit of TOI-2093.}
\label{tab:priors+posteriors}
\renewcommand{\arraystretch}{1.3}
\tabcolsep 4.0 pt
\begin{small}
\begin{tabular}{l c c c c l}
\hline
\hline
%------------------------------------------------
Parameter & Prior & \multicolumn{2}{c}{Posterior} & Unit & Description \\
          &       &  $e=0$ & Free $e$  & & \\
\hline  
%------------------------------------------------
\multicolumn{6}{c}{\textit{Stellar parameter}} \\

$\rho_{\star}$ 	& $\mathcal{N}$(2.71097, 0.45015) & 2.85$^{+0.49}_{-0.46}$ & 2.70$^{+0.43}_{-0.45}$ & g\,cm$^{-3}$ & Stellar density \\

\multicolumn{6}{c}{\textit{Photometric parameters}} \\
%------------------------------------------------

$\mu_{\rm TESS}$ & $\mathcal{N}$(0, 0.1) & \multicolumn{2}{c}{\dots} & \dots & The offset relative flux for \textit{TESS}\\
$\sigma_{\rm TESS}$ & $\mathcal{L} \mathcal{U}$(10$^{-6}$, 0.04)  & \multicolumn{2}{c}{\dots} &  \dots & The jitter for the photometric instrument \\
		
$q1_{\rm TESS}$ & $\mathcal{U}$(0, 1)  	& 0.61$^{+0.24}_{-0.25}$ & 0.59$^{+0.25}_{-0.26}$ &  \dots & Limb-darkening for photometric instrument\\
$q2_{\rm TESS}$ & $\mathcal{U}$(0, 1) & 0.44$\pm$0.29 & 0.42$^{+0.28}_{-0.25}$ &   \dots & Limb-darkening for photometric instrument\\
$D_{\rm TESS}$ 	& 1 (fixed)  	& \multicolumn{2}{c}{\dots}	& \dots	& The dilution factor\\

\multicolumn{6}{c}{\textit{RV instrumental parameters}} \\
%------------------------------------------------

$\gamma$ 	&  $\mathcal{U}$(--10, 10) & $-0.01^{+0.55}_{-0.57}$ & $-0.07^{+0.52}_{-0.59}$ & m\,$\rm s^{-1}$ & RV zero point for CARMENES\\
$\sigma$	& $\mathcal{L} \mathcal{U}$(0.001, 5) & 0.41$^{+2.22}_{-0.40}$ & 0.39$^{+2.02}_{-0.37}$ & m\,$\rm s^{-1}$ & A jitter added in quadrature \\

\multicolumn{6}{c}{\textit{RV dSHO GP parameters}} \\
%------------------------------------------------

$GP_{\sigma}$ 	& $\mathcal{U}(0.00, 15.00)$ & 5.0$^{+1.1}_{-1.2}$ & 5.0$^{+1.2}_{-1.1}$  & m\,$\rm s^{-1}$ & Standard deviation of the GP \\
$GP_{Q_{0}}$ 	& $\mathcal{U}(2.00, 20.00)$ & 8.8$^{+6.7}_{-4.0}$ & 9.9$^{+5.7}_{-5.0}$ & \dots & The quality factor \\
$GP_{\rm star\,rotation}$ 	& $\mathcal{N}(19.60, 1.20)$ & 19.57$^{+0.35}_{-0.39}$ & 19.46$^{+0.38}_{-0.35}$ & d & The primary period of variability \\
$GP_{f}$ 	& $\mathcal{U}(0.00, 1.00)$  & 0.67$^{+0.22}_{-0.26}$ & 0.64$^{+0.23}_{-0.27}$ & \dots & The fractional amplitude of the secondary mode \\
$GP_{dQ}$ 	& $\mathcal{U}(0.10, 30.00)$ & 15.4$^{+9.7}_{-10.0}$ & 14.9$^{+9.8}_{-8.8}$ & \dots & The difference between the quality factors \\

\multicolumn{6}{c}{\textit{Planet $b$ fit parameters} }\\   
%------------------------------------------------

$P$ 	& $\mathcal{U}$ (12, 13.4) &	12.836$\pm$0.021 & $12.801^{+0.036}_{-0.042}$ & d		& Period \\
$T_0$ (BJD\text{--}2,457,000) 	& $\mathcal{U}$ (2317, 2330) & 2320.53$^{+0.43}_{-0.50}$ & $2320.74^{+0.57}_{-0.48}$ &d &  Time of periastron passage\\
$e$ 		& $\mathcal{U}$ (0, 1) & 0 (fixed) & $0.17^{+0.27}_{-0.12}$ & \dots & Orbital eccentricity\\
$\omega$ 	&$\mathcal{U}$ (0, 360) & 90 (fixed)  & $119^{+146}_{-67}$   & deg & Periastron angle\\
$K$	        &  $\mathcal{U}$ (0, 15)  & 3.52$^{+0.79}_{-0.84}$ & $3.71^{+1.01}_{-0.85}$  &  m\,$\rm s^{-1}$    	 & RV semi-amplitude\\   

\multicolumn{6}{c}{\textit{Planet $c$ fit parameters} }\\   
%------------------------------------------------

$P$ 	& $\mathcal{N}$ (53.81, 0.01) &	 53.81149$\pm$0.00017 & $53.81149^{+0.00019}_{-0.00016}$ & d		& Period \\
%\noalign{\smallskip}
$T_0$ (BJD\text{--}2,457,000) 	& $\mathcal{N}$ (1751.62, 0.01) & 1751.6122$^{+0.0040}_{-0.0042}$ & $1751.6124^{+0.0044}_{-0.0048}$ & d & Central transit time\\
$e$ 		& $\mathcal{U}$ (0, 1) 	& 0 (fixed) & $0.23 \pm 0.10$ & \dots & Orbital eccentricity\\
$\omega$ 	& $\mathcal{U}$ (0, 360) & 90 (fixed) & $148 \pm 34$  & deg	& Periastron angle\\
$K$	        &  $\mathcal{U}$ (0, 15)  & 3.26$^{+0.74}_{-0.80}$ & $3.83^{+0.78}_{-0.81}$  & m\,$\rm s^{-1}$    	 & RV semi-amplitude\\   
$r_1$		&  $\mathcal{U}$ (0, 1)  & 0.706$^{+0.047}_{-0.063}$ &  0.57$\pm$0.15	& \dots  & Parameterization for $p$ and $b$  \\   
$r_2$ 		&  $\mathcal{U}$ (0, 1)  & 0.0290$^{+0.0010}_{-0.0011}$ & 0.0282$^{+0.0013}_{-0.0011}$  & \dots   & Parameterization for $p$ and $b$    \\   
\noalign{\smallskip}
\hline  
\end{tabular}
%------------------------------------------------
\renewcommand{\arraystretch}{1.}
\tablefoot{The prior labels of $\mathcal{N}$, $\mathcal{U}$, and $\mathcal{LU}$ represent normal, uniform and log-uniform distributions, respectively. The jitter was added in quadrature to the error bars of the instrument. The dilution factor was fixed to one for all TESS sectors. The offset between the different TESS sectors was modeled with a normal prior $\mathcal{N}(0, 1)$, and the upper limit on the photometric jitter term was set to three times the error bars of each individual sector following a log-uniform distribution.}
\end{small}
\end{table*} 
%--------------------------------------------------------------

%All TESS LCs folded at planet b
%------------------------------------ Fig. A.8
\begin{figure}
  \centering
  \includegraphics[width=0.45\textwidth]{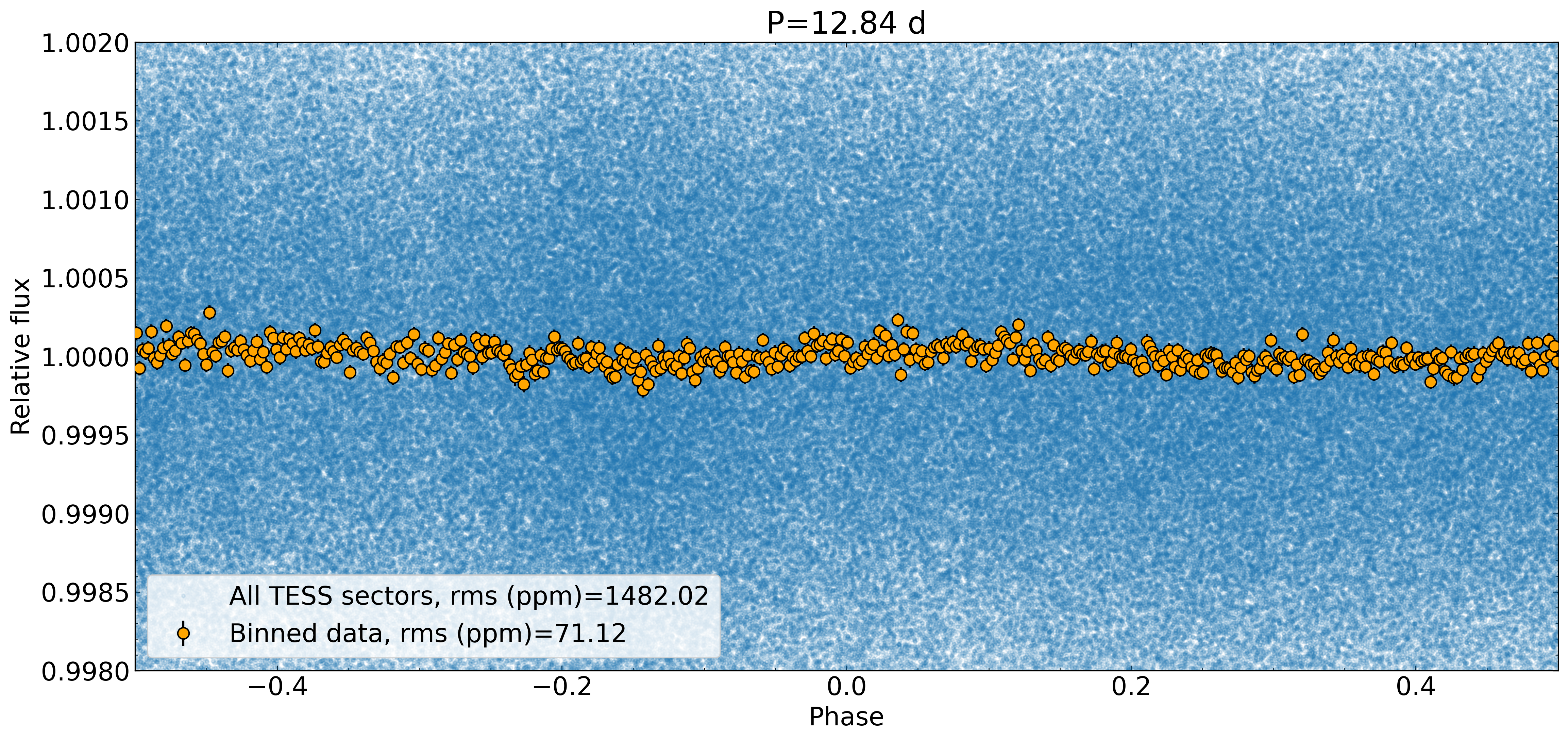}
  \caption{PDCSAP TESS light curves folded in phase at the orbital period of TOI-2093\,b. The search was made considering variations of up to 3$\sigma$ in the period. The transiting planet signal at 53\,d was previously removed.}\label{fig:lc_vs_phase_12.8} 
\end{figure}
%----------------------------------------------
%

%Corner plot
%----------------------------------  Fig. A.9
\begin{figure*}
  \centering
  \includegraphics[width=\textwidth]{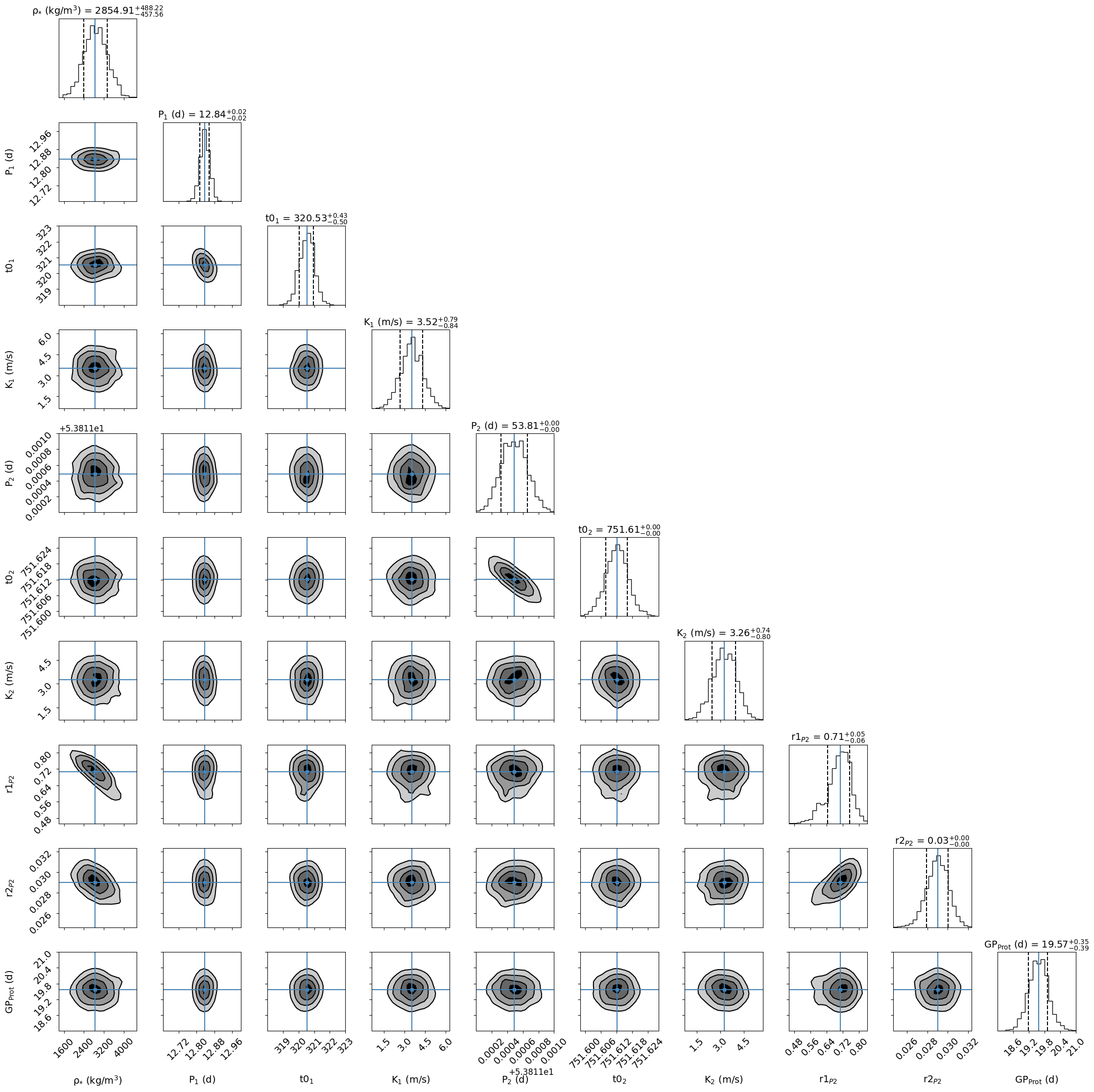}
  \caption{Posterior distributions of the principal fit planetary parameters of the TOI-2093 system as obtained from the joint combined photometric and spectroscopic fit, assuming circular orbits. The vertical dashed lines indicate the 16, 50, and 84\,\%~quantiles that were used to define the optimal values and their associated 1$\sigma$ uncertainty. The blue line stands for the median 
  values (50\% quantile) of each fit parameter.}
  \label{fig:toi2093_cornerplot}
\end{figure*}
%----------------------------------------------

%Corner plot
%----------------------------------  Fig. A.10
\begin{figure*}
  \centering
  \includegraphics[width=\textwidth]{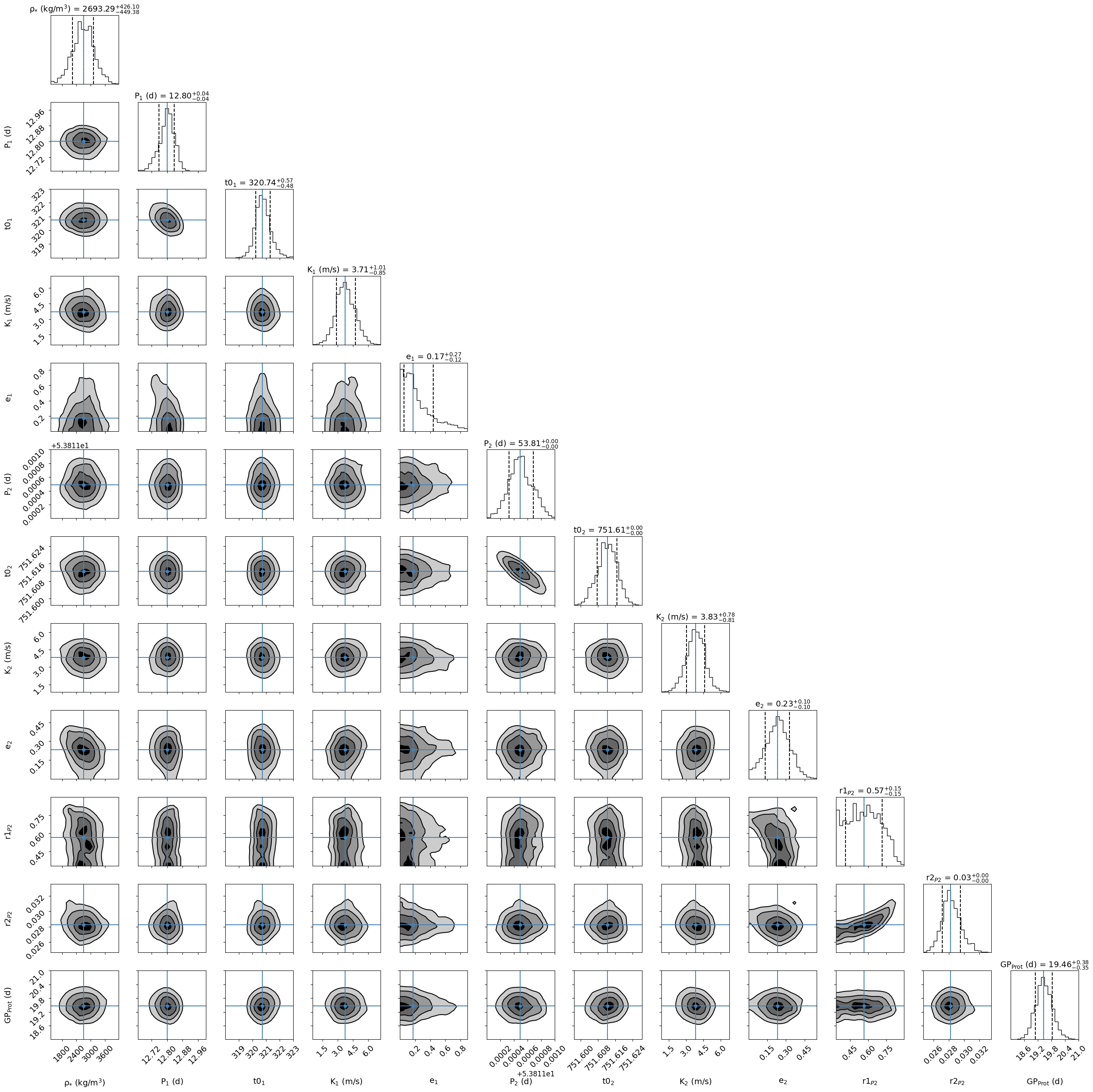}
  \caption{Same as in Fig.~\ref{fig:toi2093_cornerplot}, but including the eccentricities of the two planets as free parameters.}
  \label{fig:toi2093_cornerplot_ecc}
\end{figure*}
%----------------------------------------------

%----------------------------------  Table A.4
\begin{table*}[t]
  \caption{Planets with $R_{\rm p}<$ 5\,R$_\oplus$ and/or $M_{\rm p}<$ 24\,M$_\oplus$ in the HZ of FGK stars.}\label{tab:planetsHZ}
\vspace{-3mm}
\begin{center}
  \renewcommand{\arraystretch}{1.3}
    \setlength{\tabcolsep}{3pt} 
    \begin{scriptsize}
\begin{tabular}{lcccccccl}
\hline\hline
%--------------------------------------------------------------
Name & SpT & $T_{\rm eff}$ (K) & $T_{\rm eq}$ (K) & $R_{\rm p}$ (R$_\oplus$) & $M_{\rm p}$ (M$_\oplus$) & $a$ (au) & $P$ (d) & Reference\\
\hline
%--------------------------------------------------------------
TOI-2093\,c & K5V & 4426$\pm$85 & 329 & 2.30$\pm$0.12 & 15.8$^{+3.6}_{-3.8}$ &0.257$^{+0.017}_{-0.018}$  & 53.81149$\pm$0.00017 & This work \\
\object{Kepler-411\,A\,d} & K2V & 4837$^{+150}_{-127}$ & 368 & 3.38$^{+0.12}_{-0.12}$ & 15.2$\pm 5.1$ & 0.2757$\pm 0.0040$ & 58.02035$\pm 0.00056$ & \citet{sun19} \\
\object{Kepler-413\,(AB)\,b} & K + M & 4700/3500 & 306 & 4.35$\pm 0.10$ & 67$\pm 21$ & 0.3553$^{+0.0020}_{-0.0018}$ & 66.262$^{+0.024}_{-0.021}$ & \citet{kos14} \\
\object{Kepler-1662\,c} & G0V  & 5922$\pm 60$ & 302 & 5.44$^{+0.52}_{-0.30}$ & 15.0$^{+ 4.3}_{- 3.6}$ & 0.8539 & 284.06095$\pm 0.00119$ & \citet{vis20} \\
\object{Kepler-47\,(AB)\,d} & G6V + M3V & 5636/3357 & 292 & 7.04$^{+0.66}_{-0.49}$ & 19$^{+24}_{-12}$ & 0.6992$\pm 0.0033$ & 187.35$\pm 0.15$ & \citet{oro19} \\
\object{Kepler-90\,g} & F8V & 6080$^{+260}_{-170}$ & 349 & 8.10$\pm 0.80$ & 14.9$\pm 1.3$ & 0.710$\pm 0.080$ & 210.60697$\pm 0.00043$ & \citet{shaw25} \\
\object{HIP 41378\,f} & F7V & 6199$\pm 50$ & 276 & 9.20$\pm 0.10$ & 12.1$\pm 2.9$ & 1.370$\pm 0.020$ & 542.07975$\pm 0.00014$ & \citet{how25} \\
\object{Kepler-51\,d} & G3V & 5670$\pm 60$ & 329 & 9.70$\pm 0.50$ & 7.63$\pm 0.95$ & 0.509$\pm 0.020$ & 130.1940$^{+0.0020}_{-0.0050}$ & \citet{mas14} \\
%--------------------------------------------------------------
\hline
\end{tabular}
\renewcommand{\arraystretch}{1.}
\tablefoot{Only HIP~41378\,f and TOI-2093\,c masses were measured using RVs. The rest of planet masses were calculated using TTVs or other methods. \object{Kepler-30\,d} and \object{Kepler-55\,A\,c} were excluded because their masses are only upper limits.}
\end{scriptsize}
\end{center}
\end{table*}
%---------------------------------------------

%TTVs
%------------------------------------ Fig. A.11
\begin{figure*}
  \centering
  \includegraphics[width=0.45\textwidth]{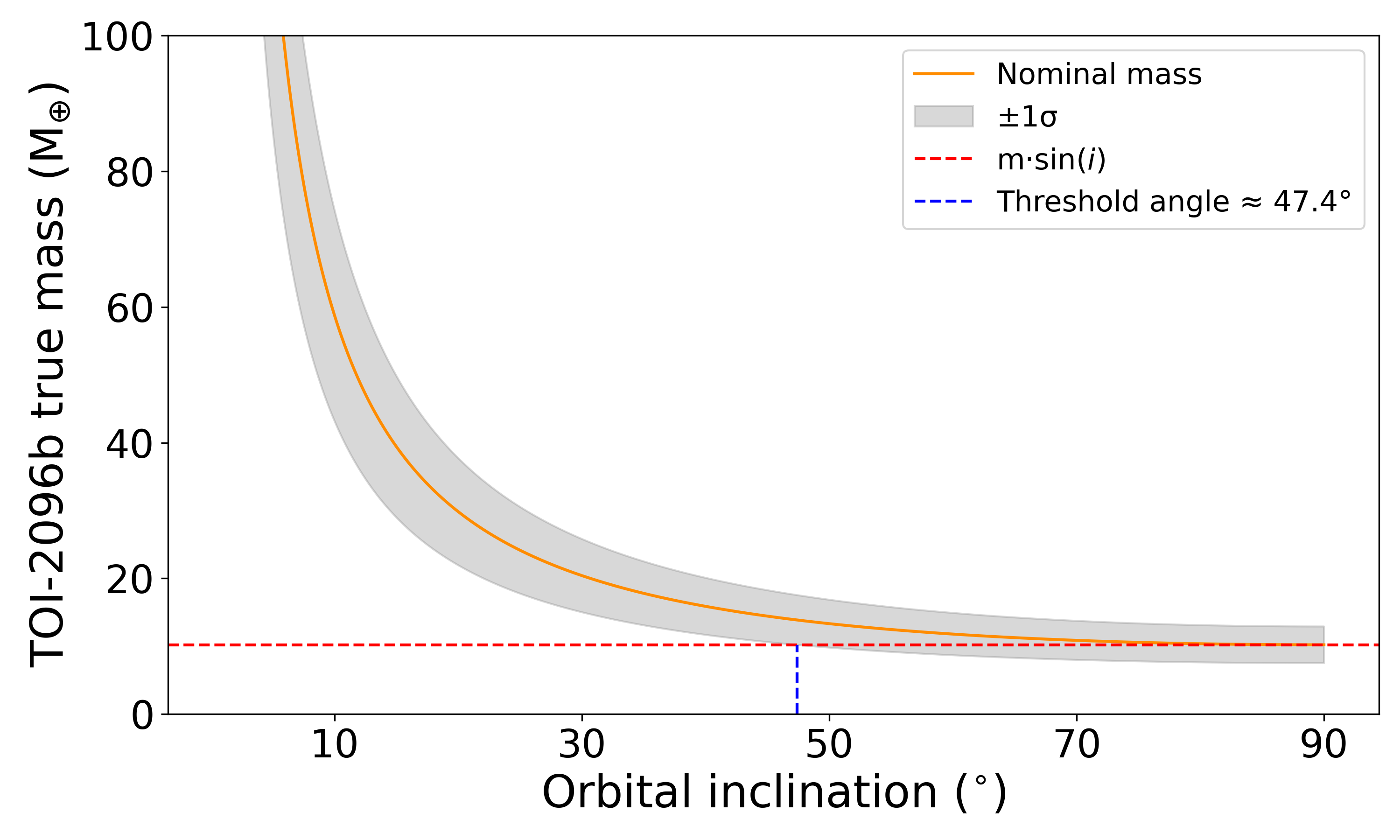}
  \caption{True mass of TOI-2093 b as a function of orbital inclination angle.
    A vertical dashed line indicates the minimal inclination that ensures dynamical stability, as described in Sect.~\ref{sec:TOI-2093 planetary system}}\label{fig:stability}
\end{figure*}
%----------------------------------------------
%

%
%----------------------------------  Fig. A.12
\begin{figure*}
  \centering
  \includegraphics[width=0.45\textwidth]{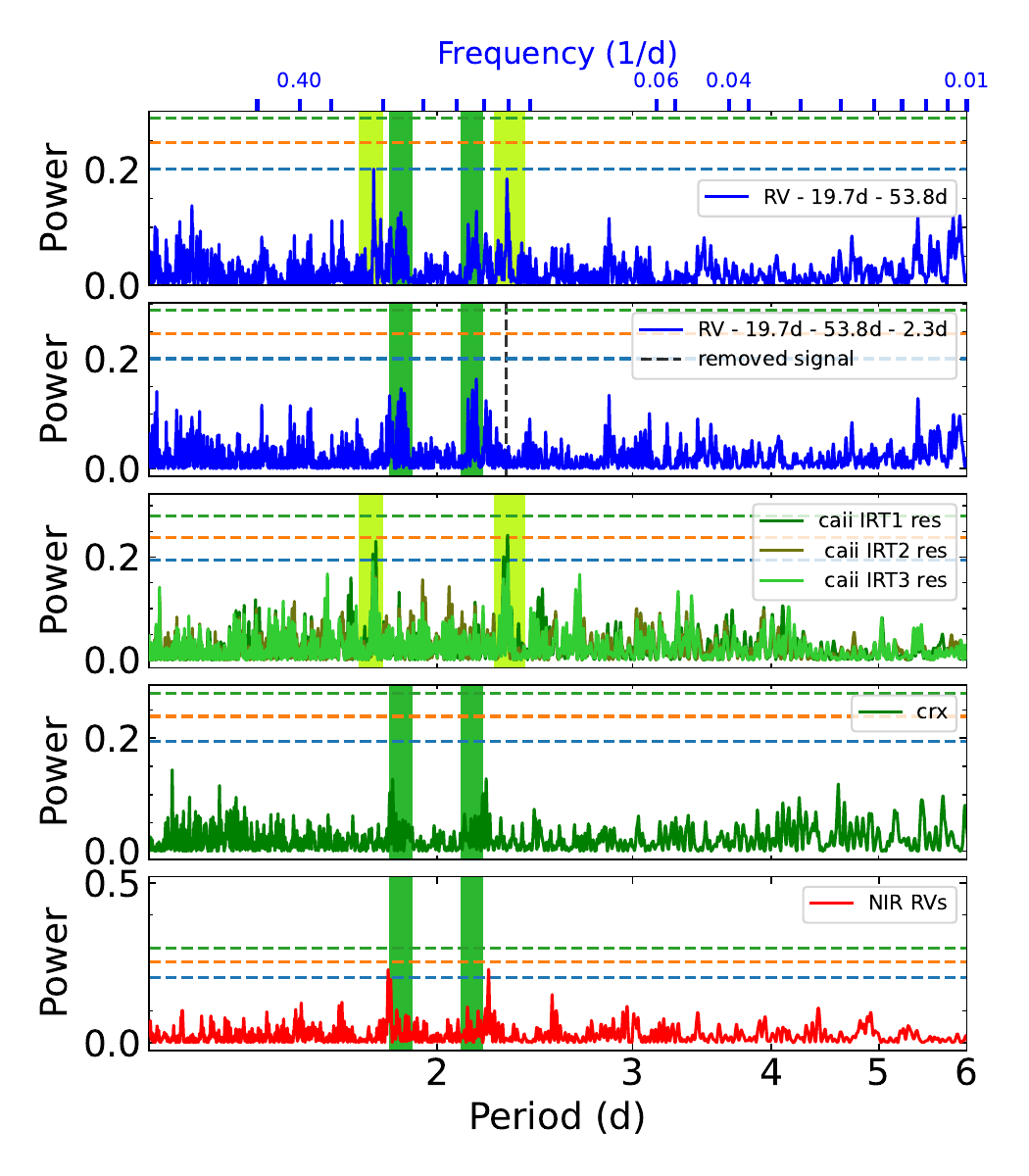}
    \caption{GLS periodogram as in Fig.~\ref{fig:glsgeneral}, focused on the region around 2\,d for different indicators. Light green bands indicate periodicities detected as
    spectroscopic and photometric signs of stellar activity. Dark green bands indicate periodicities detected in the VIS RV data, and possibly related to a planet (see Sect.~\ref{sect:constraints}).}
    \label{fig:glszoom} 
\end{figure*}
%------------------------------------------------

\end{appendix}


\begin{thebibliography}{89}
\expandafter\ifx\csname natexlab\endcsname\relax\def\natexlab#1{#1}\fi

\bibitem[{{Aller} {et~al.}(2020){Aller}, {Lillo-Box}, {Jones}, {Miranda}, \&
  {Barcel{\'o} Forteza}}]{Aller20}
{Aller}, A., {Lillo-Box}, J., {Jones}, D., {Miranda}, L.~F., \& {Barcel{\'o}
  Forteza}, S. 2020, \aap, 635, A128

\bibitem[{{Angus} {et~al.}(2018){Angus}, {Morton}, {Aigrain}, {Foreman-Mackey},
  \& {Rajpaul}}]{ang18}
{Angus}, R., {Morton}, T., {Aigrain}, S., {Foreman-Mackey}, D., \& {Rajpaul},
  V. 2018, \mnras, 474, 2094

\bibitem[{{Barnes}(2017)}]{bar17}
{Barnes}, R. 2017, Celestial Mechanics and Dynamical Astronomy, 129, 509

\bibitem[{{Bauer} {et~al.}(2020){Bauer}, {Zechmeister}, {Kaminski},
  {Rodr{\'\i}guez L{\'o}pez}, {Caballero}, {Azzaro}, {Stahl}, {Kossakowski},
  {Quirrenbach}, {Becerril Jarque}, {Rodr{\'\i}guez}, {Amado}, {Seifert},
  {Reiners}, {Sch{\"a}fer}, {Ribas}, {B{\'e}jar}, {Cort{\'e}s-Contreras},
  {Dreizler}, {Hatzes}, {Henning}, {Jeffers}, {K{\"u}rster}, {Lafarga},
  {Montes}, {Morales}, {Schmitt}, {Schweitzer}, \& {Solano}}]{Bau20}
{Bauer}, F.~F., {Zechmeister}, M., {Kaminski}, A., {et~al.} 2020, \aap, 640,
  A50

\bibitem[{{Benedict} {et~al.}(2016){Benedict}, {Henry}, {Franz}, {McArthur},
  {Wasserman}, {Jao}, {Cargile}, {Dieterich}, {Bradley}, {Nelan}, \&
  {Whipple}}]{ben16}
{Benedict}, G.~F., {Henry}, T.~J., {Franz}, O.~G., {et~al.} 2016, \aj, 152, 141

\bibitem[{{Brown} {et~al.}(2013){Brown}, {Baliber}, {Bianco}, {Bowman},
  {Burleson}, {Conway}, {Crellin}, {Depagne}, {De Vera}, {Dilday}, {Dragomir},
  {Dubberley}, {Eastman}, {Elphick}, {Falarski}, {Foale}, {Ford}, {Fulton},
  {Garza}, {Gomez}, {Graham}, {Greene}, {Haldeman}, {Hawkins}, {Haworth},
  {Haynes}, {Hidas}, {Hjelstrom}, {Howell}, {Hygelund}, {Lister}, {Lobdill},
  {Martinez}, {Mullins}, {Norbury}, {Parrent}, {Paulson}, {Petry}, {Pickles},
  {Posner}, {Rosing}, {Ross}, {Sand}, {Saunders}, {Shobbrook}, {Shporer},
  {Street}, {Thomas}, {Tsapras}, {Tufts}, {Valenti}, {Vander Horst}, {Walker},
  {White}, \& {Willis}}]{bro13}
{Brown}, T.~M., {Baliber}, N., {Bianco}, F.~B., {et~al.} 2013, \pasp, 125, 1031

\bibitem[{{Caballero} {et~al.}(2016{\natexlab{a}}){Caballero},
  {Cort{\'e}s-Contreras}, {Alonso-Floriano}, {Montes}, {Quirrenbach}, {Amado},
  {Ribas}, {Reiners}, {Abellan}, {B{\'e}jar}, {Brinkm{\"o}ller}, {Czesla},
  {Dorda}, {Gallardo}, {Gonz{\'a}lez-{\'A}lvarez}, {Hidalgo}, {Holgado},
  {Jeffers}, {Kim}, {Klutsch}, {Lamert}, {Llamas}, {L{\'o}pez-Santiago},
  {Mart{\'{\i}}nez-Rodr{\'{\i}}guez}, {Morales}, {Mundt}, {Passegger},
  {Sch{\"o}fer}, {Seifert}, \& {Zechmeister}}]{cab16}
{Caballero}, J.~A., {Cort{\'e}s-Contreras}, M., {Alonso-Floriano}, F.~J.,
  {et~al.} 2016{\natexlab{a}}, in 19th Cambridge Workshop on Cool Stars,
  Stellar Systems, and the Sun (CS19), 148

\bibitem[{{Caballero} {et~al.}(2022){Caballero}, {Gonz{\'a}lez-{\'A}lvarez},
  {Brady}, {Trifonov}, {Ellis}, {Dorn}, {Cifuentes}, {Molaverdikhani}, {Bean},
  {Boyajian}, {Rodr{\'\i}guez}, {Sanz-Forcada}, {Zapatero Osorio}, {Abia},
  {Amado}, {Anugu}, {B{\'e}jar}, {Davies}, {Dreizler}, {Dubois}, {Ennis},
  {Espinoza}, {Farrington}, {L{\'o}pez}, {Gardner}, {Hatzes}, {Henning},
  {Herrero}, {Herrero-Cisneros}, {Kaminski}, {Kasper}, {Klement}, {Kraus},
  {Labdon}, {Lanthermann}, {Le Bouquin}, {L{\'o}pez Gonz{\'a}lez}, {Luque},
  {Mann}, {Marfil}, {Monnier}, {Montes}, {Morales}, {Pall{\'e}}, {Pedraz},
  {Quirrenbach}, {Reffert}, {Reiners}, {Ribas}, {Rodr{\'\i}guez-L{\'o}pez},
  {Schaefer}, {Schweitzer}, {Seifahrt}, {Setterholm}, {Shan}, {Shulyak},
  {Solano}, {Sreenivas}, {Stef{\'a}nsson}, {St{\"u}rmer}, {Tabernero},
  {Tal-Or}, {ten Brummelaar}, {Vanaverbeke}, {von Braun}, {Youngblood}, \&
  {Zechmeister}}]{cab22}
{Caballero}, J.~A., {Gonz{\'a}lez-{\'A}lvarez}, E., {Brady}, M., {et~al.} 2022,
  \aap, 665, A120

\bibitem[{{Caballero} {et~al.}(2016{\natexlab{b}}){Caballero}, {Gu{\`a}rdia},
  {L{\'o}pez del Fresno}, {Zechmeister}, {de Juan}, {Alonso-Floriano}, {Amado},
  {Colom{\'e}}, {Cort{\'e}s-Contreras}, {Garc{\'{\i}}a-Piquer}, {Gesa}, {de
  Guindos}, {Hagen}, {Helmling}, {Hern{\'a}ndez Casta{\~n}o}, {K{\"u}rster},
  {L{\'o}pez-Santiago}, {Montes}, {Morales Mu{\~n}oz}, {Pavlov}, {Quirrenbach},
  {Reiners}, {Ribas}, {Seifert}, \& {Solano}}]{caracal}
{Caballero}, J.~A., {Gu{\`a}rdia}, J., {L{\'o}pez del Fresno}, M., {et~al.}
  2016{\natexlab{b}}, in \procspie, Vol. 9910, Observatory Operations:
  Strategies, Processes, and Systems VI, 99100E

\bibitem[{{Chen} {et~al.}(2014){Chen}, {Girardi}, {Bressan}, {Marigo},
  {Barbieri}, \& {Kong}}]{chen14}
{Chen}, Y., {Girardi}, L., {Bressan}, A., {et~al.} 2014, \mnras, 444, 2525

\bibitem[{{Cifuentes} {et~al.}(2020){Cifuentes}, {Caballero},
  {Cort{\'e}s-Contreras}, {Montes}, {Abell{\'a}n}, {Dorda}, {Holgado},
  {Zapatero Osorio}, {Morales}, {Amado}, {Passegger}, {Quirrenbach}, {Reiners},
  {Ribas}, {Sanz-Forcada}, {Schweitzer}, {Seifert}, \& {Solano}}]{cifu20}
{Cifuentes}, C., {Caballero}, J.~A., {Cort{\'e}s-Contreras}, M., {et~al.} 2020,
  \aap, 642, A115

\bibitem[{{Cincotta} \& {Sim{\'o}}(2000)}]{cin00}
{Cincotta}, P.~M. \& {Sim{\'o}}, C. 2000, \aaps, 147, 205

\bibitem[{{Collins} {et~al.}(2017){Collins}, {Kielkopf}, {Stassun}, \&
  {Hessman}}]{col17}
{Collins}, K.~A., {Kielkopf}, J.~F., {Stassun}, K.~G., \& {Hessman}, F.~V.
  2017, \aj, 153, 77

\bibitem[{{Colom{\'e}} {et~al.}(2010){Colom{\'e}}, {Casteels}, {Ribas}, \&
  {Francisco}}]{col10}
{Colom{\'e}}, J., {Casteels}, K., {Ribas}, I., \& {Francisco}, X. 2010, in
  Society of Photo-Optical Instrumentation Engineers (SPIE) Conference Series,
  Vol. 7740, Software and Cyberinfrastructure for Astronomy, ed. N.~M.
  {Radziwill} \& A.~{Bridger}, 77403K

\bibitem[{{Colome} \& {Ribas}(2006)}]{col06}
{Colome}, J. \& {Ribas}, I. 2006, IAU Special Session, 6, 11

\bibitem[{{Cort{\'e}s-Contreras} {et~al.}(2024){Cort{\'e}s-Contreras},
  {Caballero}, {Montes}, {Cardona-Guill{\'e}n}, {B{\'e}jar}, {Cifuentes},
  {Tabernero}, {Zapatero Osorio}, {Amado}, {Jeffers}, {Lafarga}, {Lodieu},
  {Quirrenbach}, {Reiners}, {Ribas}, {Sch{\"o}fer}, {Schweitzer}, \&
  {Seifert}}]{cor24}
{Cort{\'e}s-Contreras}, M., {Caballero}, J.~A., {Montes}, D., {et~al.} 2024,
  \aap, 692, A206

\bibitem[{{David} {et~al.}(2019){David}, {Petigura}, {Luger}, {Foreman-Mackey},
  {Livingston}, {Mamajek}, \& {Hillenbrand}}]{dav19}
{David}, T.~J., {Petigura}, E.~A., {Luger}, R., {et~al.} 2019, \apjl, 885, L12

\bibitem[{{Dawson} \& {Fabrycky}(2010)}]{daw10}
{Dawson}, R.~I. \& {Fabrycky}, D.~C. 2010, \apj, 722, 937

\bibitem[{{Deeming}(1975)}]{dee75}
{Deeming}, T.~J. 1975, \apss, 36, 137

\bibitem[{{Espinoza}(2018)}]{espin18}
{Espinoza}, N. 2018, Research Notes of the American Astronomical Society, 2,
  209

\bibitem[{{Espinoza} {et~al.}(2019){Espinoza}, {Kossakowski}, \&
  {Brahm}}]{juliet}
{Espinoza}, N., {Kossakowski}, D., \& {Brahm}, R. 2019, \mnras, 490, 2262

\bibitem[{{Foreman-Mackey} {et~al.}(2017){Foreman-Mackey}, {Agol},
  {Ambikasaran}, \& {Angus}}]{for17}
{Foreman-Mackey}, D., {Agol}, E., {Ambikasaran}, S., \& {Angus}, R. 2017, \aj,
  154, 220

\bibitem[{{Fulton} {et~al.}(2018){Fulton}, {Petigura}, {Blunt}, \&
  {Sinukoff}}]{ful18}
{Fulton}, B.~J., {Petigura}, E.~A., {Blunt}, S., \& {Sinukoff}, E. 2018, \pasp,
  130, 044504

\bibitem[{{Gaia Collaboration} {et~al.}(2018){Gaia Collaboration}, {Brown},
  {Vallenari}, {Prusti}, {de Bruijne}, {Babusiaux}, {Bailer-Jones}, {Biermann},
  {Evans}, {Eyer}, {Jansen}, {Jordi}, {Klioner}, {Lammers}, {Lindegren},
  {Luri}, {Mignard}, {Panem}, {Pourbaix}, {Randich}, {Sartoretti}, {Siddiqui},
  {Soubiran}, {van Leeuwen}, {Walton}, {Arenou}, {Bastian}, {Cropper},
  {Drimmel}, {Katz}, {Lattanzi}, {Bakker}, {Cacciari}, {Casta{\~n}eda},
  {Chaoul}, {Cheek}, {De Angeli}, {Fabricius}, {Guerra}, {Holl}, {Masana},
  {Messineo}, {Mowlavi}, {Nienartowicz}, {Panuzzo}, {Portell}, {Riello},
  {Seabroke}, {Tanga}, {Th{\'e}venin}, {Gracia-Abril}, {Comoretto},
  {Garcia-Reinaldos}, {Teyssier}, {Altmann}, {Andrae}, {Audard},
  {Bellas-Velidis}, {Benson}, {Berthier}, {Blomme}, {Burgess}, {Busso},
  {Carry}, {Cellino}, {Clementini}, {Clotet}, {Creevey}, {Davidson}, {De
  Ridder}, {Delchambre}, {Dell'Oro}, {Ducourant},
  {Fern{\'a}ndez-Hern{\'a}ndez}, {Fouesneau}, {Fr{\'e}mat}, {Galluccio},
  {Garc{\'\i}a-Torres}, {Gonz{\'a}lez-N{\'u}{\~n}ez}, {Gonz{\'a}lez-Vidal},
  {Gosset}, {Guy}, {Halbwachs}, {Hambly}, {Harrison}, {Hern{\'a}ndez},
  {Hestroffer}, {Hodgkin}, {Hutton}, {Jasniewicz}, {Jean-Antoine-Piccolo},
  {Jordan}, {Korn}, {Krone-Martins}, {Lanzafame}, {Lebzelter}, {L{\"o}ffler},
  {Manteiga}, {Marrese}, {Mart{\'\i}n-Fleitas}, {Moitinho}, {Mora}, {Muinonen},
  {Osinde}, {Pancino}, {Pauwels}, {Petit}, {Recio-Blanco}, {Richards},
  {Rimoldini}, {Robin}, {Sarro}, {Siopis}, {Smith}, {Sozzetti}, {S{\"u}veges},
  {Torra}, {van Reeven}, {Abbas}, {Abreu Aramburu}, {Accart}, {Aerts},
  {Altavilla}, {{\'A}lvarez}, {Alvarez}, {Alves}, {Anderson}, {Andrei},
  {Anglada Varela}, {Antiche}, {Antoja}, {Arcay}, {Astraatmadja}, {Bach},
  {Baker}, {Balaguer-N{\'u}{\~n}ez}, {Balm}, {Barache}, {Barata}, {Barbato},
  {Barblan}, {Barklem}, {Barrado}, {Barros}, {Barstow}, {Bartholom{\'e}
  Mu{\~n}oz}, {Bassilana}, {Becciani}, {Bellazzini}, {Berihuete}, {Bertone},
  {Bianchi}, {Bienaym{\'e}}, {Blanco-Cuaresma}, {Boch}, {Boeche}, {Bombrun},
  {Borrachero}, {Bossini}, {Bouquillon}, {Bourda}, {Bragaglia}, {Bramante},
  {Breddels}, {Bressan}, {Brouillet}, {Br{\"u}semeister}, {Brugaletta},
  {Bucciarelli}, {Burlacu}, {Busonero}, {Butkevich}, {Buzzi}, {Caffau},
  {Cancelliere}, {Cannizzaro}, {Cantat-Gaudin}, {Carballo}, {Carlucci},
  {Carrasco}, {Casamiquela}, {Castellani}, {Castro-Ginard}, {Charlot},
  {Chemin}, {Chiavassa}, {Cocozza}, {Costigan}, {Cowell}, {Crifo}, {Crosta},
  {Crowley}, {Cuypers}, {Dafonte}, {Damerdji}, {Dapergolas}, {David}, {David},
  {de Laverny}, \& {De Luise}}]{Gaia18}
{Gaia Collaboration}, {Brown}, A.~G.~A., {Vallenari}, A., {et~al.} 2018, \aap,
  616, A1

\bibitem[{{Gaia Collaboration} {et~al.}(2021){Gaia Collaboration}, {Brown},
  {Vallenari}, {Prusti}, {de Bruijne}, {Babusiaux}, {Biermann}, {Creevey},
  {Evans}, {Eyer}, {Hutton}, {Jansen}, {Jordi}, {Klioner}, {Lammers},
  {Lindegren}, {Luri}, {Mignard}, {Panem}, {Pourbaix}, {Randich}, {Sartoretti},
  {Soubiran}, {Walton}, {Arenou}, {Bailer-Jones}, {Bastian}, {Cropper},
  {Drimmel}, {Katz}, {Lattanzi}, {van Leeuwen}, {Bakker}, {Cacciari},
  {Casta{\~n}eda}, {De Angeli}, {Ducourant}, {Fabricius}, {Fouesneau},
  {Fr{\'e}mat}, {Guerra}, {Guerrier}, {Guiraud}, {Jean-Antoine Piccolo},
  {Masana}, {Messineo}, {Mowlavi}, {Nicolas}, {Nienartowicz}, {Pailler},
  {Panuzzo}, {Riclet}, {Roux}, {Seabroke}, {Sordo}, {Tanga}, {Th{\'e}venin},
  {Gracia-Abril}, {Portell}, {Teyssier}, {Altmann}, {Andrae}, {Bellas-Velidis},
  {Benson}, {Berthier}, {Blomme}, {Brugaletta}, {Burgess}, {Busso}, {Carry},
  {Cellino}, {Cheek}, {Clementini}, {Damerdji}, {Davidson}, {Delchambre},
  {Dell'Oro}, {Fern{\'a}ndez-Hern{\'a}ndez}, {Galluccio}, {Garc{\'\i}a-Lario},
  {Garcia-Reinaldos}, {Gonz{\'a}lez-N{\'u}{\~n}ez}, {Gosset}, {Haigron},
  {Halbwachs}, {Hambly}, {Harrison}, {Hatzidimitriou}, {Heiter},
  {Hern{\'a}ndez}, {Hestroffer}, {Hodgkin}, {Holl}, {Jan{\ss}en}, {Jevardat de
  Fombelle}, {Jordan}, {Krone-Martins}, {Lanzafame}, {L{\"o}ffler}, {Lorca},
  {Manteiga}, {Marchal}, {Marrese}, {Moitinho}, {Mora}, {Muinonen}, {Osborne},
  {Pancino}, {Pauwels}, {Petit}, {Recio-Blanco}, {Richards}, {Riello},
  {Rimoldini}, {Robin}, {Roegiers}, {Rybizki}, {Sarro}, {Siopis}, {Smith},
  {Sozzetti}, {Ulla}, {Utrilla}, {van Leeuwen}, {van Reeven}, {Abbas}, {Abreu
  Aramburu}, {Accart}, {Aerts}, {Aguado}, {Ajaj}, {Altavilla}, {{\'A}lvarez},
  {{\'A}lvarez Cid-Fuentes}, {Alves}, {Anderson}, {Anglada Varela}, {Antoja},
  {Audard}, {Baines}, {Baker}, {Balaguer-N{\'u}{\~n}ez}, {Balbinot}, {Balog},
  {Barache}, {Barbato}, {Barros}, {Barstow}, {Bartolom{\'e}}, {Bassilana},
  {Bauchet}, {Baudesson-Stella}, {Becciani}, {Bellazzini}, {Bernet}, {Bertone},
  {Bianchi}, {Blanco-Cuaresma}, {Boch}, {Bombrun}, {Bossini}, {Bouquillon},
  {Bragaglia}, {Bramante}, {Breedt}, {Bressan}, {Brouillet}, {Bucciarelli},
  {Burlacu}, {Busonero}, {Butkevich}, {Buzzi}, {Caffau}, {Cancelliere},
  {C{\'a}novas}, {Cantat-Gaudin}, {Carballo}, {Carlucci}, {Carnerero},
  {Carrasco}, {Casamiquela}, {Castellani}, {Castro-Ginard}, {Castro Sampol},
  {Chaoul}, {Charlot}, {Chemin}, {Chiavassa}, {Cioni}, {Comoretto}, {Cooper},
  {Cornez}, {Cowell}, {Crifo}, {Crosta}, {Crowley}, {Dafonte}, {Dapergolas},
  {David}, {David}, {de Laverny}, {De Luise}, {De March}, {De Ridder}, {de
  Souza}, {de Teodoro}, {de Torres}, {del Peloso}, {del Pozo}, {Delbo},
  {Delgado}, {Delgado}, {Delisle}, {Di Matteo}, {Diakite}, {Diener},
  {Distefano}, {Dolding}, {Eappachen}, {Edvardsson}, {Enke}, {Esquej}, {Fabre},
  {Fabrizio}, {Faigler}, {Fedorets}, {Fernique}, {Fienga}, {Figueras},
  {Fouron}, {Fragkoudi}, {Fraile}, {Franke}, {Gai}, {Garabato},
  {Garcia-Gutierrez}, {Garc{\'\i}a-Torres}, {Garofalo}, {Gavras}, {Gerlach},
  {Geyer}, {Giacobbe}, {Gilmore}, {Girona}, {Giuffrida}, {Gomel}, {Gomez},
  {Gonzalez-Santamaria}, {Gonz{\'a}lez-Vidal}, {Granvik},
  {Guti{\'e}rrez-S{\'a}nchez}, {Guy}, {Hauser}, {Haywood}, {Helmi}, {Hidalgo},
  {Hilger}, {H{\l}adczuk}, {Hobbs}, {Holland}, {Huckle}, {Jasniewicz},
  {Jonker}, {Juaristi Campillo}, {Julbe}, {Karbevska}, {Kervella}, {Khanna},
  {Kochoska}, {Kontizas}, {Kordopatis}, {Korn}, {Kostrzewa-Rutkowska},
  {Kruszy{\'n}ska}, {Lambert}, {Lanza}, {Lasne}, {Le Campion}, {Le Fustec},
  {Lebreton}, {Lebzelter}, {Leccia}, {Leclerc}, {Lecoeur-Taibi}, {Liao},
  {Licata}, {Lindstr{\o}m}, {Lister}, {Livanou}, {Lobel}, {Madrero Pardo},
  {Managau}, {Mann}, {Marchant}, {Marconi}, {Marcos Santos}, {Marinoni},
  {Marocco}, {Marshall}, {Martin Polo}, {Mart{\'\i}n-Fleitas}, {Masip},
  {Massari}, {Mastrobuono-Battisti}, {Mazeh}, {McMillan}, {Messina},
  {Michalik}, {Millar}, {Mints}, {Molina}, {Molinaro}, {Moln{\'a}r},
  {Montegriffo}, {Mor}, {Morbidelli}, {Morel}, {Morris}, {Mulone}, {Munoz},
  {Muraveva}, {Murphy}, {Musella}, {Noval}, {Ord{\'e}novic}, {Orr{\`u}},
  {Osinde}, {Pagani}, {Pagano}, {Palaversa}, {Palicio}, {Panahi}, {Pawlak},
  {Pe{\~n}alosa Esteller}, {Penttil{\"a}}, {Piersimoni}, {Pineau}, {Plachy},
  {Plum}, {Poggio}, {Poretti}, {Poujoulet}, {Pr{\v{s}}a}, {Pulone}, {Racero},
  {Ragaini}, {Rainer}, {Raiteri}, {Rambaux}, {Ramos}, {Ramos-Lerate}, {Re
  Fiorentin}, {Regibo}, {Reyl{\'e}}, {Ripepi}, {Riva}, {Rixon}, {Robichon},
  {Robin}, {Roelens}, {Rohrbasser}, {Romero-G{\'o}mez}, {Rowell}, {Royer},
  {Rybicki}, {Sadowski}, {Sagrist{\`a} Sell{\'e}s}, {Sahlmann}, {Salgado},
  {Salguero}, {Samaras}, {Sanchez Gimenez}, {Sanna}, {Santove{\~n}a},
  {Sarasso}, {Schultheis}, {Sciacca}, {Segol}, {Segovia}, {S{\'e}gransan},
  {Semeux}, {Shahaf}, {Siddiqui}, {Siebert}, {Siltala}, {Slezak}, {Smart},
  {Solano}, {Solitro}, {Souami}, {Souchay}, {Spagna}, {Spoto}, {Steele},
  {Steidelm{\"u}ller}, {Stephenson}, {S{\"u}veges}, {Szabados}, {Szegedi-Elek},
  {Taris}, {Tauran}, {Taylor}, {Teixeira}, {Thuillot}, {Tonello}, {Torra},
  {Torra}, {Turon}, {Unger}, {Vaillant}, {van Dillen}, {Vanel}, {Vecchiato},
  {Viala}, {Vicente}, {Voutsinas}, {Weiler}, {Wevers}, {Wyrzykowski}, {Yoldas},
  {Yvard}, {Zhao}, {Zorec}, {Zucker}, {Zurbach}, \& {Zwitter}}]{gaia}
{Gaia Collaboration}, {Brown}, A.~G.~A., {Vallenari}, A., {et~al.} 2021, \aap,
  649, A1

\bibitem[{{Gillen} {et~al.}(2020){Gillen}, {Briegal}, {Hodgkin},
  {Foreman-Mackey}, {Van Leeuwen}, {Jackman}, {McCormac}, {West}, {Queloz},
  {Bayliss}, {Goad}, {Watson}, {Wheatley}, {Belardi}, {Burleigh}, {Casewell},
  {Jenkins}, {Raynard}, {Smith}, {Tilbrook}, \& {Vines}}]{gil20}
{Gillen}, E., {Briegal}, J.~T., {Hodgkin}, S.~T., {et~al.} 2020, \mnras, 492,
  1008

\bibitem[{{Gonz{\'a}lez-{\'A}lvarez} {et~al.}(2023){Gonz{\'a}lez-{\'A}lvarez},
  {Kemmer}, {Chaturvedi}, {Caballero}, {Quirrenbach}, {Amado}, {B{\'e}jar},
  {Cifuentes}, {Herrero}, {Kossakowski}, {Reiners}, {Ribas}, {Rodr{\'\i}guez},
  {Rodr{\'\i}guez-L{\'o}pez}, {Sanz-Forcada}, {Shan}, {Stock}, {Tabernero},
  {Tal-Or}, {Osorio}, {Hatzes}, {Henning}, {L{\'o}pez-Gonz{\'a}lez}, {Montes},
  {Morales}, {Pall{\'e}}, {Pedraz}, {Perger}, {Reffert}, {Sabotta},
  {Schweitzer}, \& {Zechmeister}}]{gon23}
{Gonz{\'a}lez-{\'A}lvarez}, E., {Kemmer}, J., {Chaturvedi}, P., {et~al.} 2023,
  \aap, 675, A141

\bibitem[{{Guerrero} {et~al.}(2021){Guerrero}, {Seager}, {Huang}, {Vanderburg},
  {Garcia Soto}, {Mireles}, {Hesse}, {Fong}, {Glidden}, {Shporer}, {Latham},
  {Collins}, {Quinn}, {Burt}, {Dragomir}, {Crossfield}, {Vanderspek},
  {Fausnaugh}, {Burke}, {Ricker}, {Daylan}, {Essack}, {G{\"u}nther}, {Osborn},
  {Pepper}, {Rowden}, {Sha}, {Villanueva}, {Yahalomi}, {Yu}, {Ballard},
  {Batalha}, {Berardo}, {Chontos}, {Dittmann}, {Esquerdo}, {Mikal-Evans},
  {Jayaraman}, {Krishnamurthy}, {Louie}, {Mehrle}, {Niraula}, {Rackham},
  {Rodriguez}, {Rowden}, {Sousa-Silva}, {Watanabe}, {Wong}, {Zhan},
  {Zivanovic}, {Christiansen}, {Ciardi}, {Swain}, {Lund}, {Mullally},
  {Fleming}, {Rodriguez}, {Boyd}, {Quintana}, {Barclay}, {Col{\'o}n},
  {Rinehart}, {Schlieder}, {Clampin}, {Jenkins}, {Twicken}, {Caldwell},
  {Coughlin}, {Henze}, {Lissauer}, {Morris}, {Rose}, {Smith}, {Tenenbaum},
  {Ting}, {Wohler}, {Bakos}, {Bean}, {Berta-Thompson}, {Bieryla}, {Bouma},
  {Buchhave}, {Butler}, {Charbonneau}, {Doty}, {Ge}, {Holman}, {Howard},
  {Kaltenegger}, {Kane}, {Kjeldsen}, {Kreidberg}, {Lin}, {Minsky}, {Narita},
  {Paegert}, {P{\'a}l}, {Palle}, {Sasselov}, {Spencer}, {Sozzetti}, {Stassun},
  {Torres}, {Udry}, \& {Winn}}]{gue21}
{Guerrero}, N.~M., {Seager}, S., {Huang}, C.~X., {et~al.} 2021, \apjs, 254, 39

\bibitem[{{Hippke} \& {Heller}(2019)}]{hip19}
{Hippke}, M. \& {Heller}, R. 2019, \aap, 623, A39

\bibitem[{{H{\o}g} {et~al.}(2000){H{\o}g}, {Fabricius}, {Makarov}, {Urban},
  {Corbin}, {Wycoff}, {Bastian}, {Schwekendiek}, \& {Wicenec}}]{tyc00}
{H{\o}g}, E., {Fabricius}, C., {Makarov}, V.~V., {et~al.} 2000, \aap, 355, L27

\bibitem[{{Howard} {et~al.}(2025){Howard}, {Sinukoff}, {Blunt}, {Petigura},
  {Crossfield}, {Isaacson}, {Kosiarek}, {Rubenzahl}, {Brewer}, {Fulton},
  {Dressing}, {Hirsch}, {Knutson}, {Livingston}, {Mills}, {Roy}, {Weiss},
  {Benneke}, {Ciardi}, {Christiansen}, {Cochran}, {Crepp}, {Gonzales},
  {Hansen}, {Hardegree-Ullman}, {Howell}, {L{\'e}pine}, {Martinez}, {Rogers},
  {Schlieder}, {Werner}, {Polanski}, {Angelo}, {Beard}, {Behmard}, {Bouma},
  {Brinkman}, {Chontos}, {Dai}, {Dalba}, {Giacalone}, {Grunblatt}, {Hill},
  {Kane}, {Lubin}, {Mayo}, {Mocnik}, {Murphy}, {Rice}, {Rosenthal}, {Tyler},
  {Van Zandt}, \& {Yee}}]{how25}
{Howard}, A.~W., {Sinukoff}, E., {Blunt}, S., {et~al.} 2025, \apjs, 278, 52

\bibitem[{{Jeffries} {et~al.}(2023){Jeffries}, {Jackson}, {Wright}, {Weaver},
  {Gilmore}, {Randich}, {Bragaglia}, {Korn}, {Smiljanic}, {Biazzo}, {Casey},
  {Frasca}, {Gonneau}, {Guiglion}, {Morbidelli}, {Prisinzano}, {Sacco},
  {Tautvai{\v{s}}ien{\.{e}}}, {Worley}, \& {Zaggia}}]{jef23}
{Jeffries}, R.~D., {Jackson}, R.~J., {Wright}, N.~J., {et~al.} 2023, \mnras,
  523, 802

\bibitem[{{Jenkins}(2002)}]{jen02}
{Jenkins}, J.~M. 2002, \apj, 575, 493

\bibitem[{{Jenkins} {et~al.}(2010){Jenkins}, {Chandrasekaran}, {McCauliff},
  {Caldwell}, {Tenenbaum}, {Li}, {Klaus}, {Cote}, \& {Middour}}]{jen10}
{Jenkins}, J.~M., {Chandrasekaran}, H., {McCauliff}, S.~D., {et~al.} 2010, in
  Society of Photo-Optical Instrumentation Engineers (SPIE) Conference Series,
  Vol. 7740, Software and Cyberinfrastructure for Astronomy, ed. N.~M.
  {Radziwill} \& A.~{Bridger}, 77400D

\bibitem[{{Jenkins} {et~al.}(2020){Jenkins}, {Tenenbaum}, {Seader}, {Burke},
  {McCauliff}, {Smith}, {Twicken}, \& {Chandrasekaran}}]{jen20}
{Jenkins}, J.~M., {Tenenbaum}, P., {Seader}, S., {et~al.} 2020, {Kepler Data
  Processing Handbook: Transiting Planet Search}, Kepler Science Document
  KSCI-19081-003

\bibitem[{{Jenkins} {et~al.}(2016){Jenkins}, {Twicken}, {McCauliff},
  {Campbell}, {Sanderfer}, {Lung}, {Mansouri-Samani}, {Girouard}, {Tenenbaum},
  {Klaus}, {Smith}, {Caldwell}, {Chacon}, {Henze}, {Heiges}, {Latham},
  {Morgan}, {Swade}, {Rinehart}, \& {Vanderspek}}]{jen16}
{Jenkins}, J.~M., {Twicken}, J.~D., {McCauliff}, S., {et~al.} 2016, in
  \procspie, Vol. 9913, Software and Cyberinfrastructure for Astronomy IV,
  99133E

\bibitem[{{Kasting} {et~al.}(1993){Kasting}, {Whitmire}, \& {Reynolds}}]{kas93}
{Kasting}, J.~F., {Whitmire}, D.~P., \& {Reynolds}, R.~T. 1993, \icarus, 101,
  108

\bibitem[{{Kopparapu} {et~al.}(2013){Kopparapu}, {Ramirez}, {Kasting}, {Eymet},
  {Robinson}, {Mahadevan}, {Terrien}, {Domagal-Goldman}, {Meadows}, \&
  {Deshpande}}]{kop13}
{Kopparapu}, R.~K., {Ramirez}, R., {Kasting}, J.~F., {et~al.} 2013, \apj, 765,
  131

\bibitem[{{Kostov} {et~al.}(2014){Kostov}, {McCullough}, {Carter}, {Deleuil},
  {D{\'\i}az}, {Fabrycky}, {H{\'e}brard}, {Hinse}, {Mazeh}, {Orosz},
  {Tsvetanov}, \& {Welsh}}]{kos14}
{Kostov}, V.~B., {McCullough}, P.~R., {Carter}, J.~A., {et~al.} 2014, \apj,
  784, 14

\bibitem[{{Kreidberg}(2015)}]{Krei15}
{Kreidberg}, L. 2015, \pasp, 127, 1161

\bibitem[{{Kuzuhara} {et~al.}(2024){Kuzuhara}, {Fukui}, {Livingston},
  {Caballero}, {de Leon}, {Hirano}, {Kasagi}, {Murgas}, {Narita}, {Omiya},
  {Orell-Miquel}, {Palle}, {Changeat}, {Esparza-Borges}, {Harakawa}, {Hellier},
  {Hori}, {Ikuta}, {Ishikawa}, {Kodama}, {Kotani}, {Kudo}, {Morales}, {Mori},
  {Nagel}, {Parviainen}, {Perdelwitz}, {Reiners}, {Ribas}, {Sanz-Forcada},
  {Sato}, {Schweitzer}, {Tabernero}, {Takarada}, {Uyama}, {Watanabe},
  {Zechmeister}, {Garc{\'\i}a}, {Aoki}, {Beichman}, {B{\'e}jar}, {Brandt},
  {Calatayud-Borras}, {Carleo}, {Charbonneau}, {Collins}, {Currie}, {Doty},
  {Dreizler}, {Fern{\'a}ndez-Rodr{\'\i}guez}, {Fukuda}, {Gal{\'a}n},
  {Gerald{\'\i}a-Gonz{\'a}lez}, {Gonz{\'a}lez-Rodr{\'\i}guez}, {Hayashi},
  {Hedges}, {Henning}, {Hodapp}, {Ikoma}, {Isogai}, {Jacobson}, {Janson},
  {Jenkins}, {Kagetani}, {Kambe}, {Kawai}, {Kawauchi}, {Kokubo}, {Konishi},
  {Korth}, {Krishnamurthy}, {Kurokawa}, {Kusakabe}, {Kwon}, {Laza-Ramos},
  {Libotte}, {Luque}, {Madrigal-Aguado}, {Matsumoto}, {Mawet}, {McElwain},
  {Meni Gallardo}, {Morello}, {Mu{\~n}oz Torres}, {Nishikawa}, {Nugroho},
  {Ogihara}, {Pel{\'a}ez-Torres}, {Rapetti}, {S{\'a}nchez-Benavente},
  {Schlecker}, {Seager}, {Serabyn}, {Serizawa}, {Stangret}, {Takahashi},
  {Teng}, {Tamura}, {Terada}, {Ueda}, {Usuda}, {Vanderspek}, {Vievard},
  {Watanabe}, {Winn}, \& {Zapatero Osorio}}]{kuz24}
{Kuzuhara}, M., {Fukui}, A., {Livingston}, J.~H., {et~al.} 2024, \apjl, 967,
  L21

\bibitem[{{Lammer} {et~al.}(2009){Lammer}, {Bredeh{\"o}ft}, {Coustenis},
  {Khodachenko}, {Kaltenegger}, {Grasset}, {Prieur}, {Raulin}, {Ehrenfreund},
  {Yamauchi}, {Wahlund}, {Grie{\ss}meier}, {Stangl}, {Cockell}, {Kulikov},
  {Grenfell}, \& {Rauer}}]{lam09}
{Lammer}, H., {Bredeh{\"o}ft}, J.~H., {Coustenis}, A., {et~al.} 2009, \aapr,
  17, 181

\bibitem[{{Laskar}(1997)}]{las97}
{Laskar}, J. 1997, \aap, 317, L75

\bibitem[{{Laskar} \& {Petit}(2017)}]{Lask17}
{Laskar}, J. \& {Petit}, A.~C. 2017, \aap, 605, A72

\bibitem[{{Li} {et~al.}(2019){Li}, {Tenenbaum}, {Twicken}, {Burke}, {Jenkins},
  {Quintana}, {Rowe}, \& {Seader}}]{li19}
{Li}, J., {Tenenbaum}, P., {Twicken}, J.~D., {et~al.} 2019, \pasp, 131, 024506

\bibitem[{{Luque} {et~al.}(2019){Luque}, {Pall{\'e}}, {Kossakowski},
  {Dreizler}, {Kemmer}, {Espinoza}, {Burt}, {Anglada-Escud{\'e}}, {B{\'e}jar},
  {Caballero}, {Collins}, {Collins}, {Cort{\'e}s-Contreras},
  {D{\'\i}ez-Alonso}, {Feng}, {Hatzes}, {Hellier}, {Henning}, {Jeffers},
  {Kaltenegger}, {K{\"u}rster}, {Madden}, {Molaverdikhani}, {Montes}, {Narita},
  {Nowak}, {Ofir}, {Oshagh}, {Parviainen}, {Quirrenbach}, {Reffert}, {Reiners},
  {Rodr{\'\i}guez-L{\'o}pez}, {Schlecker}, {Stock}, {Trifonov}, {Winn},
  {Zapatero Osorio}, {Zechmeister}, {Amado}, {Anderson}, {Batalha}, {Bauer},
  {Bluhm}, {Burke}, {Butler}, {Caldwell}, {Chen}, {Crane}, {Dragomir},
  {Dressing}, {Dynes}, {Jenkins}, {Kaminski}, {Klahr}, {Kotani}, {Lafarga},
  {Latham}, {Lewin}, {McDermott}, {Monta{\~n}{\'e}s-Rodr{\'\i}guez}, {Morales},
  {Murgas}, {Nagel}, {Pedraz}, {Ribas}, {Ricker}, {Rowden}, {Seager},
  {Shectman}, {Tamura}, {Teske}, {Twicken}, {Vanderspeck}, {Wang}, \&
  {Wohler}}]{luq19}
{Luque}, R., {Pall{\'e}}, E., {Kossakowski}, D., {et~al.} 2019, \aap, 628, A39

\bibitem[{{Mamajek} \& {Hillenbrand}(2008)}]{mam08}
{Mamajek}, E.~E. \& {Hillenbrand}, L.~A. 2008, \apj, 687, 1264

\bibitem[{{Mann} {et~al.}(2019){Mann}, {Dupuy}, {Kraus}, {Gaidos}, {Ansdell},
  {Ireland}, {Rizzuto}, {Hung}, {Dittmann}, {Factor}, {Feiden}, {Martinez},
  {Ru{\'\i}z-Rodr{\'\i}guez}, \& {Thao}}]{mann19}
{Mann}, A.~W., {Dupuy}, T., {Kraus}, A.~L., {et~al.} 2019, \apj, 871, 63

\bibitem[{{Mann} {et~al.}(2015){Mann}, {Feiden}, {Gaidos}, {Boyajian}, \& {von
  Braun}}]{mann15}
{Mann}, A.~W., {Feiden}, G.~A., {Gaidos}, E., {Boyajian}, T., \& {von Braun},
  K. 2015, \apj, 804, 64

\bibitem[{{Marfil} {et~al.}(2021){Marfil}, {Tabernero}, {Montes}, {Caballero},
  {L{\'a}zaro}, {Gonz{\'a}lez Hern{\'a}ndez}, {Nagel}, {Passegger},
  {Schweitzer}, {Ribas}, {Reiners}, {Quirrenbach}, {Amado}, {Cifuentes},
  {Cort{\'e}s-Contreras}, {Dreizler}, {Duque-Arribas},
  {Galad{\'\i}-Enr{\'\i}quez}, {Henning}, {Jeffers}, {Kaminski}, {K{\"u}rster},
  {Lafarga}, {L{\'o}pez-Gallifa}, {Morales}, {Shan}, \& {Zechmeister}}]{mar21}
{Marfil}, E., {Tabernero}, H.~M., {Montes}, D., {et~al.} 2021, \aap, 656, A162

\bibitem[{{Mart{\'\i}nez-Rodr{\'\i}guez}
  {et~al.}(2019){Mart{\'\i}nez-Rodr{\'\i}guez}, {Caballero}, {Cifuentes},
  {Piro}, \& {Barnes}}]{mar19}
{Mart{\'\i}nez-Rodr{\'\i}guez}, H., {Caballero}, J.~A., {Cifuentes}, C.,
  {Piro}, A.~L., \& {Barnes}, R. 2019, \apj, 887, 261

\bibitem[{{Masuda}(2014)}]{mas14}
{Masuda}, K. 2014, \apj, 783, 53

\bibitem[{{Morris} {et~al.}(2020){Morris}, {Twicken}, {Smith}, {Clarke},
  {Jenkins}, {Bryson}, {Girouard}, \& {Klaus}}]{mor20}
{Morris}, R.~L., {Twicken}, J.~D., {Smith}, J.~C., {et~al.} 2020, {Kepler Data
  Processing Handbook: Photometric Analysis}, Kepler Science Document
  KSCI-19081-003

\bibitem[{{Mortier} {et~al.}(2015){Mortier}, {Faria}, {Correia}, {Santerne}, \&
  {Santos}}]{mor15}
{Mortier}, A., {Faria}, J.~P., {Correia}, C.~M., {Santerne}, A., \& {Santos},
  N.~C. 2015, \aap, 573, A101

\bibitem[{{Nagel} {et~al.}(2023){Nagel}, {Czesla}, {Kaminski}, {Zechmeister},
  {Tal-Or}, {Schmitt}, {Reiners}, {Quirrenbach}, {Garc{\'\i}a L{\'o}pez},
  {Caballero}, {Ribas}, {Amado}, {B{\'e}jar}, {Cort{\'e}s-Contreras},
  {Dreizler}, {Hatzes}, {Henning}, {Jeffers}, {K{\"u}rster}, {Lafarga},
  {L{\'o}pez-Puertas}, {Montes}, {Morales}, {Pedraz}, \& {Schweitzer}}]{nag23}
{Nagel}, E., {Czesla}, S., {Kaminski}, A., {et~al.} 2023, \aap, 680, A73

\bibitem[{{Nicholson} \& {Aigrain}(2022)}]{nic22}
{Nicholson}, B.~A. \& {Aigrain}, S. 2022, \mnras, 515, 5251

\bibitem[{{Orosz} {et~al.}(2019){Orosz}, {Welsh}, {Haghighipour}, {Quarles},
  {Short}, {Mills}, {Satyal}, {Torres}, {Agol}, {Fabrycky}, {Jontof-Hutter},
  {Windmiller}, {M{\"u}ller}, {Hinse}, {Cochran}, {Endl}, {Ford}, {Mazeh}, \&
  {Lissauer}}]{oro19}
{Orosz}, J.~A., {Welsh}, W.~F., {Haghighipour}, N., {et~al.} 2019, \aj, 157,
  174

\bibitem[{{Pecaut} \& {Mamajek}(2013)}]{pec13}
{Pecaut}, M.~J. \& {Mamajek}, E.~E. 2013, \apjs, 208, 9

\bibitem[{{Pozuelos} {et~al.}(2023){Pozuelos}, {Timmermans}, {Rackham},
  {Garcia}, {Burgasser}, {Kane}, {G{\"u}nther}, {Stassun}, {Van Grootel},
  {D{\'e}vora-Pajares}, {Luque}, {Edwards}, {Niraula}, {Schanche}, {Wells},
  {Ducrot}, {Howell}, {Sebastian}, {Barkaoui}, {Waalkes}, {Cadieux}, {Doyon},
  {Boyle}, {Dietrich}, {Burdanov}, {Delrez}, {Demory}, {de Wit}, {Dransfield},
  {Gillon}, {G{\'o}mez Maqueo Chew}, {Hooton}, {Jehin}, {Murray}, {Pedersen},
  {Queloz}, {Thompson}, {Triaud}, {Z{\'u}{\~n}iga-Fern{\'a}ndez}, {Collins},
  {Fausnaugh}, {Hedges}, {Hesse}, {Jenkins}, {Kunimoto}, {Latham}, {Shporer},
  {Ting}, {Torres}, {Amado}, {Rod{\'o}n}, {Rodr{\'\i}guez-L{\'o}pez},
  {Su{\'a}rez}, {Alonso}, {Benkhaldoun}, {Berta-Thompson}, {Chinchilla},
  {Ghachoui}, {G{\'o}mez-Mu{\~n}oz}, {Rebolo}, {Sabin}, {Schroffenegger},
  {Furlan}, {Gnilka}, {Lester}, {Scott}, {Aganze}, {Gerasimov}, {Hsu},
  {Theissen}, {Apai}, {Chen}, {Gabor}, {Henning}, \& {Mancini}}]{poz23}
{Pozuelos}, F.~J., {Timmermans}, M., {Rackham}, B.~V., {et~al.} 2023, \aap,
  672, A70

\bibitem[{{Quirrenbach} {et~al.}(2014){Quirrenbach}, {Amado}, {Caballero},
  {Mundt}, {Reiners}, {Ribas}, {Seifert}, {Abril}, {Aceituno},
  {Alonso-Floriano}, {Ammler-von Eiff}, {Antona Jim{\'e}nez},
  {Anwand-Heerwart}, {Azzaro}, {Bauer}, {Barrado}, {Becerril}, {B{\'e}jar},
  {Ben{\'\i}tez}, {Berdi{\~n}as}, {C{\'a}rdenas}, {Casal}, {Claret},
  {Colom{\'e}}, {Cort{\'e}s-Contreras}, {Czesla}, {Doellinger}, {Dreizler},
  {Feiz}, {Fern{\'a}ndez}, {Galad{\'\i}}, {G{\'a}lvez-Ortiz},
  {Garc{\'\i}a-Piquer}, {Garc{\'\i}a-Vargas}, {Garrido}, {Gesa}, {G{\'o}mez
  Galera}, {Gonz{\'a}lez {\'A}lvarez}, {Gonz{\'a}lez Hern{\'a}ndez},
  {Gr{\"o}zinger}, {Gu{\`a}rdia}, {Guenther}, {de Guindos},
  {Guti{\'e}rrez-Soto}, {Hagen}, {Hatzes}, {Hauschildt}, {Helmling}, {Henning},
  {Hermann}, {Hern{\'a}ndez Casta{\~n}o}, {Herrero}, {Hidalgo}, {Holgado},
  {Huber}, {Huber}, {Jeffers}, {Joergens}, {de Juan}, {Kehr}, {Klein},
  {K{\"u}rster}, {Lamert}, {Lalitha}, {Laun}, {Lemke}, {Lenzen}, {L{\'o}pez del
  Fresno}, {L{\'o}pez Mart{\'\i}}, {L{\'o}pez-Santiago}, {Mall}, {Mandel},
  {Mart{\'\i}n}, {Mart{\'\i}n-Ruiz}, {Mart{\'\i}nez-Rodr{\'\i}guez}, {Marvin},
  {Mathar}, {Mirabet}, {Montes}, {Morales Mu{\~n}oz}, {Moya}, {Naranjo},
  {Ofir}, {Oreiro}, {Pall{\'e}}, {Panduro}, {Passegger}, {P{\'e}rez-Calpena},
  {P{\'e}rez Medialdea}, {Perger}, {Pluto}, {Ram{\'o}n}, {Rebolo}, {Redondo},
  {Reffert}, {Reinhardt}, {Rhode}, {Rix}, {Rodler}, {Rodr{\'\i}guez},
  {Rodr{\'\i}guez-L{\'o}pez}, {Rodr{\'\i}guez-P{\'e}rez}, {Rohloff}, {Rosich},
  {S{\'a}nchez-Blanco}, {S{\'a}nchez Carrasco}, {Sanz-Forcada}, {Sarmiento},
  {Sch{\"a}fer}, {Schiller}, {Schmidt}, {Schmitt}, {Solano}, {Stahl}, {Storz},
  {St{\"u}rmer}, {Su{\'a}rez}, {Ulbrich}, {Veredas}, {Wagner}, {Winkler},
  {Zapatero Osorio}, {Zechmeister}, {Abell{\'a}n de Paco},
  {Anglada-Escud{\'e}}, {del Burgo}, {Klutsch}, {Lizon}, {L{\'o}pez-Morales},
  {Morales}, {Perryman}, {Tulloch}, \& {Xu}}]{spie14}
{Quirrenbach}, A., {Amado}, P.~J., {Caballero}, J.~A., {et~al.} 2014, in
  Society of Photo-Optical Instrumentation Engineers (SPIE) Conference Series,
  Vol. 9147, Ground-based and Airborne Instrumentation for Astronomy V, ed.
  S.~K. {Ramsay}, I.~S. {McLean}, \& H.~{Takami}, 91471F

\bibitem[{{Randich} \& {Magrini}(2021)}]{ran21}
{Randich}, S. \& {Magrini}, L. 2021, Frontiers in Astronomy and Space Sciences,
  8, 6

\bibitem[{{Rein} \& {Liu}(2012)}]{reb12}
{Rein}, H. \& {Liu}, S.~F. 2012, \aap, 537, A128

\bibitem[{{Rein} \& {Tamayo}(2015)}]{meg15}
{Rein}, H. \& {Tamayo}, D. 2015, \mnras, 452, 376

\bibitem[{{Reiners} {et~al.}(2014){Reiners}, {Sch{\"u}ssler}, \&
  {Passegger}}]{rei14}
{Reiners}, A., {Sch{\"u}ssler}, M., \& {Passegger}, V.~M. 2014, \apj, 794, 144

\bibitem[{{Reiners} {et~al.}(2018){Reiners}, {Zechmeister}, {Caballero},
  {Ribas}, {Morales}, {Jeffers}, {Sch{\"o}fer}, {Tal-Or}, {Quirrenbach},
  {Amado}, {Kaminski}, {Seifert}, {Abril}, {Aceituno}, {Alonso-Floriano},
  {Ammler-von Eiff}, {Antona}, {Anglada-Escud{\'e}}, {Anwand-Heerwart},
  {Arroyo-Torres}, {Azzaro}, {Baroch}, {Barrado}, {Bauer}, {Becerril},
  {B{\'e}jar}, {Ben{\'\i}tez}, {Berdinas}, {Bergond}, {Bl{\"u}mcke},
  {Brinkm{\"o}ller}, {del Burgo}, {Cano}, {C{\'a}rdenas V{\'a}zquez}, {Casal},
  {Cifuentes}, {Claret}, {Colom{\'e}}, {Cort{\'e}s-Contreras}, {Czesla},
  {D{\'\i}ez-Alonso}, {Dreizler}, {Feiz}, {Fern{\'a}ndez}, {Ferro},
  {Fuhrmeister}, {Galad{\'\i}-Enr{\'\i}quez}, {Garcia-Piquer}, {Garc{\'\i}a
  Vargas}, {Gesa}, {G{\'o}mez Galera}, {Gonz{\'a}lez Hern{\'a}ndez},
  {Gonz{\'a}lez-Peinado}, {Gr{\"o}zinger}, {Grohnert}, {Gu{\`a}rdia},
  {Guenther}, {Guijarro}, {de Guindos}, {Guti{\'e}rrez-Soto}, {Hagen},
  {Hatzes}, {Hauschildt}, {Hedrosa}, {Helmling}, {Henning}, {Hermelo},
  {Hern{\'a}ndez Arab{\'\i}}, {Hern{\'a}ndez Casta{\~n}o}, {Hern{\'a}ndez
  Hernando}, {Herrero}, {Huber}, {Huke}, {Johnson}, {de Juan}, {Kim}, {Klein},
  {Kl{\"u}ter}, {Klutsch}, {K{\"u}rster}, {Lafarga}, {Lamert}, {Lamp{\'o}n},
  {Lara}, {Laun}, {Lemke}, {Lenzen}, {Launhardt}, {L{\'o}pez del Fresno},
  {L{\'o}pez-Gonz{\'a}lez}, {L{\'o}pez-Puertas}, {L{\'o}pez Salas},
  {L{\'o}pez-Santiago}, {Luque}, {Mag{\'a}n Madinabeitia}, {Mall}, {Mancini},
  {Mandel}, {Marfil}, {Mar{\'\i}n Molina}, {Maroto Fern{\'a}ndez},
  {Mart{\'\i}n}, {Mart{\'\i}n-Ruiz}, {Marvin}, {Mathar}, {Mirabet}, {Montes},
  {Moreno-Raya}, {Moya}, {Mundt}, {Nagel}, {Naranjo}, {Nortmann}, {Nowak},
  {Ofir}, {Oreiro}, {Pall{\'e}}, {Panduro}, {Pascual}, {Passegger}, {Pavlov},
  {Pedraz}, {P{\'e}rez-Calpena}, {P{\'e}rez Medialdea}, {Perger}, {Perryman},
  {Pluto}, {Rabaza}, {Ram{\'o}n}, {Rebolo}, {Redondo}, {Reffert}, {Reinhart},
  {Rhode}, {Rix}, {Rodler}, {Rodr{\'\i}guez}, {Rodr{\'\i}guez-L{\'o}pez},
  {Rodr{\'\i}guez Trinidad}, {Rohloff}, {Rosich}, {Sadegi},
  {S{\'a}nchez-Blanco}, {S{\'a}nchez Carrasco}, {S{\'a}nchez-L{\'o}pez},
  {Sanz-Forcada}, {Sarkis}, {Sarmiento}, {Sch{\"a}fer}, {Schmitt}, {Schiller},
  {Schweitzer}, {Solano}, {Stahl}, {Strachan}, {St{\"u}rmer}, {Su{\'a}rez},
  {Tabernero}, {Tala}, {Trifonov}, {Tulloch}, {Ulbrich}, {Veredas}, {Vico
  Linares}, {Vilardell}, {Wagner}, {Winkler}, {Wolthoff}, {Xu}, {Yan}, \&
  {Zapatero Osorio}}]{rein18}
{Reiners}, A., {Zechmeister}, M., {Caballero}, J.~A., {et~al.} 2018, \aap, 612,
  A49

\bibitem[{{Ribas} {et~al.}(2023){Ribas}, {Reiners}, {Zechmeister}, {Caballero},
  {Morales}, {Sabotta}, {Baroch}, {Amado}, {Quirrenbach}, {Abril}, {Aceituno},
  {Anglada-Escud{\'e}}, {Azzaro}, {Barrado}, {B{\'e}jar}, {Ben{\'\i}tez de
  Haro}, {Bergond}, {Bluhm}, {Calvo Ortega}, {Cardona Guill{\'e}n},
  {Chaturvedi}, {Cifuentes}, {Colom{\'e}}, {Cont}, {Cort{\'e}s-Contreras},
  {Czesla}, {D{\'\i}ez-Alonso}, {Dreizler}, {Duque-Arribas}, {Espinoza},
  {Fern{\'a}ndez}, {Fuhrmeister}, {Galad{\'\i}-Enr{\'\i}quez},
  {Garc{\'\i}a-L{\'o}pez}, {Gonz{\'a}lez-{\'A}lvarez}, {Gonz{\'a}lez
  Hern{\'a}ndez}, {Guenther}, {de Guindos}, {Hatzes}, {Henning}, {Herrero},
  {Hintz}, {Huelmo}, {Jeffers}, {Johnson}, {de Juan}, {Kaminski}, {Kemmer},
  {Khaimova}, {Khalafinejad}, {Kossakowski}, {K{\"u}rster}, {Labarga},
  {Lafarga}, {Lalitha}, {Lamp{\'o}n}, {Lillo-Box}, {Lodieu}, {L{\'o}pez
  Gonz{\'a}lez}, {L{\'o}pez-Puertas}, {Luque}, {Mag{\'a}n}, {Mancini},
  {Marfil}, {Mart{\'\i}n}, {Mart{\'\i}n-Ruiz}, {Molaverdikhani}, {Montes},
  {Nagel}, {Nortmann}, {Nowak}, {Pall{\'e}}, {Passegger}, {Pavlov}, {Pedraz},
  {Perdelwitz}, {Perger}, {Ram{\'o}n-Ballesta}, {Reffert}, {Revilla},
  {Rodr{\'\i}guez}, {Rodr{\'\i}guez-L{\'o}pez}, {Sadegi}, {S{\'a}nchez
  Carrasco}, {S{\'a}nchez-L{\'o}pez}, {Sanz-Forcada}, {Sch{\"a}fer},
  {Schlecker}, {Schmitt}, {Sch{\"o}fer}, {Schweitzer}, {Seifert}, {Shan},
  {Skrzypinski}, {Solano}, {Stahl}, {Stangret}, {Stock}, {St{\"u}rmer},
  {Tabernero}, {Tal-Or}, {Trifonov}, {Vanaverbeke}, {Yan}, \& {Zapatero
  Osorio}}]{rib23}
{Ribas}, I., {Reiners}, A., {Zechmeister}, M., {et~al.} 2023, \aap, 670, A139

\bibitem[{{Ricker} {et~al.}(2015){Ricker}, {Winn}, {Vanderspek}, {Latham},
  {Bakos}, {Bean}, {Berta-Thompson}, {Brown}, {Buchhave}, {Butler}, {Butler},
  {Chaplin}, {Charbonneau}, {Christensen-Dalsgaard}, {Clampin}, {Deming},
  {Doty}, {De Lee}, {Dressing}, {Dunham}, {Endl}, {Fressin}, {Ge}, {Henning},
  {Holman}, {Howard}, {Ida}, {Jenkins}, {Jernigan}, {Johnson}, {Kaltenegger},
  {Kawai}, {Kjeldsen}, {Laughlin}, {Levine}, {Lin}, {Lissauer}, {MacQueen},
  {Marcy}, {McCullough}, {Morton}, {Narita}, {Paegert}, {Palle}, {Pepe},
  {Pepper}, {Quirrenbach}, {Rinehart}, {Sasselov}, {Sato}, {Seager},
  {Sozzetti}, {Stassun}, {Sullivan}, {Szentgyorgyi}, {Torres}, {Udry}, \&
  {Villasenor}}]{ric15}
{Ricker}, G.~R., {Winn}, J.~N., {Vanderspek}, R., {et~al.} 2015, Journal of
  Astronomical Telescopes, Instruments, and Systems, 1, 014003

\bibitem[{{Sanz-Forcada} {et~al.}(2025){Sanz-Forcada}, {L{\'o}pez-Puertas},
  {Lamp{\'o}n}, {Czesla}, {Nortmann}, {Caballero}, {Zapatero Osorio}, {Amado},
  {Murgas}, {Orell-Miquel}, {Pall{\'e}}, {Quirrenbach}, {Reiners}, {Ribas},
  {S{\'a}nchez-L{\'o}pez}, \& {Solano}}]{san25}
{Sanz-Forcada}, J., {L{\'o}pez-Puertas}, M., {Lamp{\'o}n}, M., {et~al.} 2025,
  \aap, 693, A285

\bibitem[{{Sanz-Forcada} {et~al.}(2011){Sanz-Forcada}, {Micela}, {Ribas},
  {Pollock}, {Eiroa}, {Velasco}, {Solano}, \&
  {Garc{\'{\i}}a-{\'A}lvarez}}]{san11}
{Sanz-Forcada}, J., {Micela}, G., {Ribas}, I., {et~al.} 2011, \aap, 532, A6

\bibitem[{{Schweitzer} {et~al.}(2019){Schweitzer}, {Passegger}, {Cifuentes},
  {B{\'e}jar}, {Cort{\'e}s-Contreras}, {Caballero}, {del Burgo}, {Czesla},
  {K{\"u}rster}, {Montes}, {Zapatero Osorio}, {Ribas}, {Reiners},
  {Quirrenbach}, {Amado}, {Aceituno}, {Anglada-Escud{\'e}}, {Bauer},
  {Dreizler}, {Jeffers}, {Guenther}, {Henning}, {Kaminski}, {Lafarga},
  {Marfil}, {Morales}, {Schmitt}, {Seifert}, {Solano}, {Tabernero}, \&
  {Zechmeister}}]{sch19}
{Schweitzer}, A., {Passegger}, V.~M., {Cifuentes}, C., {et~al.} 2019, \aap,
  625, A68

\bibitem[{{Shappee} {et~al.}(2014){Shappee}, {Prieto}, {Grupe}, {Kochanek},
  {Stanek}, {De Rosa}, {Mathur}, {Zu}, {Peterson}, {Pogge}, {Komossa}, {Im},
  {Jencson}, {Holoien}, {Basu}, {Beacom}, {Szczygie{\l}}, {Brimacombe},
  {Adams}, {Campillay}, {Choi}, {Contreras}, {Dietrich}, {Dubberley},
  {Elphick}, {Foale}, {Giustini}, {Gonzalez}, {Hawkins}, {Howell}, {Hsiao},
  {Koss}, {Leighly}, {Morrell}, {Mudd}, {Mullins}, {Nugent}, {Parrent},
  {Phillips}, {Pojmanski}, {Rosing}, {Ross}, {Sand}, {Terndrup}, {Valenti},
  {Walker}, \& {Yoon}}]{asassn}
{Shappee}, B.~J., {Prieto}, J.~L., {Grupe}, D., {et~al.} 2014, \apj, 788, 48

\bibitem[{{Shaw} {et~al.}(2025){Shaw}, {Weiss}, {Agol}, {Collins}, {Barkaoui},
  {Watkins}, {Schwarz}, {Relles}, {Stockdale}, {Kielkopf}, {Rodriguez
  Frustaglia}, {Bieryla}, {Gregorio}, {Mitchem}, {Linnenkohl}, {Popowicz},
  {Narita}, {Fukui}, {Gillon}, {Sefako}, {Shporer}, {Lark}, {Heying}, {Khan},
  {Chen}, {Carden}, {Terndrup}, {Taylor}, {Crocker}, {Ballard}, \&
  {Fabrycky}}]{shaw25}
{Shaw}, D.~E., {Weiss}, L.~M., {Agol}, E., {et~al.} 2025, \aj, 170, 146

\bibitem[{{Skrutskie} {et~al.}(2006){Skrutskie}, {Cutri}, {Stiening},
  {Weinberg}, {Schneider}, {Carpenter}, {Beichman}, {Capps}, {Chester},
  {Elias}, {Huchra}, {Liebert}, {Lonsdale}, {Monet}, {Price}, {Seitzer},
  {Jarrett}, {Kirkpatrick}, {Gizis}, {Howard}, {Evans}, {Fowler}, {Fullmer},
  {Hurt}, {Light}, {Kopan}, {Marsh}, {McCallon}, {Tam}, {Van Dyk}, \&
  {Wheelock}}]{skr06}
{Skrutskie}, M.~F., {Cutri}, R.~M., {Stiening}, R., {et~al.} 2006, \aj, 131,
  1163

\bibitem[{{Smith} {et~al.}(2012){Smith}, {Stumpe}, {Van Cleve}, {Jenkins},
  {Barclay}, {Fanelli}, {Girouard}, {Kolodziejczak}, {McCauliff}, {Morris}, \&
  {Twicken}}]{smi12}
{Smith}, J.~C., {Stumpe}, M.~C., {Van Cleve}, J.~E., {et~al.} 2012, \pasp, 124,
  1000

\bibitem[{{Stassun} {et~al.}(2018){Stassun}, {Oelkers}, {Pepper}, {Paegert},
  {De Lee}, {Torres}, {Latham}, {Charpinet}, {Dressing}, {Huber}, {Kane},
  {L{\'e}pine}, {Mann}, {Muirhead}, {Rojas-Ayala}, {Silvotti}, {Fleming},
  {Levine}, \& {Plavchan}}]{sta18}
{Stassun}, K.~G., {Oelkers}, R.~J., {Pepper}, J., {et~al.} 2018, \aj, 156, 102

\bibitem[{{Stock} {et~al.}(2023){Stock}, {Kemmer}, {Kossakowski}, {Sabotta},
  {Reffert}, \& {Quirrenbach}}]{sto23}
{Stock}, S., {Kemmer}, J., {Kossakowski}, D., {et~al.} 2023, \aap, 674, A108

\bibitem[{{Stock} {et~al.}(2020){Stock}, {Kemmer}, {Reffert}, {Trifonov},
  {Kaminski}, {Dreizler}, {Quirrenbach}, {Caballero}, {Reiners}, {Jeffers},
  {Anglada-Escud{\'e}}, {Ribas}, {Amado}, {Barrado}, {Barnes}, {Bauer},
  {Berdi{\~n}as}, {B{\'e}jar}, {Coleman}, {Cort{\'e}s-Contreras},
  {D{\'\i}ez-Alonso}, {Dom{\'\i}nguez-Fern{\'a}ndez}, {Espinoza}, {Haswell},
  {Hatzes}, {Henning}, {Jenkins}, {Jones}, {Kossakowski}, {K{\"u}rster},
  {Lafarga}, {Lee}, {L{\'o}pez Gonz{\'a}lez}, {Montes}, {Morales}, {Morales},
  {Pall{\'e}}, {Pedraz}, {Rodr{\'\i}guez}, {Rodr{\'\i}guez-L{\'o}pez}, \&
  {Zechmeister}}]{sto20}
{Stock}, S., {Kemmer}, J., {Reffert}, S., {et~al.} 2020, \aap, 636, A119

\bibitem[{{Stumpe} {et~al.}(2014){Stumpe}, {Smith}, {Catanzarite}, {Van Cleve},
  {Jenkins}, {Twicken}, \& {Girouard}}]{stu14}
{Stumpe}, M.~C., {Smith}, J.~C., {Catanzarite}, J.~H., {et~al.} 2014, \pasp,
  126, 100

\bibitem[{{Stumpe} {et~al.}(2012){Stumpe}, {Smith}, {Van Cleve}, {Twicken},
  {Barclay}, {Fanelli}, {Girouard}, {Jenkins}, {Kolodziejczak}, {McCauliff}, \&
  {Morris}}]{stu12}
{Stumpe}, M.~C., {Smith}, J.~C., {Van Cleve}, J.~E., {et~al.} 2012, \pasp, 124,
  985

\bibitem[{{Sun} {et~al.}(2019){Sun}, {Ioannidis}, {Gu}, {Schmitt}, {Wang}, \&
  {Kouwenhoven}}]{sun19}
{Sun}, L., {Ioannidis}, P., {Gu}, S., {et~al.} 2019, \aap, 624, A15

\bibitem[{{Tabernero} {et~al.}(2022){Tabernero}, {Marfil}, {Montes}, \&
  {Gonz{\'a}lez Hern{\'a}ndez}}]{tab22}
{Tabernero}, H.~M., {Marfil}, E., {Montes}, D., \& {Gonz{\'a}lez
  Hern{\'a}ndez}, J.~I. 2022, \aap, 657, A66

\bibitem[{{Trifonov} {et~al.}(2021){Trifonov}, {Caballero}, {Morales},
  {Seifahrt}, {Ribas}, {Reiners}, {Bean}, {Luque}, {Parviainen}, {Pall{\'e}},
  {Stock}, {Zechmeister}, {Amado}, {Anglada-Escud{\'e}}, {Azzaro}, {Barclay},
  {B{\'e}jar}, {Bluhm}, {Casasayas-Barris}, {Cifuentes}, {Collins}, {Collins},
  {Cort{\'e}s-Contreras}, {de Leon}, {Dreizler}, {Dressing}, {Esparza-Borges},
  {Espinoza}, {Fausnaugh}, {Fukui}, {Hatzes}, {Hellier}, {Henning}, {Henze},
  {Herrero}, {Jeffers}, {Jenkins}, {Jensen}, {Kaminski}, {Kasper},
  {Kossakowski}, {K{\"u}rster}, {Lafarga}, {Latham}, {Mann}, {Molaverdikhani},
  {Montes}, {Montet}, {Murgas}, {Narita}, {Oshagh}, {Passegger}, {Pollacco},
  {Quinn}, {Quirrenbach}, {Ricker}, {Rodr{\'\i}guez L{\'o}pez}, {Sanz-Forcada},
  {Schwarz}, {Schweitzer}, {Seager}, {Shporer}, {Stangret}, {St{\"u}rmer},
  {Tan}, {Tenenbaum}, {Twicken}, {Vanderspek}, \& {Winn}}]{tri21}
{Trifonov}, T., {Caballero}, J.~A., {Morales}, J.~C., {et~al.} 2021, Science,
  371, 1038

\bibitem[{{Trotta}(2008)}]{tro08}
{Trotta}, R. 2008, Contemporary Physics, 49, 71

\bibitem[{{Twicken} {et~al.}(2018){Twicken}, {Catanzarite}, {Clarke},
  {Girouard}, {Jenkins}, {Klaus}, {Li}, {McCauliff}, {Seader}, {Tenenbaum},
  {Wohler}, {Bryson}, {Burke}, {Caldwell}, {Haas}, {Henze}, \&
  {Sanderfer}}]{twi18}
{Twicken}, J.~D., {Catanzarite}, J.~H., {Clarke}, B.~D., {et~al.} 2018, \pasp,
  130, 064502

\bibitem[{{Twicken} {et~al.}(2010){Twicken}, {Clarke}, {Bryson}, {Tenenbaum},
  {Wu}, {Jenkins}, {Girouard}, \& {Klaus}}]{twi10}
{Twicken}, J.~D., {Clarke}, B.~D., {Bryson}, S.~T., {et~al.} 2010, in
  \procspie, Vol. 7740, Software and Cyberinfrastructure for Astronomy, 774023

\bibitem[{{Vissapragada} {et~al.}(2020){Vissapragada}, {Jontof-Hutter},
  {Shporer}, {Knutson}, {Liu}, {Thorngren}, {Lee}, {Chachan}, {Mawet},
  {Millar-Blanchaer}, {Nilsson}, {Tinyanont}, {Vasisht}, \& {Wright}}]{vis20}
{Vissapragada}, S., {Jontof-Hutter}, D., {Shporer}, A., {et~al.} 2020, \aj,
  159, 108

\bibitem[{{Wright} {et~al.}(2011){Wright}, {Drake}, {Mamajek}, \&
  {Henry}}]{wri11}
{Wright}, N.~J., {Drake}, J.~J., {Mamajek}, E.~E., \& {Henry}, G.~W. 2011,
  \apj, 743, 48

\bibitem[{{Zechmeister} {et~al.}(2018){Zechmeister}, {Reiners}, {Amado},
  {Azzaro}, {Bauer}, {B{\'e}jar}, {Caballero}, {Guenther}, {Hagen}, {Jeffers},
  {Kaminski}, {K{\"u}rster}, {Launhardt}, {Montes}, {Morales}, {Quirrenbach},
  {Reffert}, {Ribas}, {Seifert}, {Tal-Or}, \& {Wolthoff}}]{serval}
{Zechmeister}, M., {Reiners}, A., {Amado}, P.~J., {et~al.} 2018, \aap, 609, A12

\bibitem[{{Zeng} {et~al.}(2016){Zeng}, {Sasselov}, \& {Jacobsen}}]{zen16}
{Zeng}, L., {Sasselov}, D.~D., \& {Jacobsen}, S.~B. 2016, \apj, 819, 127

\end{thebibliography}
\end{document}